\documentclass[manuscript]{aastex631}
\usepackage{natbib}
\usepackage{multirow}
\usepackage{multirow}
\usepackage{graphicx}
\usepackage{verbatim}%%note
\graphicspath{{./}{figures/}}

\begin{document}

\title{\textit{Insight}-HXMT dedicated 33-day observation of SGR J1935+2154 
\uppercase\expandafter{\romannumeral2}. Burst Spectral Catalog}

%A new page starts with each \ include section.
\author{Ce Cai\textsuperscript}
\affiliation{Key Laboratory of Particle Astrophysics, Institute of High Energy Physics, Chinese Academy of Sciences, 19B Yuquan Road, Beijing 100049, China}
\affiliation{University of Chinese Academy of Sciences, Chinese Academy of Sciences, Beijing 100049, China}

\author{Shao-Lin Xiong\textsuperscript{*}}
\email{xiongsl@ihep.ac.cn}
\affiliation{Key Laboratory of Particle Astrophysics, Institute of High Energy Physics, Chinese Academy of Sciences, 19B Yuquan Road, Beijing 100049, China}

\author{Lin Lin\textsuperscript{*}}
\email{llin@bnu.edu.cn}
\affiliation{Department of Astronomy, Beijing Normal University, Beijing 100088, People’s Republic of China}

\author{Cheng-Kui Li\textsuperscript{*}}
\email{lick@ihep.ac.cn}
\affiliation{Key Laboratory of Particle Astrophysics, Institute of High Energy Physics, Chinese Academy of Sciences, 19B Yuquan Road, Beijing 100049, China}

\author{Shuang-Nan Zhang}
\affiliation{Key Laboratory of Particle Astrophysics, Institute of High Energy Physics, Chinese Academy of Sciences, 19B Yuquan Road, Beijing 100049, China}
\affiliation{University of Chinese Academy of Sciences, Chinese Academy of Sciences, Beijing 100049, China}

\author{Wang-Chne Xue\textsuperscript}
\affiliation{Key Laboratory of Particle Astrophysics, Institute of High Energy Physics, Chinese Academy of Sciences, 19B Yuquan Road, Beijing 100049, China}
\affiliation{University of Chinese Academy of Sciences, Chinese Academy of Sciences, Beijing 100049, China}

\author{You-Li Tuo}
\affiliation{Key Laboratory of Particle Astrophysics, Institute of High Energy Physics, Chinese Academy of Sciences, 19B Yuquan Road, Beijing 100049, China}

\author{Xiao-Bo Li}
\affiliation{Key Laboratory of Particle Astrophysics, Institute of High Energy Physics, Chinese Academy of Sciences, 19B Yuquan Road, Beijing 100049, China}

\author{Ming-Yu Ge}
\affiliation{Key Laboratory of Particle Astrophysics, Institute of High Energy Physics, Chinese Academy of Sciences, 19B Yuquan Road, Beijing 100049, China}

\author{Hai-Sheng Zhao}
\affiliation{Key Laboratory of Particle Astrophysics, Institute of High Energy Physics, Chinese Academy of Sciences, 19B Yuquan Road, Beijing 100049, China}

\author{Li-Ming Song}
\affiliation{Key Laboratory of Particle Astrophysics, Institute of High Energy Physics, Chinese Academy of Sciences, 19B Yuquan Road, Beijing 100049, China}
\affiliation{University of Chinese Academy of Sciences, Chinese Academy of Sciences, Beijing 100049, China}

\author{Fang-Jun Lu}
\affiliation{Key Laboratory of Particle Astrophysics, Institute of High Energy Physics, Chinese Academy of Sciences, 19B Yuquan Road, Beijing 100049, China}
\affiliation{University of Chinese Academy of Sciences, Chinese Academy of Sciences, Beijing 100049, China}

\author{Shu Zhang}
\affiliation{Key Laboratory of Particle Astrophysics, Institute of High Energy Physics, Chinese Academy of Sciences, 19B Yuquan Road, Beijing 100049, China}

\author{Qing-Xin Li}
\affiliation{Department of Astronomy, Beijing Normal University, Beijing 100088, People’s Republic of China}

\author{Shuo Xiao}
\affiliation{Key Laboratory of Particle Astrophysics, Institute of High Energy Physics, Chinese Academy of Sciences, 19B Yuquan Road, Beijing 100049, China}
\affiliation{University of Chinese Academy of Sciences, Chinese Academy of Sciences, Beijing 100049, China}

\author{Zhi-Wei Guo}
\affiliation{Key Laboratory of Particle Astrophysics, Institute of High Energy Physics, Chinese Academy of Sciences, 19B Yuquan Road, Beijing 100049, China}
\affiliation{College of physics Sciences Technology, Hebei University, No. 180 Wusi Dong Road, Lian Chi District, Baoding City, Hebei Province 071002, China}

\author{Sheng-Lun Xie}
\affiliation{Key Laboratory of Particle Astrophysics, Institute of High Energy Physics, Chinese Academy of Sciences, 19B Yuquan Road, Beijing 100049, China}
\affiliation{School of Physical Science and Technology, Central China Normal University, Wuhan 430097, China}

\author{Yan-Qiu Zhang}
\affiliation{Key Laboratory of Particle Astrophysics, Institute of High Energy Physics, Chinese Academy of Sciences, 19B Yuquan Road, Beijing 100049, China}
\affiliation{University of Chinese Academy of Sciences, Chinese Academy of Sciences, Beijing 100049, China}

\author{Qi-Bin Yi}
\affiliation{Key Laboratory of Particle Astrophysics, Institute of High Energy Physics, Chinese Academy of Sciences, 19B Yuquan Road, Beijing 100049, China}
\affiliation{School of Physics and Optoelectronics, Xiangtan University, Xiangtan 411105, Hunan, China}

\author{Yi Zhao}
\affiliation{Key Laboratory of Particle Astrophysics, Institute of High Energy Physics, Chinese Academy of Sciences, 19B Yuquan Road, Beijing 100049, China}
\affiliation{Department of Astronomy, Beijing Normal University, Beijing 100088, People’s Republic of China}

\author{Zhen Zhang}
\affiliation{Key Laboratory of Particle Astrophysics, Institute of High Energy Physics, Chinese Academy of Sciences, 19B Yuquan Road, Beijing 100049, China}

\author{Jia-Cong Liu}
\affiliation{Key Laboratory of Particle Astrophysics, Institute of High Energy Physics, Chinese Academy of Sciences, 19B Yuquan Road, Beijing 100049, China}
\affiliation{University of Chinese Academy of Sciences, Chinese Academy of Sciences, Beijing 100049, China}

\author{Chao Zheng}
\affiliation{Key Laboratory of Particle Astrophysics, Institute of High Energy Physics, Chinese Academy of Sciences, 19B Yuquan Road, Beijing 100049, China}
\affiliation{University of Chinese Academy of Sciences, Chinese Academy of Sciences, Beijing 100049, China}

\author{Ping Wang}
\affiliation{Key Laboratory of Particle Astrophysics, Institute of High Energy Physics, Chinese Academy of Sciences, 19B Yuquan Road, Beijing 100049, China}

\begin{abstract}
%The \textit{Insight}-Hard X-ray Modulation Telescope (\textit{Insight}-HXMT) is China's first X-ray astronomy satellite, launched on June 15, 2017. 
Since April 28, 2020, \textit{Insight}-HXMT has implemented a dedicated observation on the magnetar SGR J1935+2154. Thanks to the wide energy band (1$-$250 keV) and high sensitivity of \textit{Insight}-HXMT, we obtained 75 bursts from SGR J1935+2154 during a month-long activity episode after the emission of FRB 200428.
%Here, we report the detailed time-integrated spectral analysis of these bursts from SGR J1935+2154. We provide spectral parameters for various models, fluences and flux for bursts in our sample. 
Here, we report the detailed time-integrated spectral analysis of these bursts and the statistical distribution of the spectral parameters.
We find that for $\sim15\%$ (11/75) of SGR J1935+2154 bursts, the CPL model is preferred, and most of them occurred in the later part of this active epoch.
In the cumulative fluence distribution, we find that the fluence of bursts in our sample is about an order of magnitude weaker than that of \textit{Fermi}/GBM, but follow the same power law distribution. %as the results of \textit{Fermi}/GBM. 
Finally, we find a burst with similar peak energy to the time-integrated spectrum of the X-ray burst associated with FRB 200428 (FRB 200428-Associated Burst), but the low energy index is harder. 
\end{abstract}

%\keywords{Magnetars (992)}

\section{Introduction}
%要强调这次连续33天监测的价值
Magnetars are a group of neutron stars with extremely strong magnetic fields ($\sim$ 10$^{14}$ $-$ 10$^{15}$ G) \citep{1995MNRAS.275..255T}. They are usually identified with long spin periods ($\sim2-12$ s ) and fast spin-down rates ($\sim$ 10$^{-13}$ $-$ 10$^{-11}$ s s$^{-1}$) \citep{1998Natur.393..235K,2014ApJS..212....6O}. Some special emission phenomena show strong connections with magnetars, including Soft Gamma Repeaters (SGRs) and Anomaly X-ray Pulsars (AXPs). In fact, those unpredictable soft gamma-ray or hard X-ray bursts have been detected from more than two-thirds of the magnetar population \citep{2014ApJS..212....6O} \footnote{http://www.physics.mcgill.ca/~pulsar/magnetar/main.html}. Based on the duration and the peak luminosity, these soft gamma-ray bursts can be classified into three categories. Giant flares are the most energetic and rarest events. They start with a sub-second hard spike which is brighter than 10$^{44}$ erg s$^{-1}$, followed by the emission decay lasting for several minutes \citep{1979Natur.282..587M,1999Natur.397...41H,2005Natur.434.1107P,2021Natur.589..207R}. Intermediate flares present broader peaks (1$-$40 s) with the peak luminosity of 10$^{41}$ $-$ 10$^{43}$ erg s$^{-1}$ \citep{2015}. Short bursts are the most common ones. Typically they last for $\sim$ 0.1$-$1 s and have a peak luminosity of 10$^{39}$ $-$ 10$^{41}$ erg s$^{-1}$ \citep{2015}.

As important clues for understanding the emission mechanism, the spectral properties of magnetar bursts have been extensively studied. Our knowledge of the burst spectrum goes deeper thanks to X-/gamma-ray telescopes which are sensitive to a broad energy range. Using \textit{Fermi}/GBM data (8-200 keV), \citet{2011ApJ...739...87L} found the cutoff power law model (COMPT) and the sum of two blackbody functions (BB+BB) can describe the burst spectrum equally well for magnetar SGR J0501+4501. Similar results are also reported for other magnetars (e.g., \cite{2012ApJ...749..122V,2015ApJS..218...11C}). %(e.g. van der Horst 2012, gbmcatalog, \textcolor{red}{cite?}).
However, these two spectral models have distinct implications: The BB+BB model indicates that photons come from thermalized plasma, while COMPT points to the opposite direction. Occasionally, some bursts are detected by both soft X-ray and gamma-ray telescopes, enabling a joint broader band spectral analysis (e.g. 1-200 keV) which statistically prefer the BB+BB model in most cases \citep{2008ApJ...685.1114I, 2012Lin}. Unfortunately, the vast majority of magnetar bursts were not observed in such a broad energy range, thus whether their spectra are thermal or non-thermal is somewhat debatable.

%Magnetars are neutron stars with extremely strong magnetic fields of $\sim$ 10$^{14}$ $-$ 10$^{15}$ G \citep{1995MNRAS.275..255T,1998Natur.393..235K}. They have relatively long rotational periods (P $\sim$ 2 $-$ 12 s ) and large secular spin-down rate (\.P) of  $\sim$ 10$^{-13}$ $-$ 10$^{-11}$ s s$^{-1}$. The number of discovered magnetars is $\sim$30, including six candidates \footnote{http://www.physics.mcgill.ca/~pulsar/magnetar/main.html}. Bursts from more than two-thirds of the magnetar population have been detected by  space-based hard X-ray/soft gamma-ray instruments to-date \citep{2014ApJS..212....6O}. Magnetar bursts appear in broad duration and energetic varieties, falling into three types: Giant Flares with a duration of several minutes and energies in excess of 10$^{44}$ erg s$^{-1}$ \citep{1979Natur.282..587M,1999Natur.397...41H,2005Natur.434.1107P}; intermediate flares with a duration of 1$-$40 s and peak luminosity of 10$^{41}$ $-$ 10$^{43}$ erg s$^{-1}$; short bursts with typical duration of $\sim$ 0.1 s$-$1 s and peak luminosity of 10$^{39}$ $-$ 10$^{41}$ erg s$^{-1}$ \citep{2015}. 
%\textcolor{red}{\cite{2008ApJ...685.1114I,2011ApJ...739...87L,2012Lin} found complex models (i.e., Comptonization model or the sum of two blackbody functions (BB+BB)) are preferred in fitting the spectra over broad energy ranges.}

SGR J1935+2154 %as a Galactic magnetar, 
was discovered by the \textit{Swift}/Burst Alert Telescope (BAT) on 2014 July 5 \citep{2014GCN.16520....1S}. It was confirmed as a magnetar with a spin period P = 3.25 s and spin-down rate  \.P = 1.43 $\times$ 10$^{-11}$ s s$^{-1}$ \citep{2016MNRAS.457.3448I}. 
%, which was measured by an extensive observational campaign with \textit{Chandra} and \textit{XMM-Newton} \citep{2016MNRAS.457.3448I}. 
Since its discovery, SGR J1935+2154 has experienced at least seven outbursts in 2014, 2015, 2016, 2019, 2020 and 2021, respectively \citep{2017ApJ...847...85Y, 2020ApJ...893..156L, Lin_2020}, making it the most prolific magnetar. During its active episode in 2020, a fast radio burst (FRB 200428) was detected from the general direction of SGR J1935+2154 by CHIME and STARE2 \citep{collaboration2020bright,2020Bochenek}. At the dispersion corrected burst time of FRB~200428, a hard X-ray burst from SGR J1935+2154 was detected by \textit{Insight}-Hard X-ray Modulation Telescope (\textit{Insight}-HXMT hereafter) \citep{2021NatAs...5..378L}, INTEGRAL \citep{Mereghetti_2020}, Konus-Wind \citep{2021NatAs...5..372R} and AGILE \citep{tavani2020xray}. Thanks to the accurate localization provided by \textit{Insight}-HXMT and INTEGRAL \citep{2021NatAs...5..378L, Mereghetti_2020}, this hard X-ray burst provides an unambiguous evidence that FRB~200428 originates from the Galactic magnetar SGR J1935+2154. Moreover, we first suggested that these two narrow peaks in this hard X-ray burst aligned well with these two radio pulses of FRB~200428 is probably the high energy counterpart of FRB~200428 \citep{2021NatAs...5..378L}. 

%During the recent activation of 2020, a X-ray burst from SGR J1935+2154 was detected on April 28$^{th}$ by International Gamma-Ray Astrophysics Laboratory (\textit{INTEGRAL}), Konus-Wind, \textit{Insight}-Hard X-ray Modulation Telescope (\textit{Insight}-HXMT) and AGILE \citep{Mereghetti_2020, 2021NatAs...5..372R, 2021NatAs...5..378L, tavani2020xray}, which is associated with a Fast Radio Burst (FRB), FRB 200428 \citep{collaboration2020bright,2020Bochenek}, lending support to the magnetospheric models of FRBs. The spectral characteristics of this x-ray burst were unusual with a harder, non-thermal profile, compared to other bursts from previous activations, which was demonstrated by \cite{2021NatAs.tmp...33Y}.

On the contrary, FAST unveiled a rare connection between magnetar X-ray bursts and FRBs by placing the most stringent radio upper limit to 29 X-ray bursts from SGR J1935+2154 shortly before FRB~200428 \citep{2020Natur.587...63L}. The reason of missing radio emission of magnetar X-ray bursts is unclear. It can be intrinsic (e.g. the X-ray burst associated with FRB is atypical) or due to some selection effects (e.g. beaming effects or narrow bandwidth of FAST) or both \citep{2020Natur.587...63L}. Nevertheless, using \textit{Insight}-HXMT data in the wide energy range of 1 to 250 keV, \citet{2021NatAs...5..378L} confirmed that the X-ray burst associated with FRB~200428 (denoted as FRB~200428-Associated Burst hereafter) prefers a non-thermal origin, while the spectrum of typical magnetar burst is more thermalized or curved in the similar energy range \citep{2012Lin,2021NatAs.tmp...33Y}.
%(Lin 2012,Younes 2021, \textcolor{red}{cite}). 
Therefore, multi-wavelength observations of active magnetars are crucial to expand the sample of the FRB and magnetar X-ray burst association pairs and to deepen the understanding of the physics behind the SGR-FRB connection.

The discovery of the X-ray burst associated with FRB 200428 demonstrated that \textit{Insight}-HXMT, with a broad energy coverage (1-250 keV) and high sensitivity, is very powerful to explore the SGR-FRB relation and constrain the emission properties of magnetar bursts. It also motivated a dedicated long Target of Opportunity (ToO) observation to continuously monitor SGR J1935+2154 until its bursting activity ceased. Complimentary by other gamma-ray burst monitors (e.g., \textit{Fermi}/GBM), \textit{Insight}-HXMT provided the most complete monitoring coverage of SGR J1935+2154 bursts after the emission of FRB 200428. 

From this 33-day dedicated observation of \textit{Insight}-HXMT, we identified 75 bursts ($\sim 90\%$ of all bursts detected from SGR J1935+2154 during this period) in the exposure time of 1.65 Ms, as described in the first paper of this special series (\textcolor{red}{\cite{cai2022insighthxmt}}, hereafter paper \uppercase\expandafter{\romannumeral1}). This is an exceptional burst sample with a broad energy range (1-250 keV) which could enable detailed studies of the burst properties and the search for potential association between SGR bursts and FRB. 
%\footnote{\cite{cai2022insighthxmt}Cai et al. submitted}

In order to investigate how special the FRB 200428-Associated Burst is in terms of its non-thermal spectrum and whether SGR J1935+2154 is unique or not among the SGR population, we analyzed the spectra of the 75 bursts detected by \textit{Insight}-HXMT during this dedicated ToO observation, which are presented in detail in this paper.

\begin{figure*}
\centering
\begin{tabular}{cc}
\includegraphics[width=0.50\textwidth]{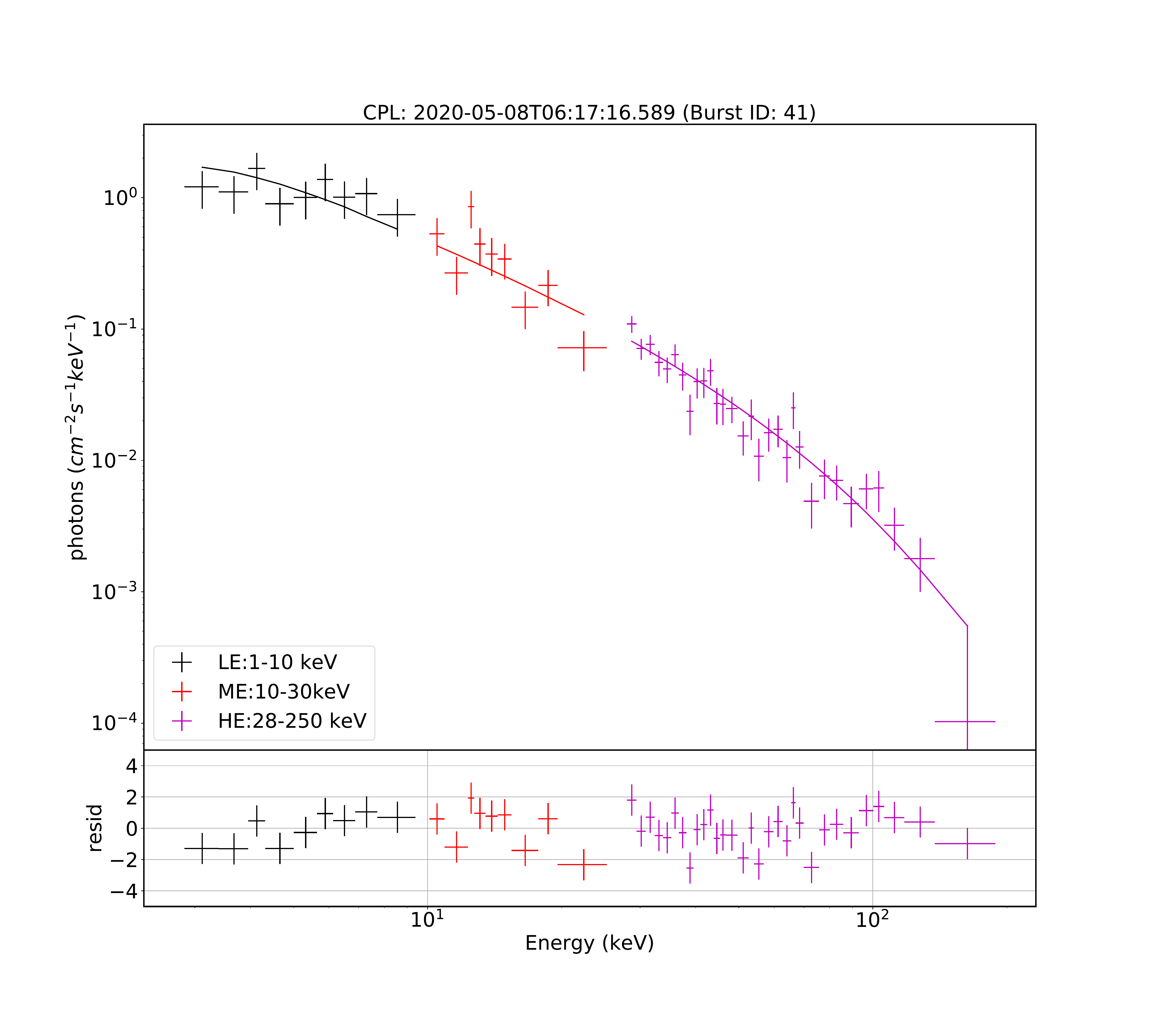} &
\includegraphics[width=0.50\textwidth]{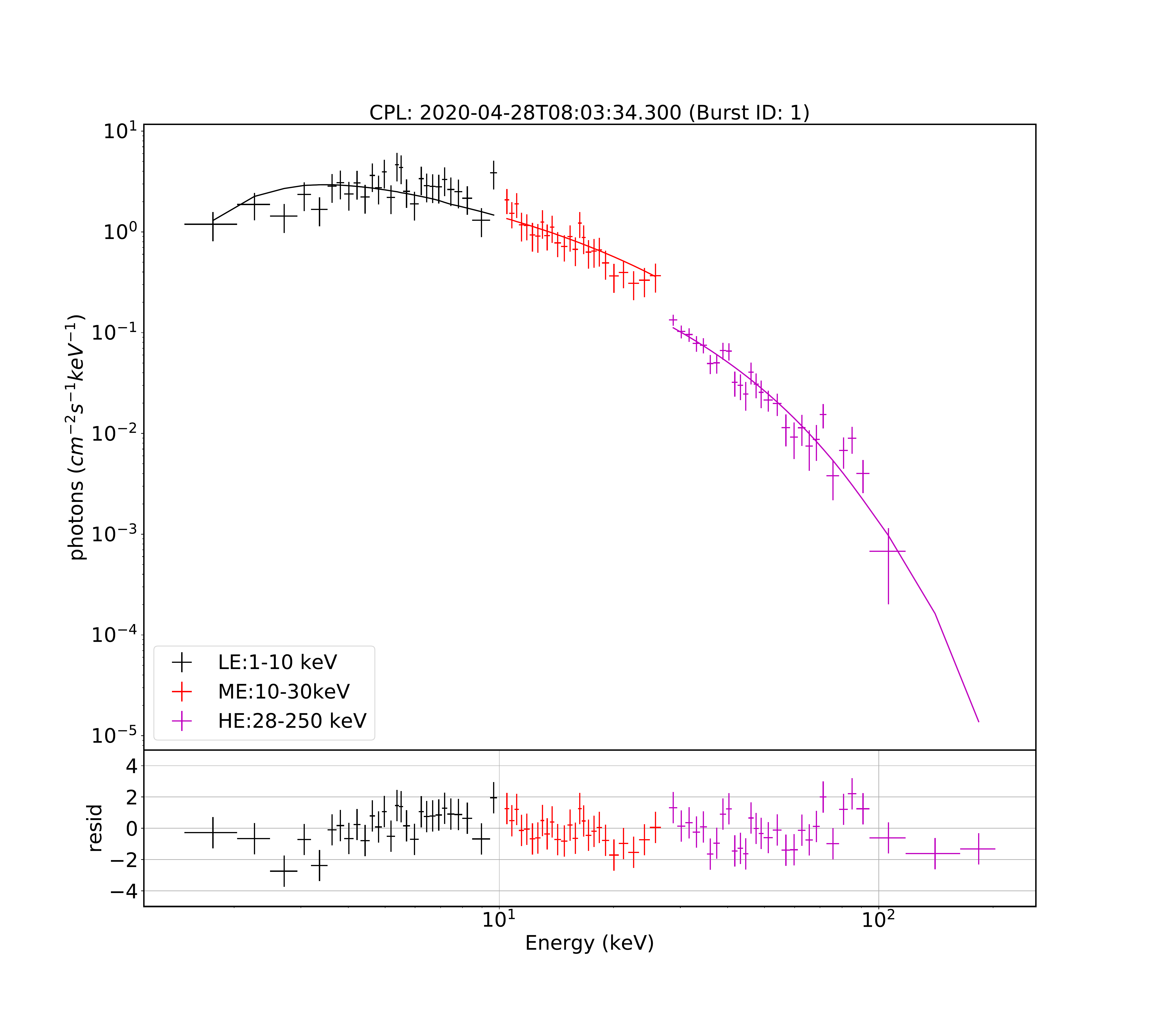}\\
\end{tabular}
\caption{\label{fig:SpecFit} The spectra of an unsaturated burst (left panel) and a saturated burst (right panel) from SGR J1935+2154 observed by \textit{Insight}-HXMT. These spectra are fitted with the CPL model in the energy band of 1$-$250 keV. The lower panels show the fit residuals. For the burst suffered saturation in HE telescope, the flux normalization parameter ($f$) of HE is allowed to vary while that of ME and LE are frozen to be 1 during the fit.}
%\textit{Left}: $wabs*cutoffpowerlaw$. \textit{Right}: $const*wabs*cutoffpowerlaw$.
\end{figure*}

\begin{table*}
\centering
\caption{Summary of spectral models employed in fitting SGR J1935+2154 burst spectra.}
\label{tab: model_Results}
\begin{tabular}{ p{3.7cm}<{\centering} p{2.4cm}<{\centering}  p{2.cm}<{\centering}  p{1.7cm}<{\centering} p{1.7cm}<{\centering} p{1.7cm}<{\centering} p{1.7cm}<{\centering}} 
\hline
\hline
Samples & Number of burst  &  CPL$^{\rm b}$  &  BB+BB$^{\rm b}$ & BB+PL$^{\rm b}$  & PL$^{\rm b}$ & BB$^{\rm b}$ \\  
\hline
Unsaturated & 68 &  23 (4) &  25 (3) &  18 (0)  & 33(18) & 19 (9) \\
\hline
Saturated  & 6 &  6 (6)  &  1 (0) &  0   & 0 & 0 \\
\hline
FRB 200428-Associated $^{\rm a}$ & 1 &  1 (1)  &  0  &  1 (1)  & 0 & 0 \\
\hline
All  & 75 &  30 (11)  &  26 (3) &  19 (1)  & 33 (18) & 19 (9) \\
\hline
\end{tabular}
\tablecomments{$^a$ The X-ray burst associated with FRB 200428. \\
$^b$ The number of bursts whose adequate (preferred) model is CPL, BB+BB, BB+PL, PL and BB respectively.\\}
%$^b$ The number of bursts that can be adequately fit with different models . \\
%$^c$ The number of fitting spectra with preferred models.}
\end{table*}

\section{Observations} \label{sec:ANALYSIS_METHOD}
\subsection{Instrument} \label{Instru}
As China's first X-ray astronomy satellite launched on June 15th, 2017, \textit{Insight}-HXMT \citep{Zhang:2019xpt,2018SPIE10699E..1UZ,LI202064} consists of three telescopes: the High Energy X-ray telescope (HE) covering the energies of 20$-$250 keV \citep{2019arXiv191004955L}, the Medium Energy X-ray telescope (ME) with energies of 5$-$30 keV \citep{2019arXiv191004451C} and the Low Energy X-ray telescope (LE) in 1$-$15 keV \citep{2019arXiv191008319C}. The time resolution of LE, ME and HE are 0.98 ms, 255 $\mu$s and less than 10 $\mu$s, respectively. The energy resolution of LE, ME and HE are about $\sim2.4\%$ at 5.9 keV, $\sim16.9\%$ at 17.8 keV and $\sim15\%$ at 60 keV \citep{2019arXiv191004955L,2019arXiv191004451C,2019arXiv191008319C}.
For HE, data from 17 NaI detectors (rather than CsI detector) are used to do spectral analysis, whereas the blind NaI detector is used for background modeling \citep{2019arXiv191004955L}.
%\textcolor{orange}{The HE is composed of 17 NaI detectors including XX with a small field of view (FOV), XX with a wide FOV and XX blind detector dedicated to measure the instrumental background. The CsI detectors of the HE are excluded in this work, because magentar bursts are soft.  }

\subsection{Burst Sample} \label{DataSample} 
About 13 hours after the first \textit{Fermi}/GBM and \textit{Swift}/BAT trigger of a burst from SGR J1935+2154 on 2020 April 27$^{th}$, \textit{Insight}-HXMT launched a 33-day dedicated ToO monitor campaign of this source. This observation covered from April 28th 07:14:51 to April 29th 12:02:36 and from April 30th 06:58:23 to June 1st 00:00:01 (all time quoted in this paper is in UTC). Using a multi-detector trigger method and a careful identification of fake bursts from instrumental effects and other sources (see paper \uppercase\expandafter{\romannumeral1} for details), we found 75 short bursts from SGR J1935+2154 (paper \uppercase\expandafter{\romannumeral1}). In this burst sample, the LE data are filtered out due to the bright Earth for 14 bursts, whereas both the LE and ME data are lost on board for 1 burst (bursts \#67). 7 bright bursts suffer from data saturation (i.e. data loss caused by huge amount of data in a short time period) in HE or LE, while ME does not have data saturation issue.

Because the majority of bursts, including the burst forests \citep{2021}, are concentrated in the first $\sim10$ hours of this episode \citep{Lin_2020}, which is before the \textit{Insight}-HXMT ToO observation, the number of bursts found in this \textit{Insight}-HXMT observation is not as large as that reported by \citet{Lin_2020} and \citet{2020NICER} which monitored the source in a more active episode with \textit{Fermi}/GBM and NICER respectively. However, \textit{Insight}-HXMT has a higher sensitivity and broader energy range coverage for hard X-ray bursts and thus provides a unique data set for this 33-day observation embracing FRB 200428. Among 75 bursts detected by \textit{Insight}-HXMT, only 7 bright bursts were jointly detected by \textit{Fermi}/GBM.
%We analyzed the data of 2020 outbursts observed by \textit{Insight}-HXMT with a dedicated and long Time of Opportunity (ToO) from 2020-04-28T07:14:51 UTC to 2020-04-29T12:02:36 UTC and from 2020-04-30T06:58:23 UTC to 2020-06-01T00:00:01 UTC.A multi-detector trigger method is used to search for magnetar-like short bursts in the event-by-event (EVT) data. The 75 bursts which trigger in the search pipeline are shown in paper \uppercase\expandafter{\romannumeral1}, including fourteen bursts of which the raw data are lost in LE due to bright earth and one burst of which the LE and ME raw data are lost on board. There are also some bursts in our catalog that suffered from saturation because of the extremely high flux. In summary, our samples comprise 75 bursts, including 7 saturated bursts and 68 unsaturated bursts.
%{\color{orange}[Are golden and silver required? same with saturated and non-saturated.]}
%So we define two types of burst samples: 'GOLDEN' bursts without saturation, 'SILVER' bursts with saturation.

\begin{table}
\begin{center}
\caption{Results of the Gaussian fits to the spectral parameter distributions of SGR J1935+2154 bursts.}
\label{parameter_table}
\begin{tabular}{ p{1cm}<{\centering} p{2.5cm}<{\centering}  p{1.7cm}<{\centering}  p{1.7cm}<{\centering} } 
\hline\hline
model & Parameter &  $\mu^a$  &  $\sigma^b$ \\  
\hline
BB & $kT$ (keV) & $12.03{\pm0.33}$& $4.15{\pm0.27}$  \\
\hline
PL & photon Index  & $1.76{\pm0.01}$& $0.26{\pm0.01}$  \\
\hline
CPL& photon Index & $1.04{\pm0.07}$& $0.34{\pm0.06}$  \\
%CPL $E_{cut}$ (keV) & - & -  \\
%COMP $E_{peak}$ (keV) & - & -  \\
\hline
BB+BB & $kT_{\rm low}$ (keV)  & $2.91{\pm0.003}$& $1.10{\pm0.001}$  \\
BB+BB & $kT_{\rm high}$ (keV) & $12.14{\pm0.37}$& $2.90{\pm0.30}$  \\
\hline
BB+PL & $kT$ (keV) & $7.72{\pm0.39}$& $3.13{\pm0.32}$  \\
BB+PL& photon Index & $1.80{\pm0.02}$& $0.12{\pm0.01}$  \\
\hline
\end{tabular}
\tablecomments{$^a$ The mean value of Gaussian fit. \\
$^b$ The error of Gaussian fit. }
\end{center}
\end{table}

\section{Spectral Analysis Method} \label{sec:SpecAnalysisMethod}
\subsection{Spectral Models} \label{Models}
For consistency with the previous spectral analyses of the FRB 200428-Associated Burst of \textit{Insight}-HXMT \citep{2021NatAs...5..378L}, we choose five different models to fit the spectra of bursts in our sample: (1) single power-law (PL), (2) cutoff power-law (CPL), (3) single blackbody (BB), (4) two blackbodies (BB+BB), (5) blackbody plus power-law (BB+PL). All models are formulated in units of photon flux with energy
($E$) in keV and multiplied by a normalization factor $N$ in units of photons cm$^{-2}$ s$^{-1}$ keV$^{-1}$. 

\begin{enumerate}
\item  \textit{PL}: A simple photon power law with photon index of $\alpha$ and normalization of $N$
\begin{equation}
{f}_{\rm PL}(E) = N{E}^{-\alpha}.
\end{equation}

\item  \textit{CPL}: A power law with high energy exponential rolloff parameterized by power law photon index $\alpha$, e-folding energy of exponential rolloff (in keV) $E_{\rm cut}$  and normalization $N$
\begin{equation}
{f}_{\rm CPL}(E) = N{E}^{-\alpha}{\rm exp}(-\frac{E}{E_{\rm cut}}),
\end{equation}
In this model, the peak energy ($E_{\rm peak}$) in the $\nu F_\nu$ spectrum is defined as:
\begin{equation}
E_{\rm peak} = {E_{\rm cut}}(2-\alpha).
\end{equation}

%\item  \textit{COMPT}: The same as CPL with different parameters: characteristic energy $E_{\rm peak}$
%\begin{equation}
%{f}_{\rm compt}(E) = N{E}^{-\alpha}exp(-\frac{(2-\alpha)E}{E_{\rm peak}})
%\end{equation}

\item  \textit{BB}: A blackbody spectrum with normalization proportional to the surface area parameterized with temperature $kT$ (in keV) and normalization $K$ 
\begin{equation}
{f}_{\rm BB}(E) = \frac{K\times1.0344\times10^{-3}E^2}{{\rm exp}(E/kT)-1},
\end{equation}
\begin{equation}
K = \frac{R^{2}_{\rm km}}{D^2_{10}},
\end{equation}
where $R_{\rm km}$ is the radius in km of the blackbody emission region and $D_{10}$ is the distance to the source in unit of 10 kpc.
\end{enumerate}

Besides, a $wabs$ model is used to account for the absorption effect of the interstellar medium. The equivalent hydrogen column in the model for interstellar absorption could not be constrained well due to the low flux in the majority bursts of this sample. Therefore,
we fix the interstellar absorption term ($N_{\rm H}$) to 2.79 $\times 10^{22}$ cm$^{-2}$ in the following analyses\footnote{The results do not change significantly when varying the $N_{\rm H}$}, which is the best fit value derived from the bright FRB 200428-Associated Burst \citep{2021NatAs...5..378L}.

\begin{figure*}
\centering
\begin{tabular}{cc}
\includegraphics[width=0.50\textwidth]{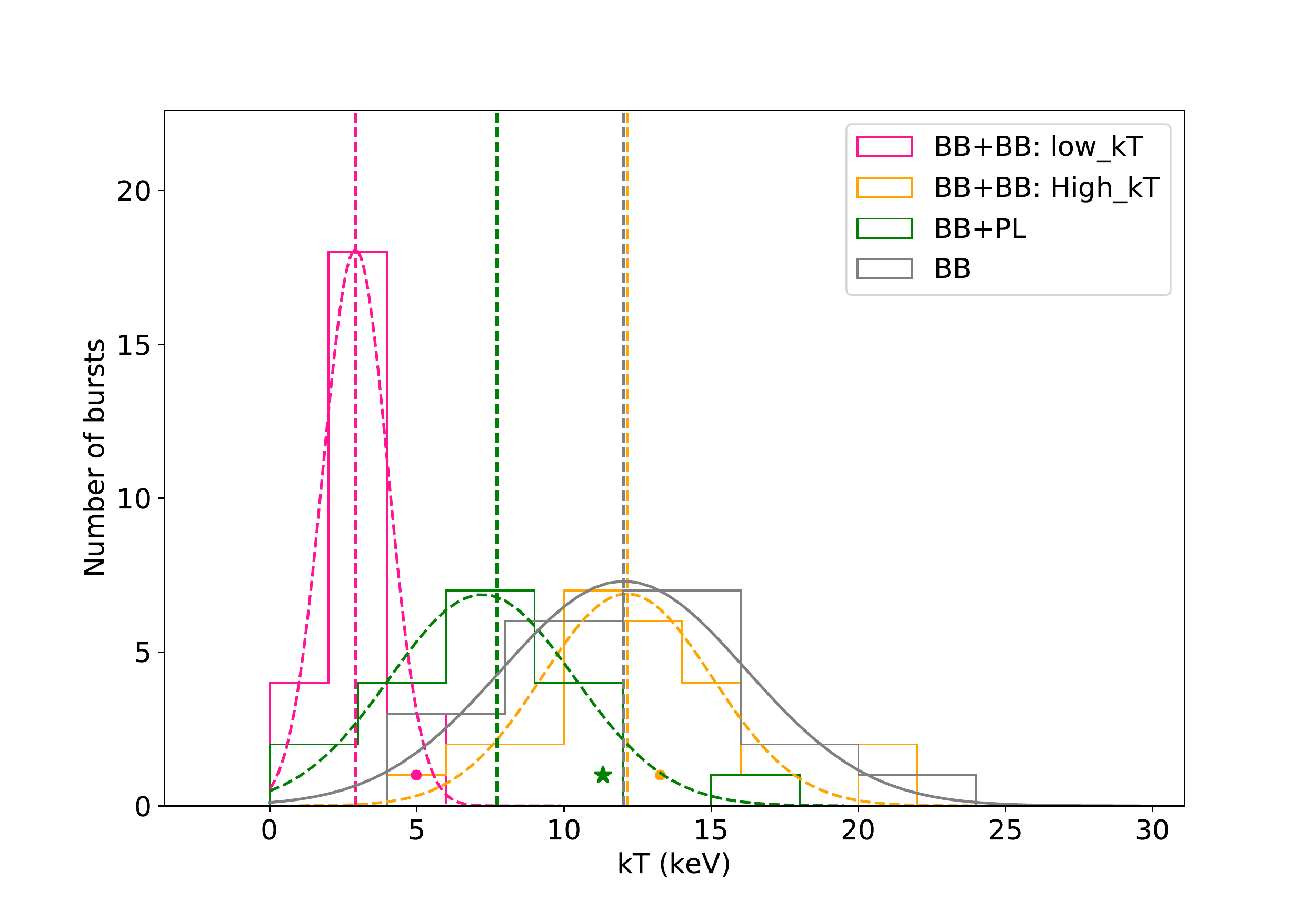} &
\includegraphics[width=0.50\textwidth]{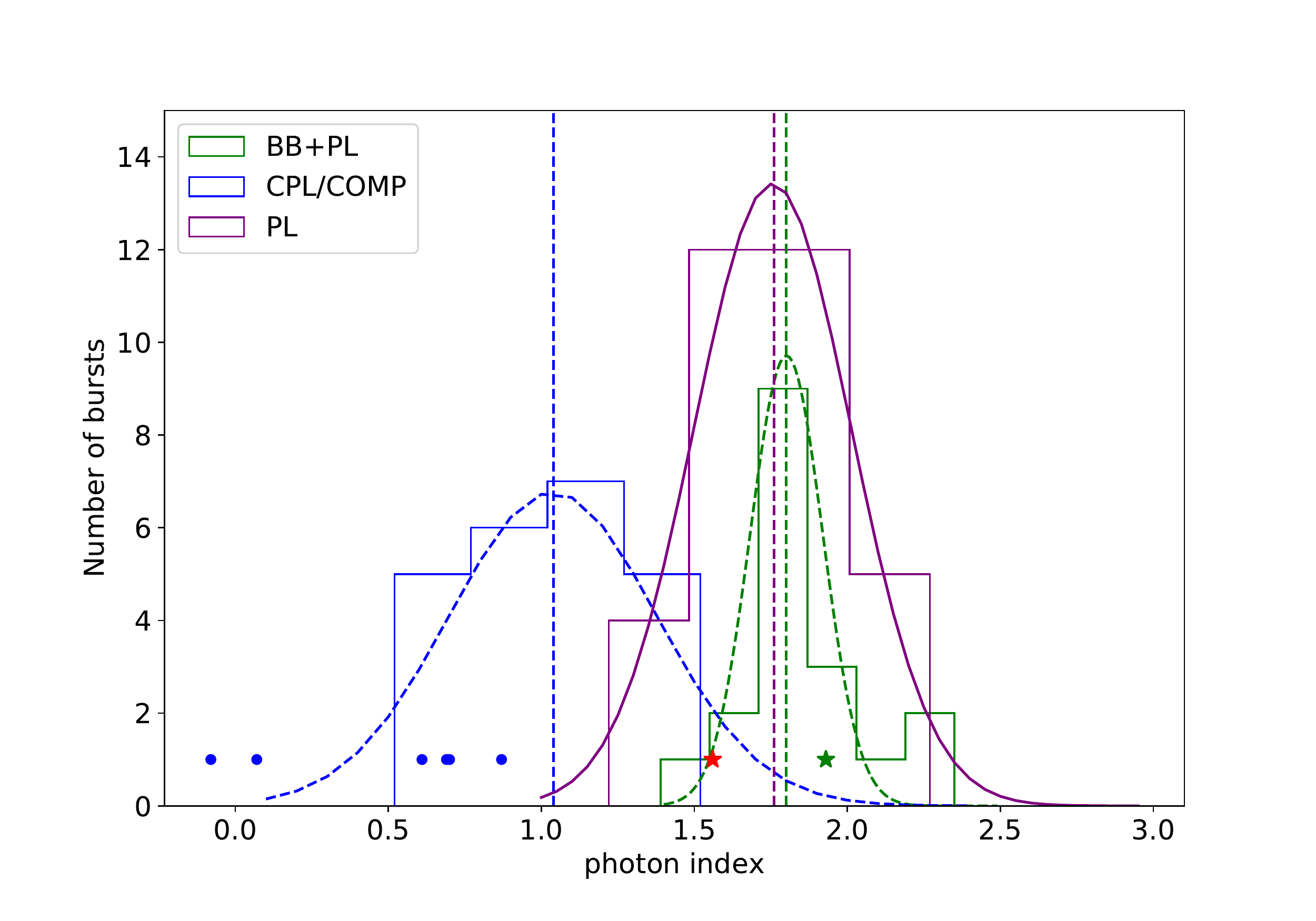} \\
\end{tabular}
\caption{\label{fig:kT_index} \textit{Left}: Distributions of the BB temperatures derived with the BB+BB (pink and orange represent low and high BB temperatures respectively), BB+PL (green) and BB (gay) models. \textit{Right}: Distributions of the photon index of BB+PL (green), CPL (blue) and PL (purple) models. The dots and stars in both panels represent the saturated bursts and the FRB 200428-Associated Burst with different models, respectively. The green and red stars are the FRB 200428-Associated Burst with BB+PL model and CPL model, respectively.
The FRB 200428-Associated Burst is labelled as red star in all figures in this paper.}
\end{figure*}

\begin{table*}
\centering
\caption{Results of power law (PL) fits to parameter correlations and Spearman test results of the correlations.}
\label{correlations_table}
\begin{tabular}{ p{3.5cm}<{\centering}   p{3cm}<{\centering}   p{3.8cm} <{\centering} p{4cm}<{\centering}  } 
\hline
\hline
Correlation $^*$ &  PL fit index ($\alpha$)   &  correlation coefficient ($\rho$) & chance probability ($P$)  \\  
\hline
$R^2_{\rm high}$ $\propto$ $(R^2_{\rm low})^\alpha$ & $0.60{\pm0.10}$ & 0.56 & 2.00E-03 \\
$F_{\rm high}$ $\propto$ $(F_{\rm low})^\alpha$ & $0.74{\pm0.09}$ & 0.84 & 1.09E-08 \\
$R^2_{\rm low}$ $\propto$ $(kT_{\rm low})^\alpha$ & $-3.11{\pm0.68}$ & -0.64 & 4.74E-04 \\
$R^2_{\rm high}$ $\propto$ $(kT_{\rm high})^\alpha$ & $-3.24{\pm0.60}$ & -0.60 & 1.42E-03 \\
$R^2$ $\propto$ $(kT)^\alpha$ & $-3.45{\pm0.23}$ & -0.90 & 5.10E-20 \\

\hline
\end{tabular}
\tablecomments{$^*$ $R^2$, $F$, $kT$ are the emitting area, flux and temperature of a BB component.\\}
\end{table*}

\subsection{Spectral Analysis Procedure} \label{Spectral analyses}
The spectral analysis of the burst sample is based on the \textit{Insight}-HXMT 1L data using the current version (version 2.04) of \textit{Insight}-HXMT Data Analysis Software package (\textit{HXMTDAS}) \footnote{http://www.hxmt.cn/}.
We generate the spectra using the commands \textit{hespecgen}, \textit{mespecgen}, \textit{lespecgen}. The time-integrated spectra are derived jointly using HE, ME and LE data, excluding the blinded detector unit of HE, ME and LE, and the wide field of view (FOV) detector unit of ME and LE. The data of 1$-$10 keV with LE, 10$-$30 keV with ME, 28$-$250 keV with HE, are used to fit the spectra.
There are some bursts for which the LE or ME data are not available due to bright Earth or data loss, so only the HE data or HE and ME data are used to fit their spectra. The time-integrated burst spectra are extracted for the burst duration derived from the Bayesian block method, ($T_{\rm bb}$), which is the common burst interval of all 3 telescopes (see paper \uppercase\expandafter{\romannumeral1} for details)\footnote{We note that performing spectral analysis in the total burst duration of all three telescopes does not alter any of our main conclusions.}. 
For weak bursts, the \textit{GRPPHA} command\footnote{https://heasarc.gsfc.nasa.gov/ftools/} is used to group the observed data (e.g., GROUP MIN RCNTS) to ensure a sufficient statistics.
The background spectra are accumulated from the data events during the pre- and post- the burst time intervals (i.e., $T_0-7$ s $\sim T_0-2$ s and $T_0+2$ s $\sim T_0+7$ s, where $T_0$ is the trigger time of the burst).

We perform spectral fitting with Xspec 12.11.1 using Cash statistics (C-stat) \citep{1979ApJ...228..939C}. We call a model (${\rm M}_{1}$) as adequate fit when all model parameters fall in the physically meaningful region with a confidence level of no less than 1 $\sigma$ (e.g., kT, $E_{\rm cut}$ and normalization are positive). To identify which model can better describe the spectrum, we calculate the Bayesian Information Criterion (BIC) for each spectral model as presented in \cite{2021}. A smaller BIC indicates that the model fits the data better. If the reduction in BIC is more than 10, then the model with smaller value is considered to be significantly better than the other. Therefore, for one burst, we may have either a significantly preferred model or more than one adequate models. We calculate the unabsorbed fluence and flux using the parameters of the model with the smallest BIC. 

%and the model with minimum BIC is called as the best-fit model. The unabsorbed fluence and flux are calculated using the parameters of the best-fit spectral models. 
%To quantify the difference between two spectral fits, we calculate the $\Delta$ of BIC of two models, which is defined as:
%\begin{equation}
%\Delta = B({\rm M}_{1})-B({\rm M}_{2}), \label{f_6}
%\end{equation}
%where ${B}$ is the value of BIC, and ${\rm M}_{1}$ and ${\rm M}_{2}$ represent two models to be compared. \textcolor{orange}{(LIN: don't need to redefine BIC as B.)}
%In the present work, we call the best-fit model as the favorite model only when it has $\Delta>10$ with respect to all other models. Thus, for some bursts, we may find several adequate models and one best-fit model with minimum BIC, but none of them is a favorite one.

%\subsection{Spectral Analysis for Saturated Bursts} 

There are six bursts in our sample which suffered from the data saturation problem (FRB200428-Associated burst excluded), that some of the events can not be recorded. The data lose is due to the limited transmission bandwidth. But for each of them, at least one telescope is free from the saturation.  Therefore we multiply a constant to the spectral models during the spectral fitting process in order to correct the saturation effect ($const*wabs*(bb+bb)$ in Xspec for example). This constant parameter represents the relative data loss ratio. We fix the constant at one for the telescopes which are not affected by the saturation and leave it free for others.

There are seven bursts in our catalog jointly observed by \textit{Insight}-HXMT and \textit{Fermi}/GBM. In order to verify our spectral analysis and compare between different instruments, we perform a joint spectral analysis using both data. We extract burst spectra using data of NaI detectors of GBM with an angle to the source less than 30$^{\circ}$. The integration time of each burst spectrum is the maximum burst interval of the two instruments. We also include a multiplicative factor to account for the systematic difference between these two instruments.

\begin{figure*}
\centering
\begin{tabular}{cc}
\includegraphics[width=0.50\textwidth]{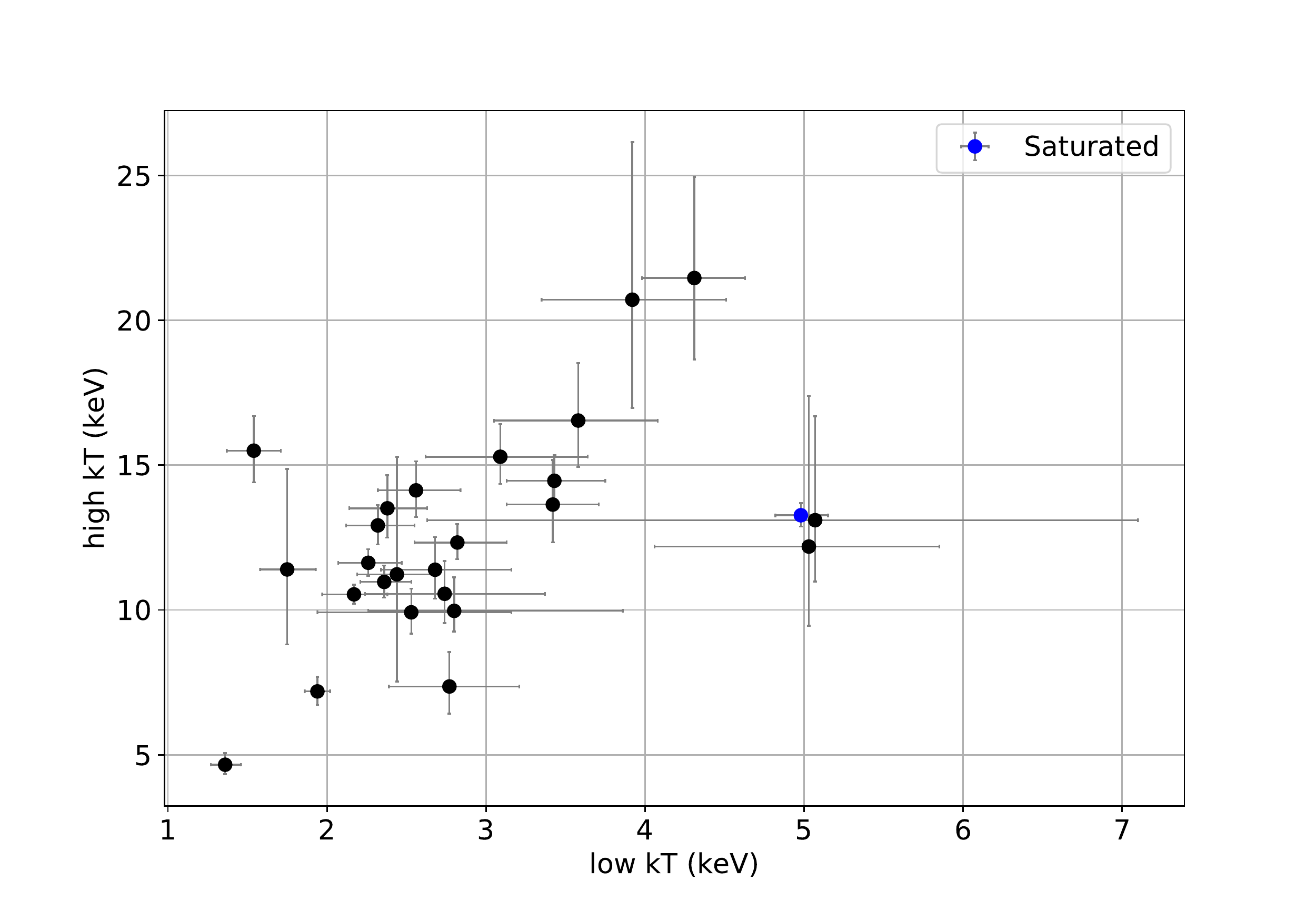} &
\includegraphics[width=0.50\textwidth]{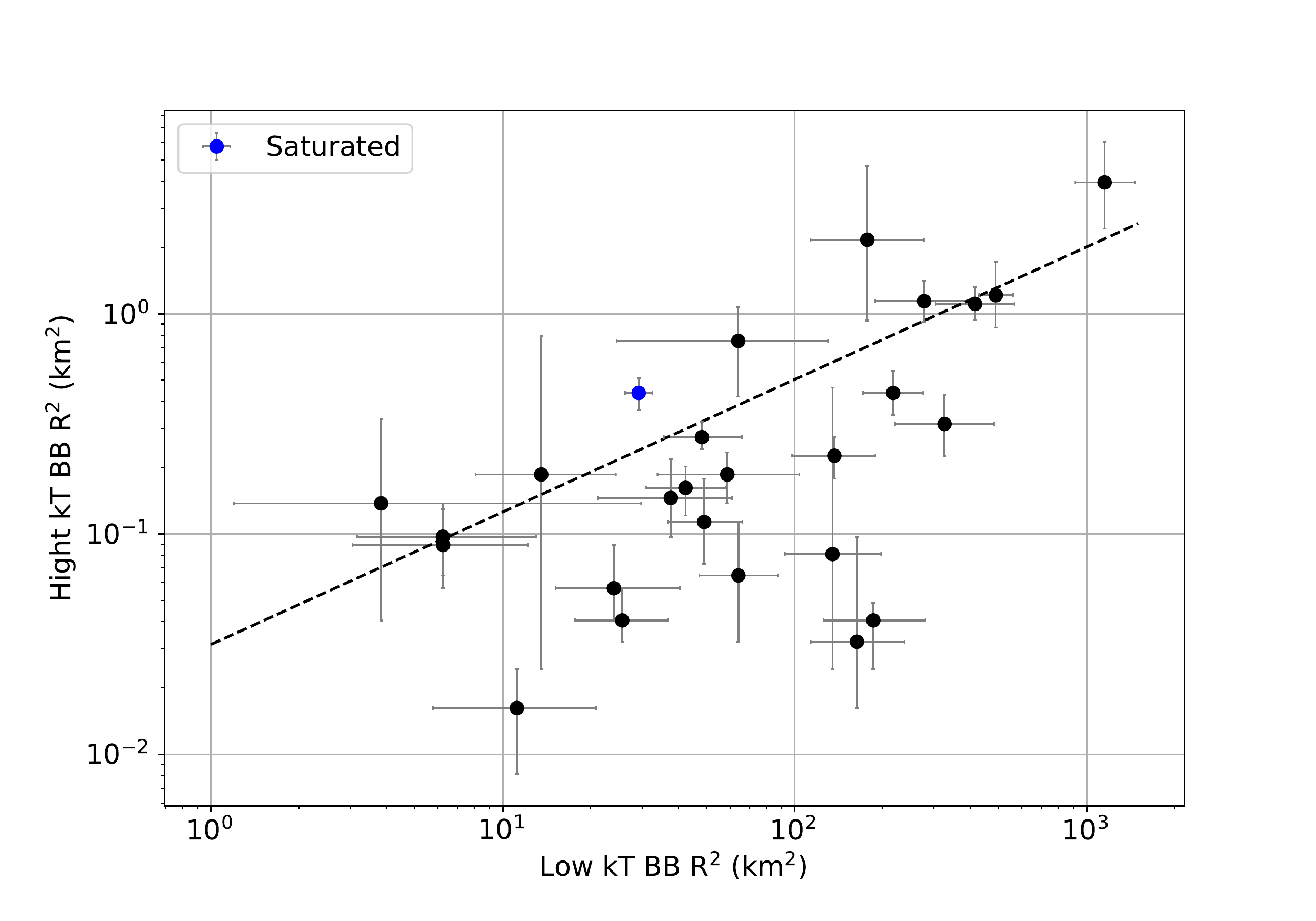} \\
\includegraphics[width=0.50\textwidth]{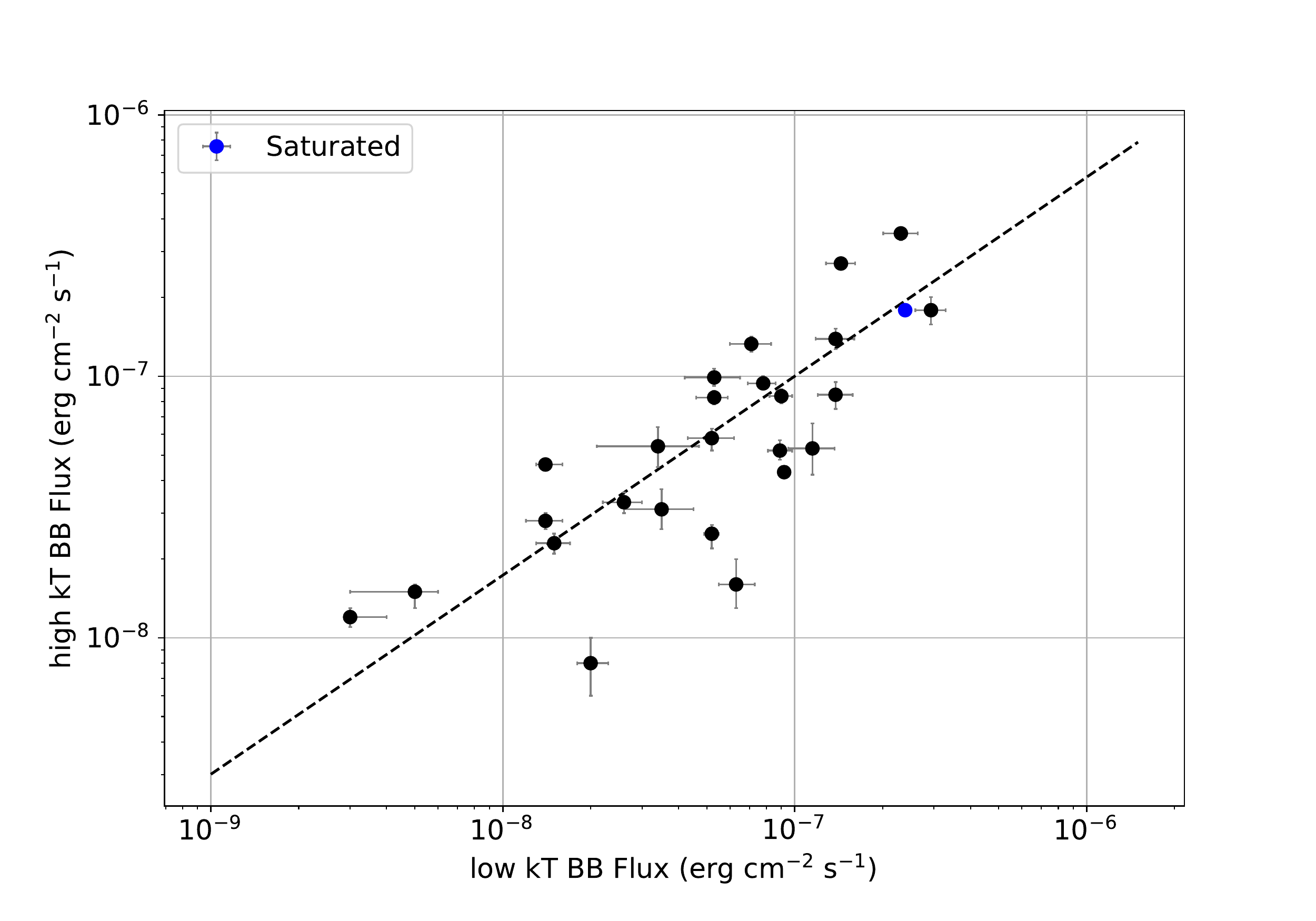} &
\includegraphics[width=0.50\textwidth]{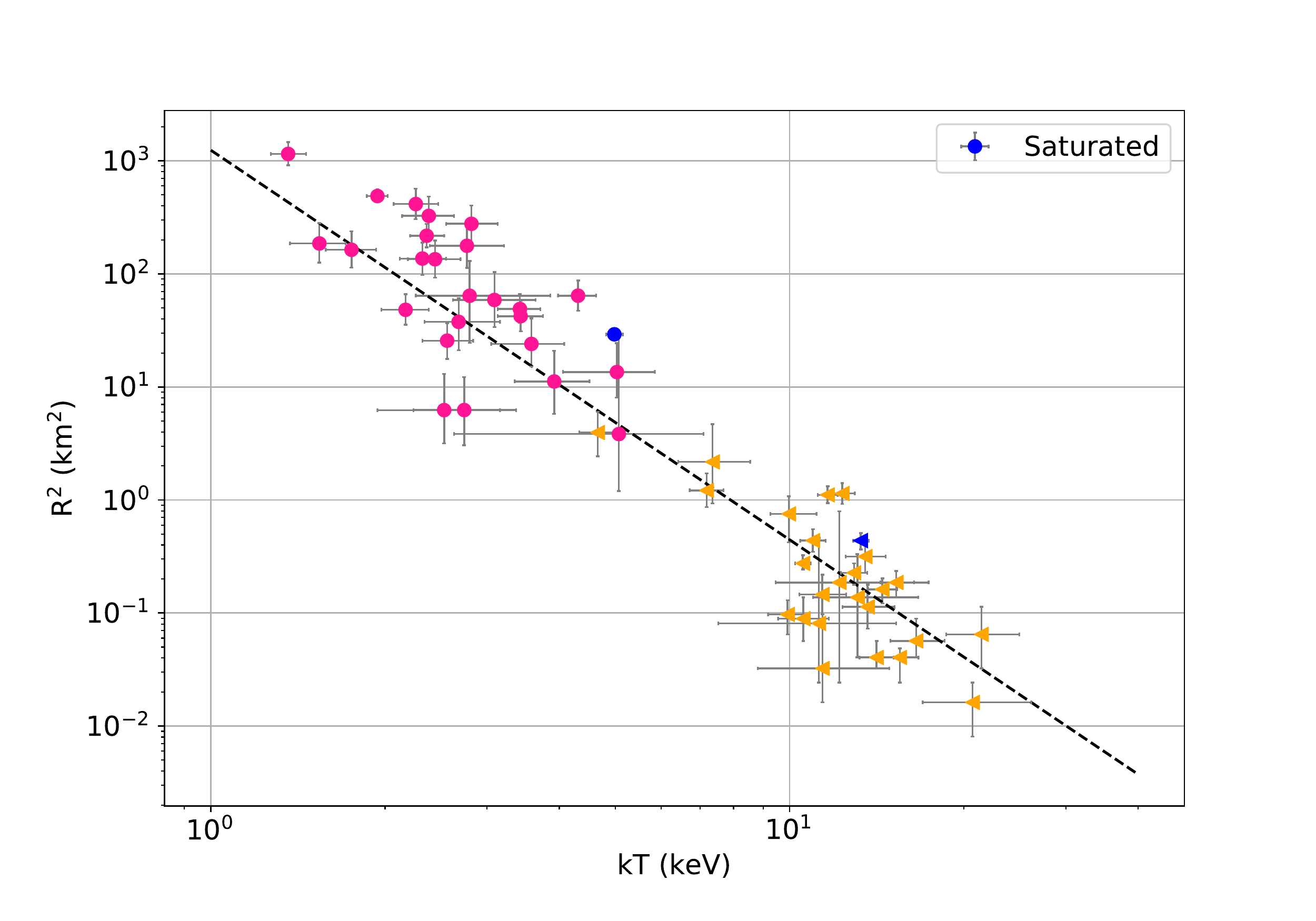} \\
\end{tabular}
\caption{\label{fig:par_relation} \textit{Top left}: Relation between $kT$ of the two BB components in the BB+BB model. \textit{Top right}: Correlations of the radius of BB emission regions of the two BB components in BB+BB model.
\textit{Bottom left}: The flux of both BB components, measured in the 1-250 keV energy band. \textit{Bottom right}: The emission areas ($R^2$) as a function of the low (pink circles) and high (orange triangle) BB temperatures. The blue dots and black-dashed lines of those panels represent the saturated bursts and the PL fit of both BB components.
}
\end{figure*}

\section{Results}

\subsection{Spectral results of the burst sample} \label{sec:result}

\begin{figure}
\centering
\begin{tabular}{c}
\includegraphics[width=0.50\textwidth]{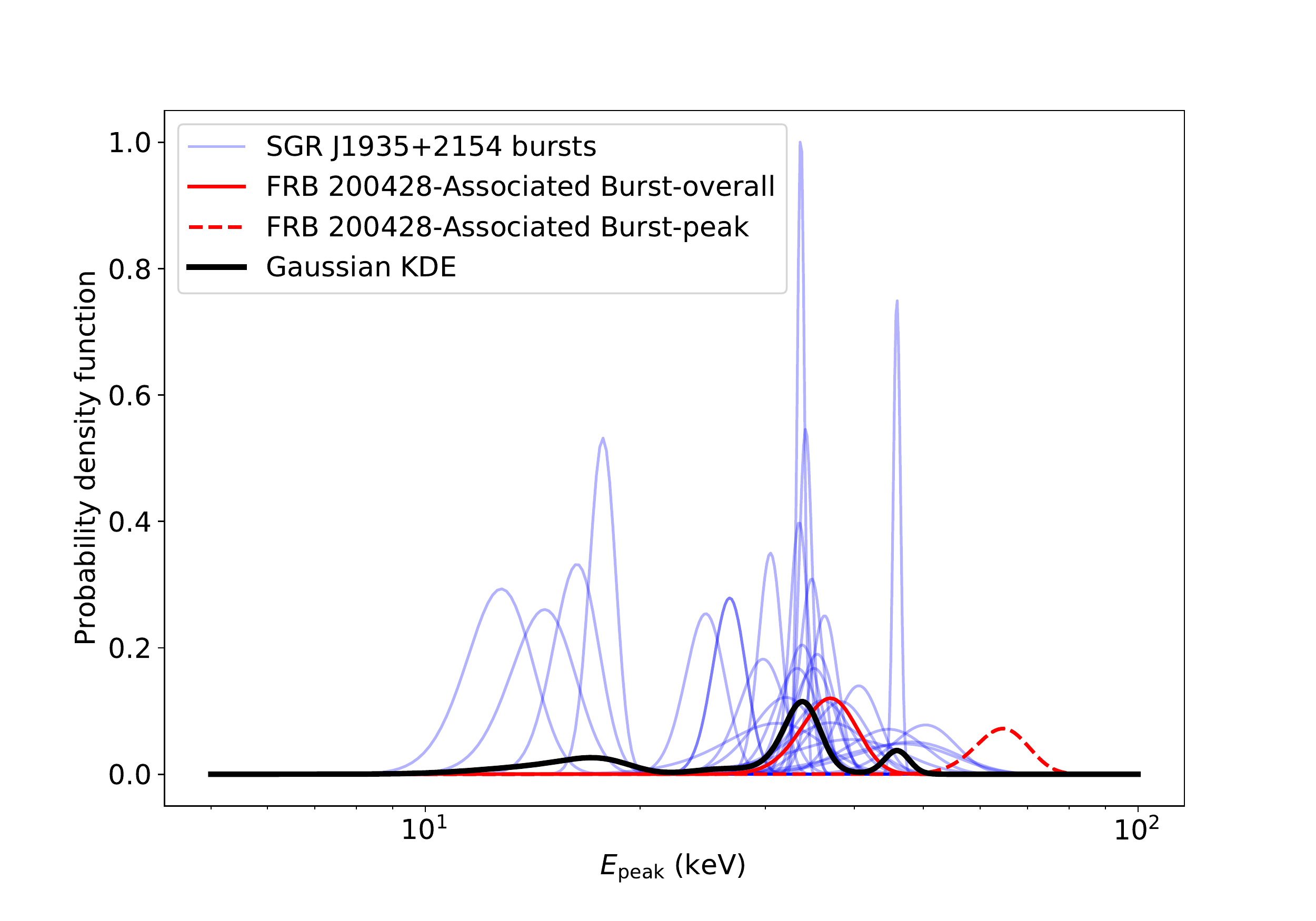} \\
\end{tabular}
\caption{\label{fig:Epeak} The distribution of $E_{\rm peak}$ of CPL model for 29 bursts and comparison to the FRB 200428-Associated Burst. The blue lines are the probability density function (PDF) of $E_{\rm peak}$. The black line represents the PDF of a Gaussian kernel for the PDFs of 29 bursts. The red solid line (time-averaged spectrum from $T_0-0.2$ s $\sim T_0+1.0$ s) and dashed line (peak spectrum from $T_0+0.41$ s $\sim T_0+0.47$ s) are the PDF of the $E_{\rm peak}$ of the FRB 200428-Associated Burst as measured with \textit{Insight}-HXMT (see supplementary table 4 and table 6 in \cite{2021NatAs...5..378L} for more detail).}
\end{figure}

\begin{figure}
\centering
\begin{tabular}{c}
\includegraphics[width=0.50\textwidth]{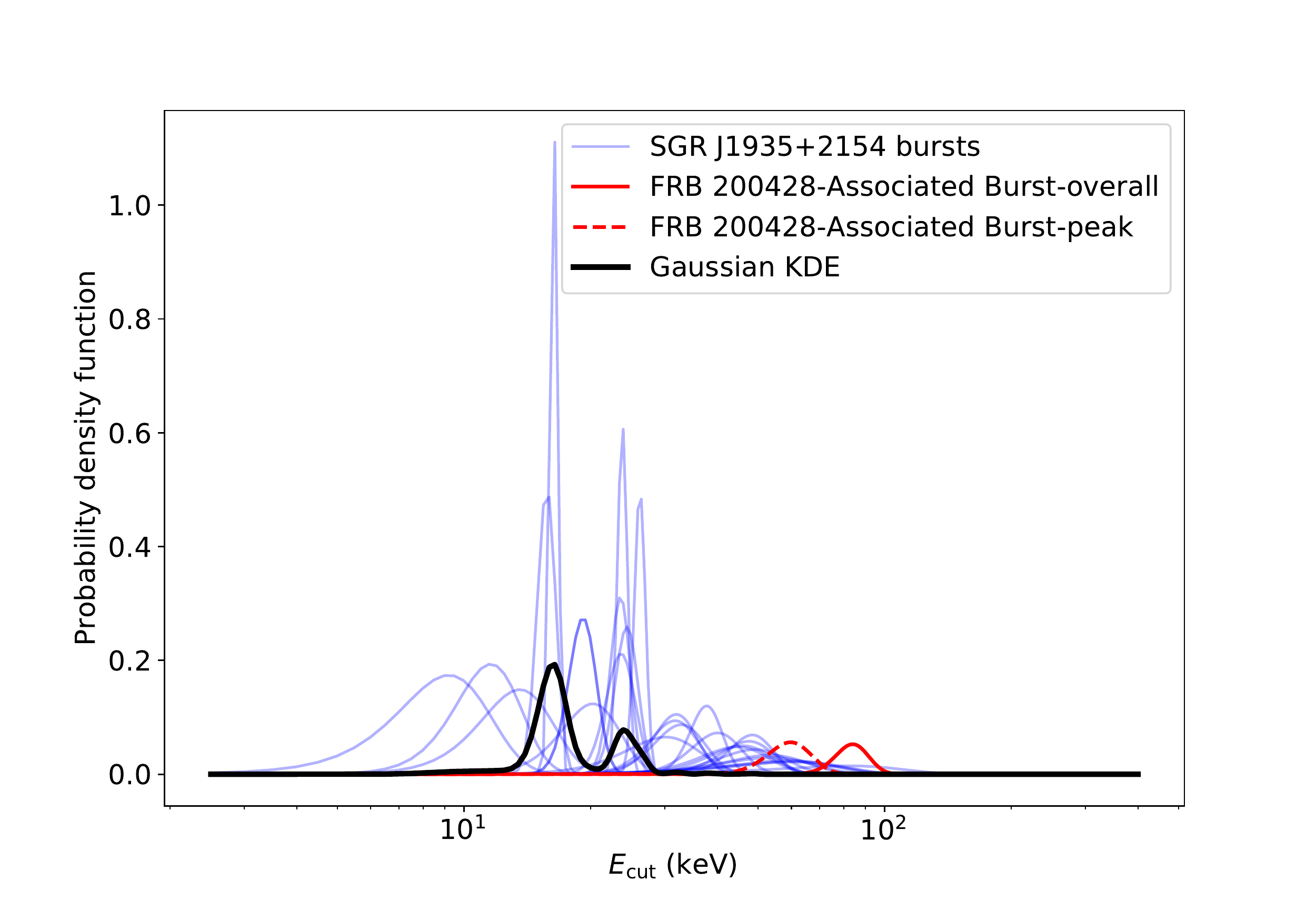} \\
\end{tabular}
\caption{\label{fig:Ecut}The distribution of $E_{\rm cut}$ derived with the CPL model for 29 bursts and comparison to the FRB 200428-Associated Burst. The red line is the PDF of $E_{\rm cut}$ of the FRB 200428-Associated Burst as measured with \textit{Insight}-HXMT \citep{2021NatAs...5..378L}. Other captions are the same as Figure \ref{fig:Epeak}.}.
\end{figure}

\begin{figure*}
\centering
\begin{tabular}{cc}
\includegraphics[width=0.50\textwidth]{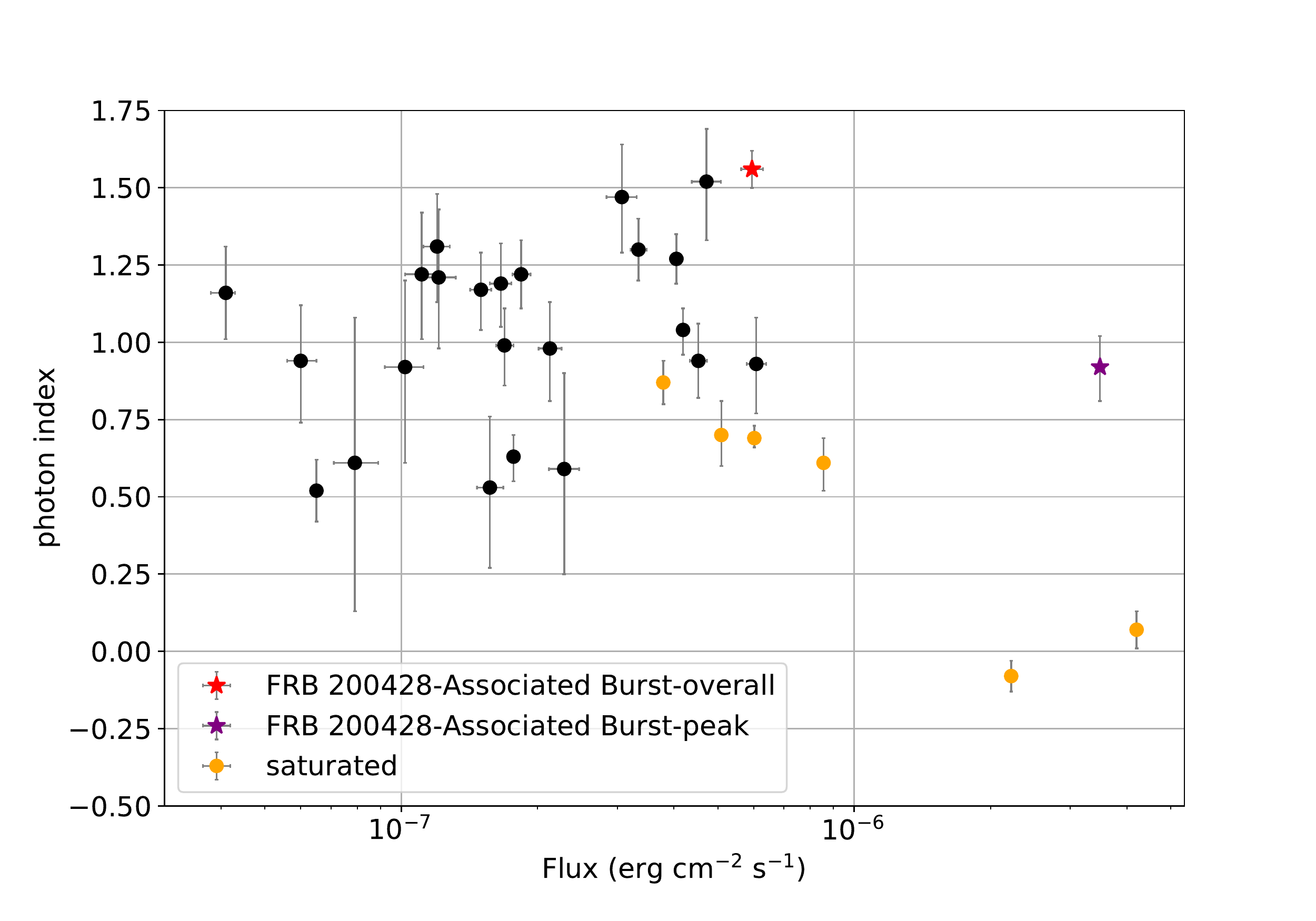} &
\includegraphics[width=0.50\textwidth]{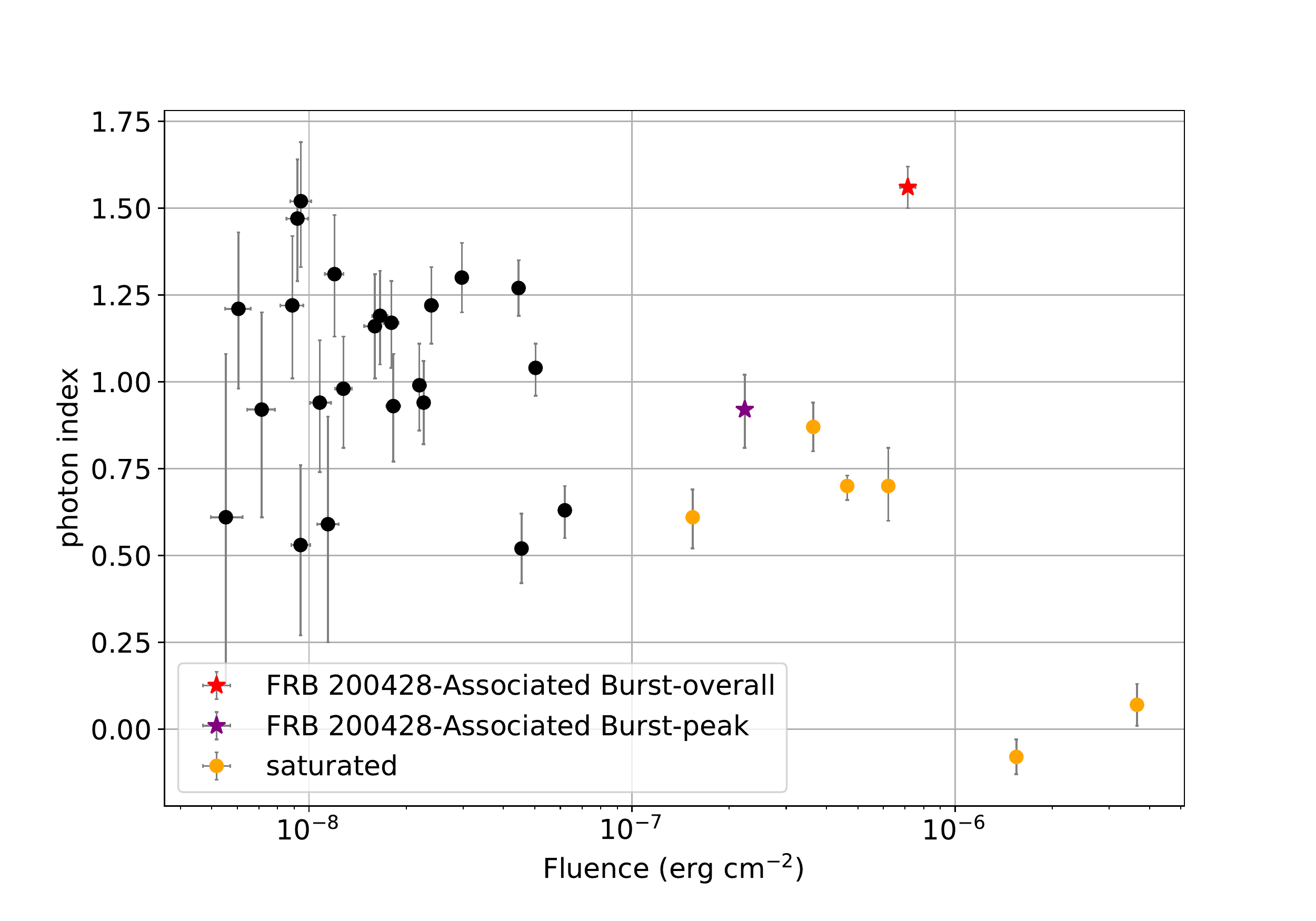} \\
\end{tabular}
\caption{\label{fig:index_f_cpl} \textit{Left}: Scatter plot of the photon index and flux in 1$-$250 keV derived with the CPL model. \textit{Right}: Scatter plot of the photon index and fluence in 1$-$250 keV derived with the CPL model. The red and purple stars are the time-integral and peak spectra of FRB 200428-Associated Burst \citep{2021NatAs...5..378L}, respectively. The orange points represent the saturated bursts.}
\end{figure*}

\begin{figure*}
\centering
\begin{tabular}{cc}
\includegraphics[width=0.50\textwidth]{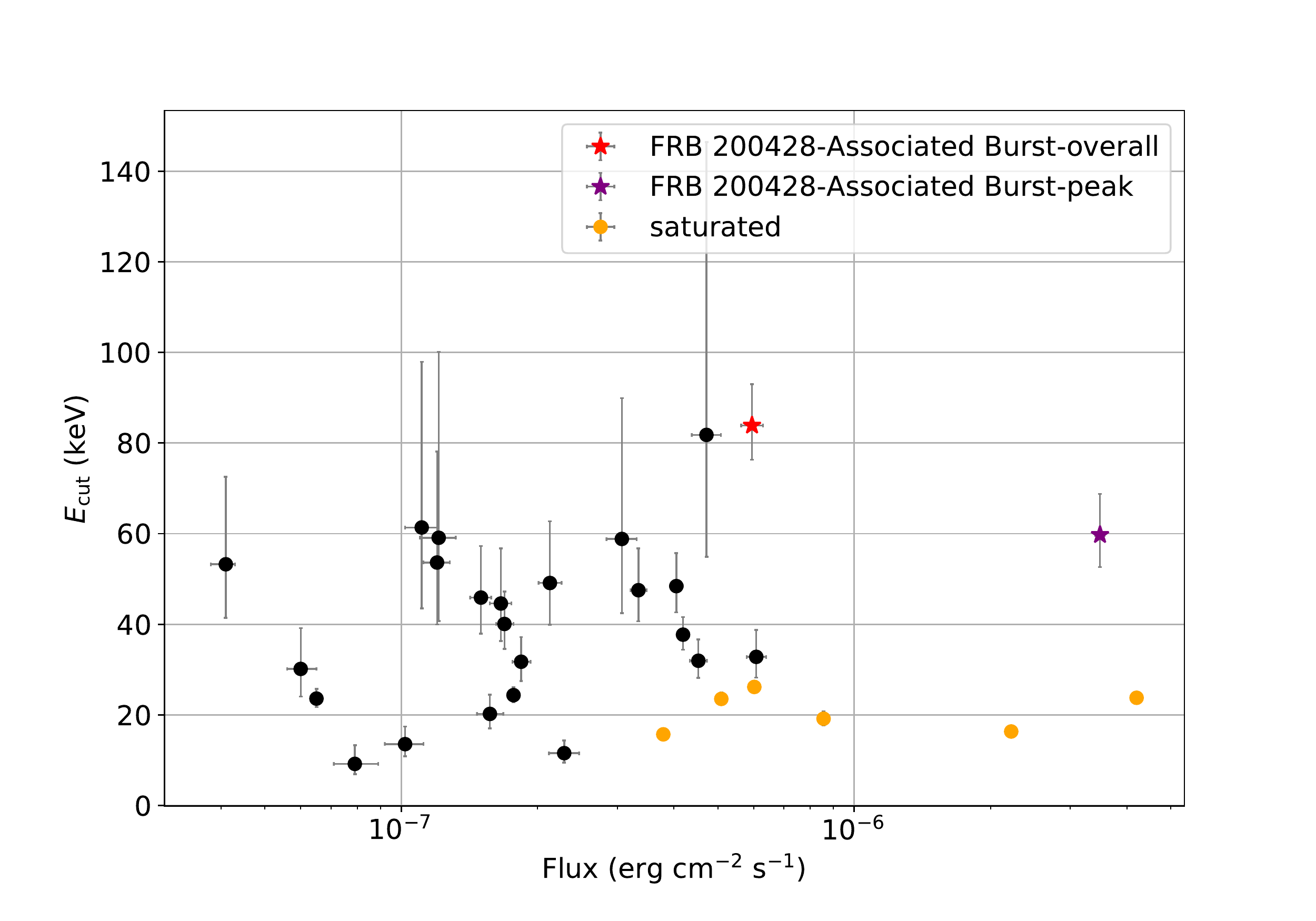} &
\includegraphics[width=0.50\textwidth]{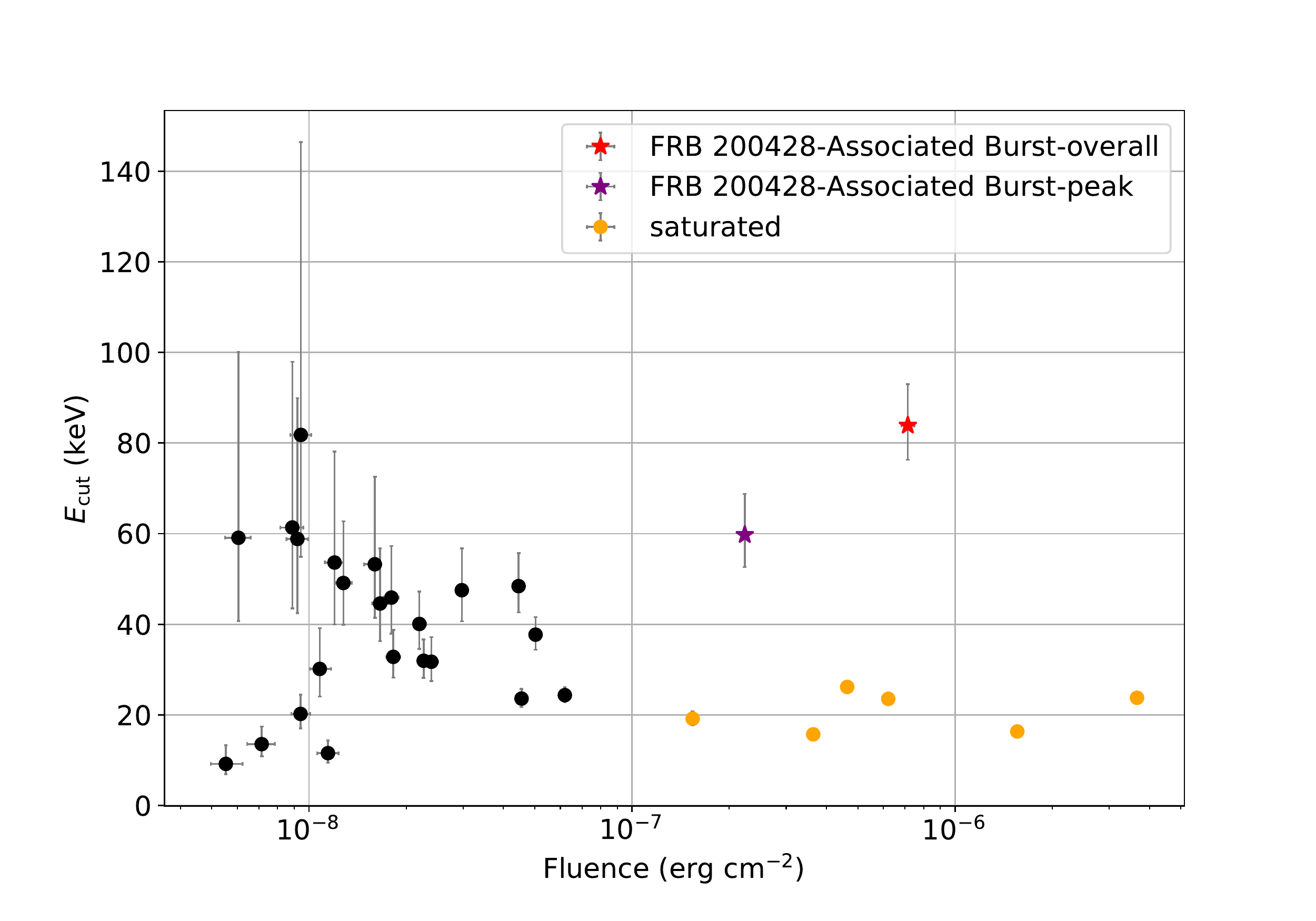} \\
\end{tabular}
\caption{\label{fig:Ecut_f_cpl} \textit{Left}: Scatter plot of $E_{\rm cut}$ and flux in 1$-$250 keV derived with the CPL model. \textit{Right}: Scatter plot of $E_{\rm cut}$ and fluence in 1$-$250 keV  derived with the CPL model. The red and purple stars are the time-integral and peak spectra of FRB 200428-Associated Burst \citep{2021NatAs...5..378L} respectively. The orange points represent the saturated bursts.}
\end{figure*}

\begin{figure*}
\centering
\begin{tabular}{cc}
\includegraphics[width=0.50\textwidth]{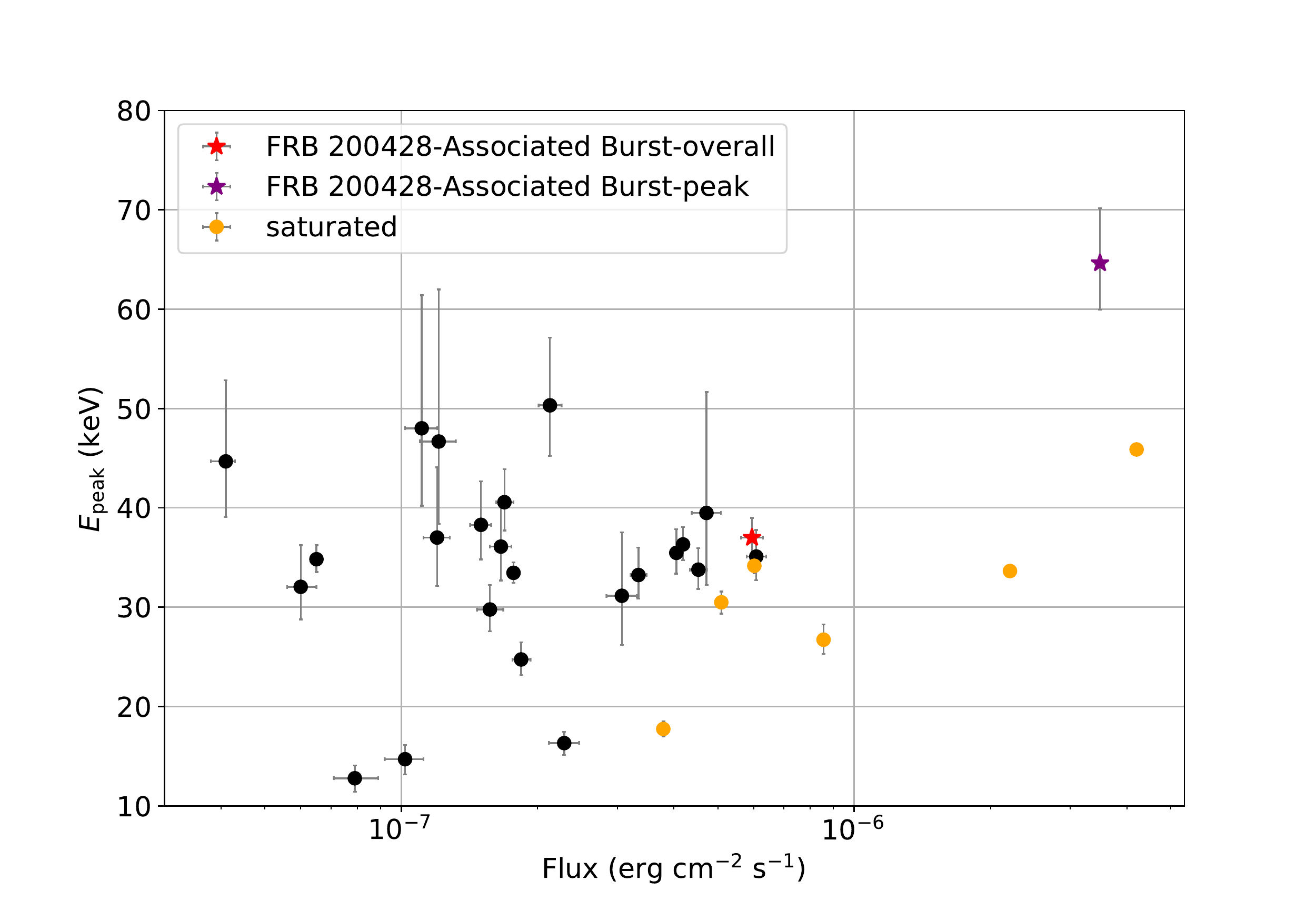} &
\includegraphics[width=0.50\textwidth]{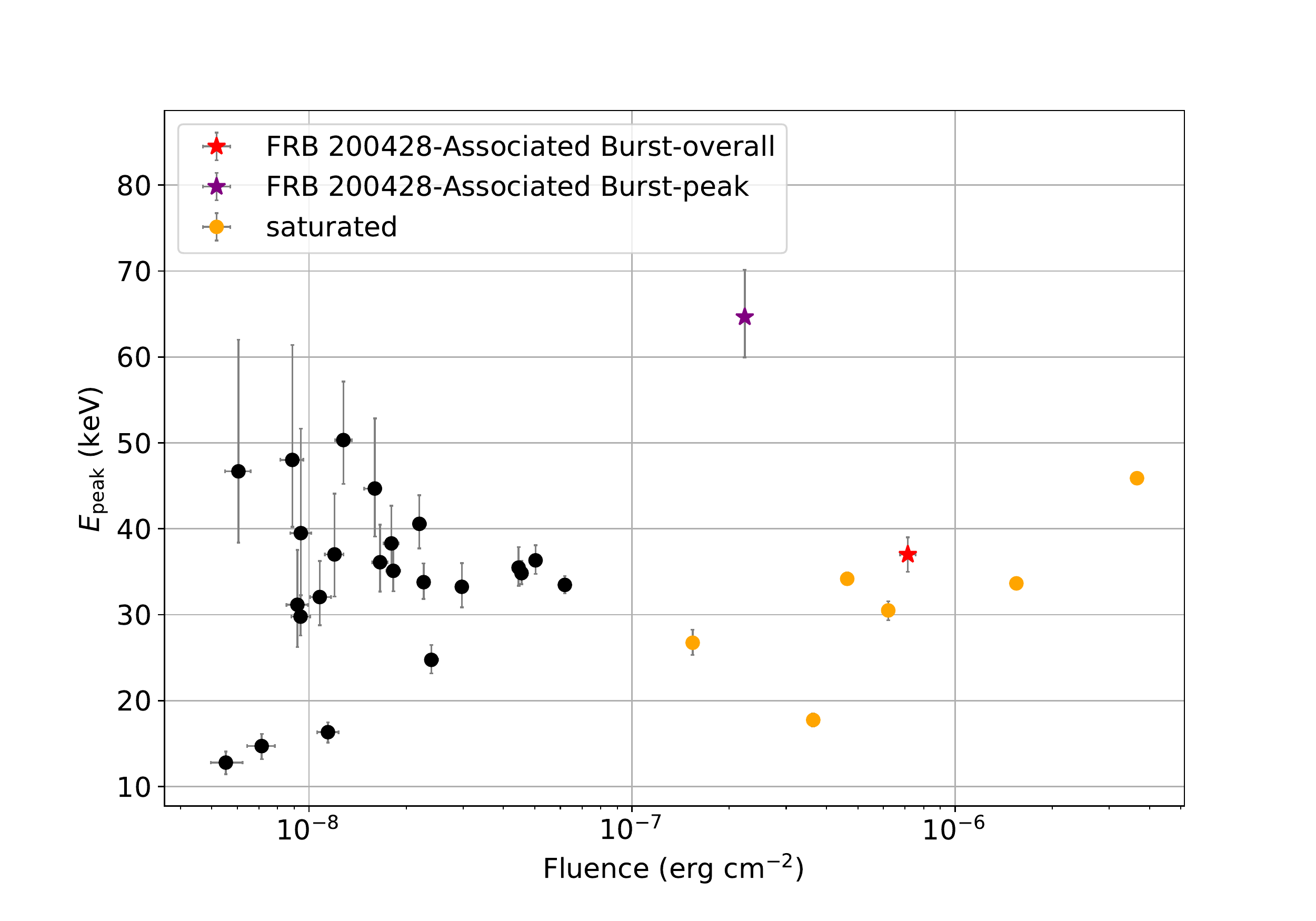} \\
\end{tabular}
\caption{\label{fig:Epeak_f} \textit{Left}: Scatter plot of the $E_{\rm peak}$ and flux in 1$-$250 keV derived with the CPL model. \textit{Right}: Scatter plot of the $E_{\rm peak}$ and fluence in 1$-$250 keV derived with the CPL model. The red and purple stars are the time-integral and peak spectra of FRB 200428-Associated Burst \citep{2021NatAs...5..378L}, respectively. The orange points represent the saturated bursts.}
\end{figure*}

\begin{figure*}
\centering
\begin{tabular}{cccc}
\includegraphics[width=0.50\textwidth]{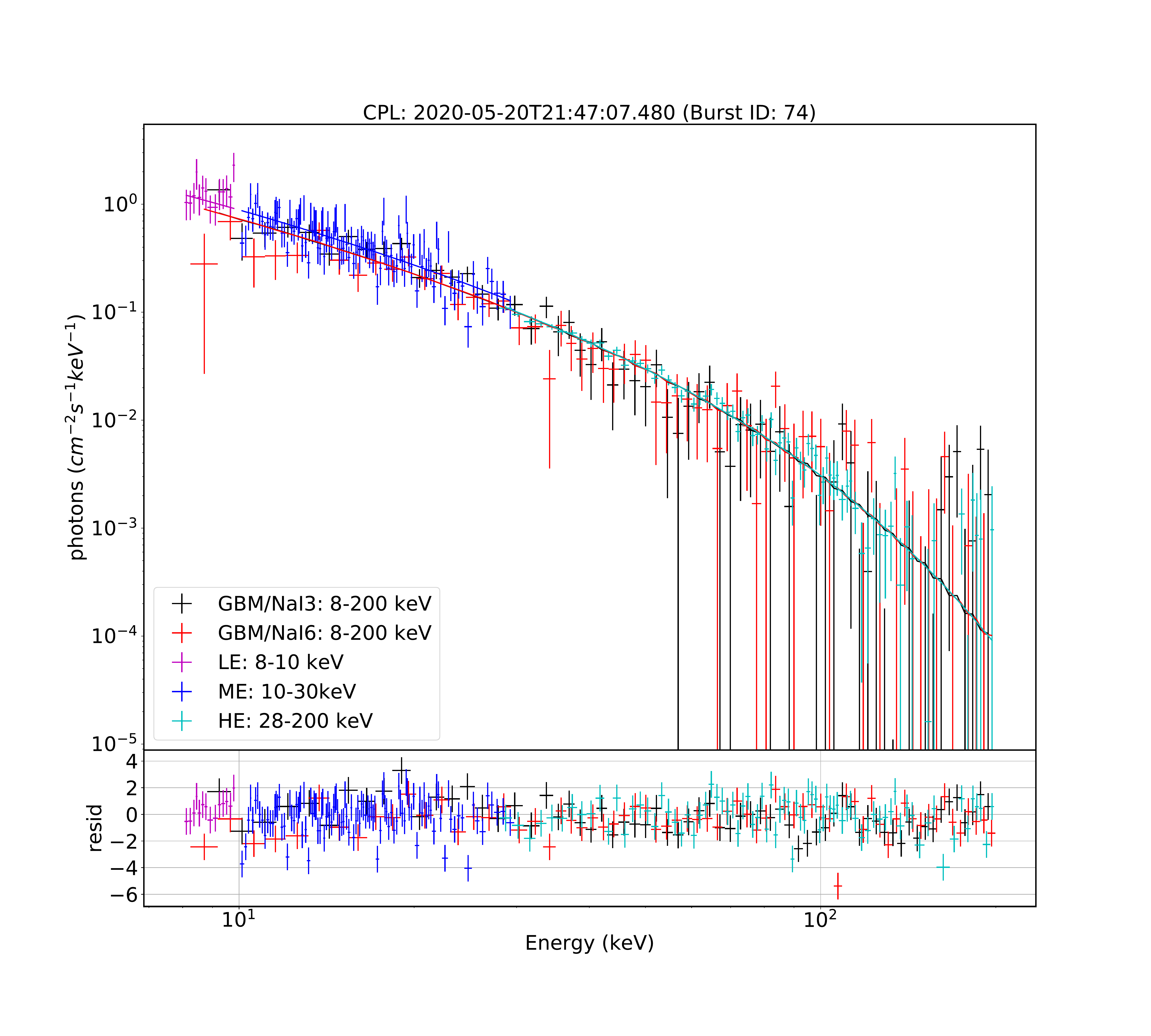} &
\includegraphics[width=0.50\textwidth]{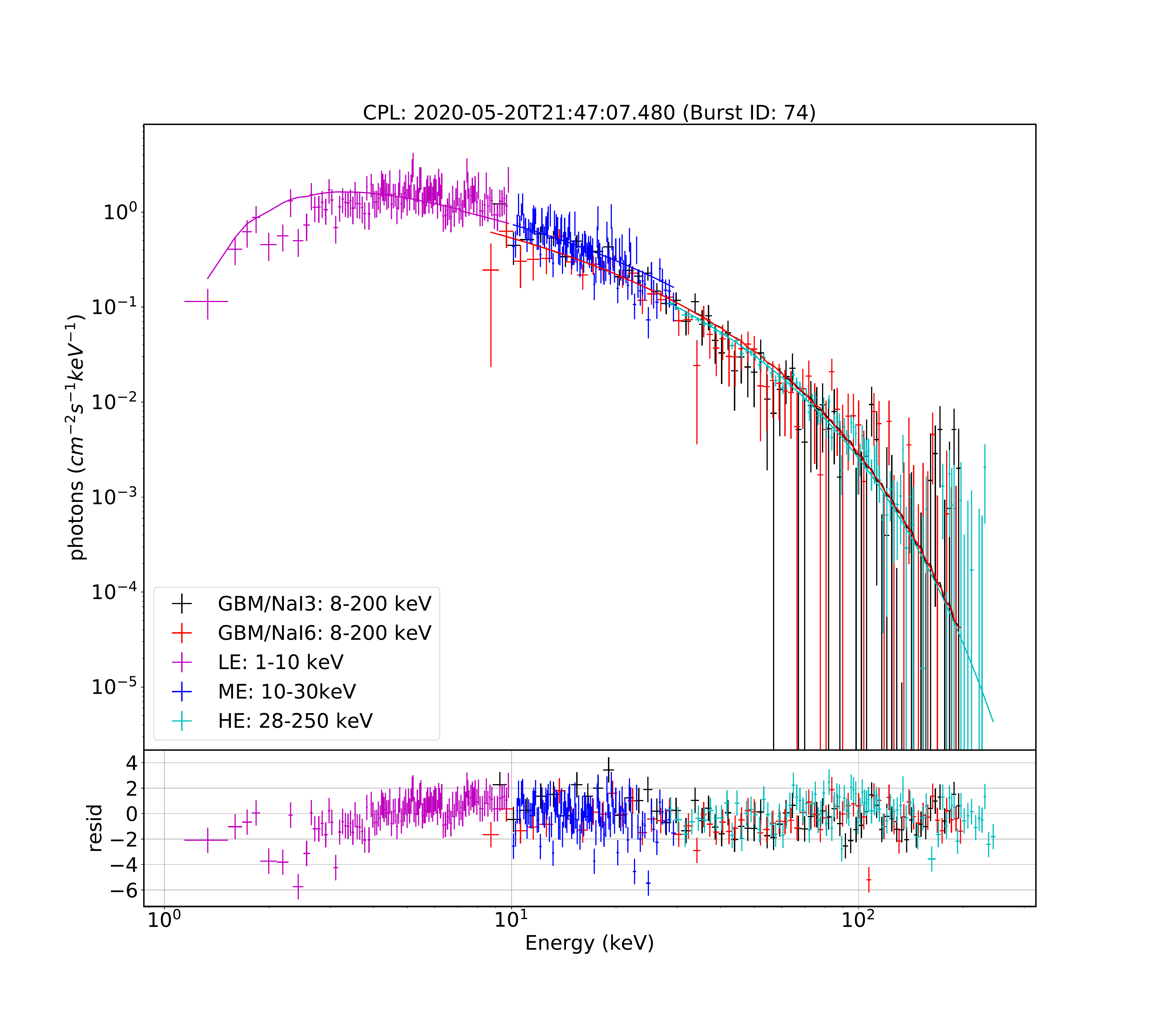} \\
\includegraphics[width=0.50\textwidth]{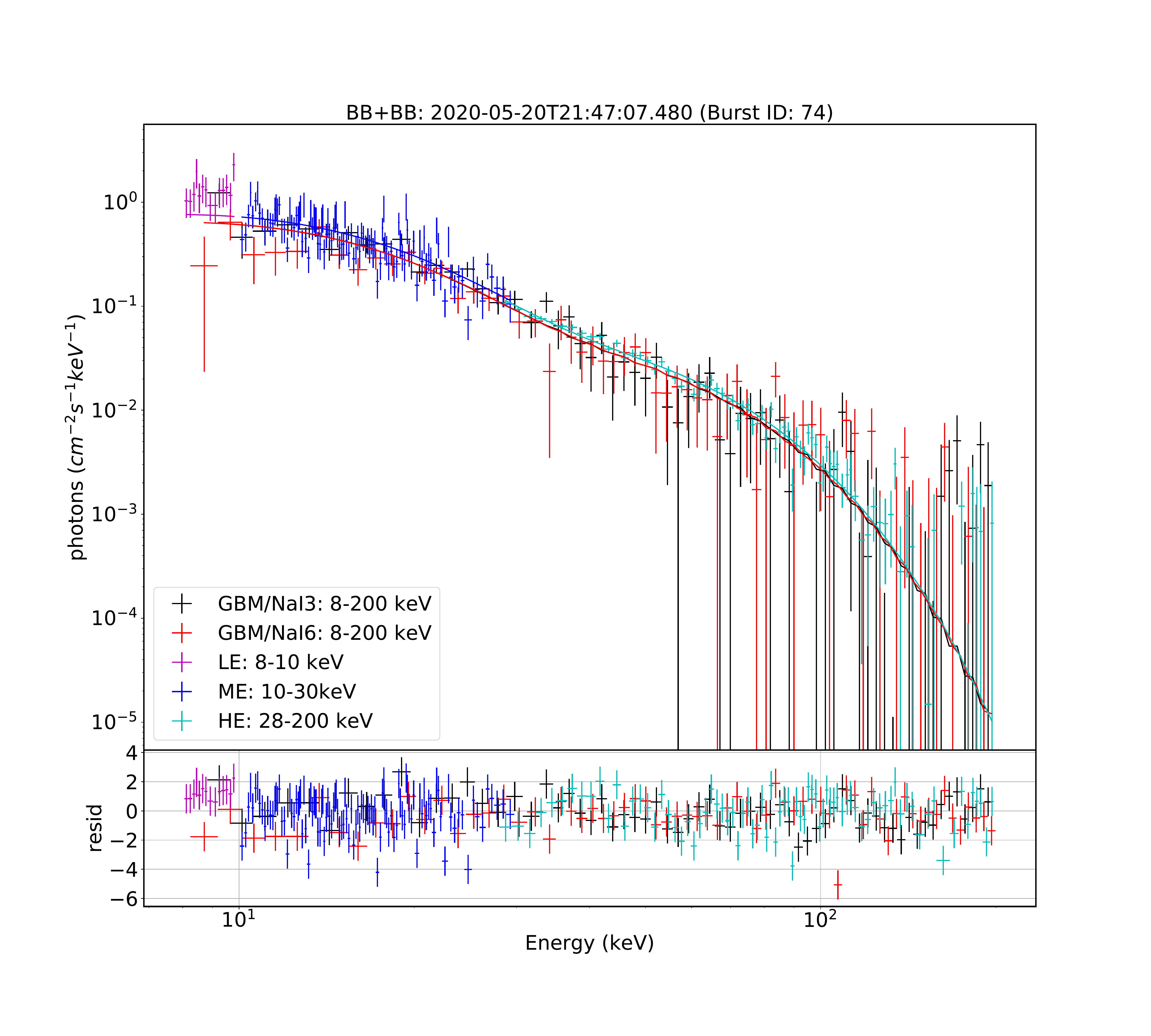} &
\includegraphics[width=0.50\textwidth]{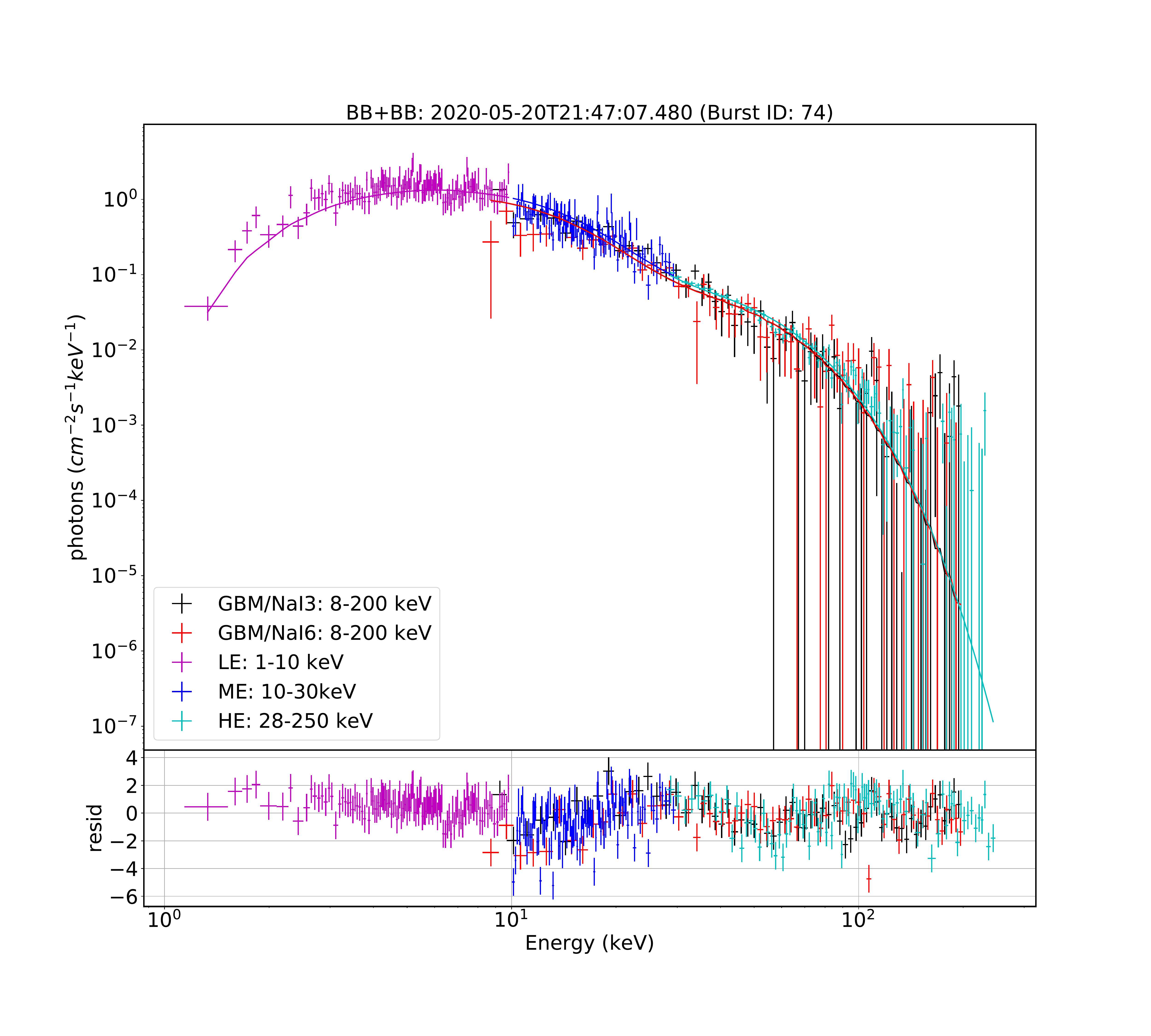} \\
\end{tabular}
\caption{\label{fig:JointFit} The \textit{Insight}-HXMT and \textit{Fermi}/GBM joint spectral analysis of SGR J1935+2154 bursts. Data from HE, ME and LE of \textit{Insight}-HXMT and NaI detector \#3 and \#6 of \textit{Fermi}/GBM are used. Interstellar absorption parameters of $2.79\times10^{22}$ cm$^{-2}$ is adopted. 
\textit{Left}: \textit{Insight}-HXMT and \textit{Fermi}/GBM data in the same energy range of 8$-$200 keV are used in the joint spectral analysis. The CPL model fit ($\alpha$ = $1.34{\pm0.07}$, $E_{\rm cut}$ = $37.28{\pm2.44}$ keV and C-Stat/DOF = $642.18/337$) is shown in the top left panel, while the BB+BB fit ($kT_{\rm low}$ = $4.31{\pm0.09}$ keV, $kT_{\rm high}$ = $13.80{\pm0.26}$ keV and C-Stat/DOF = $714.18/336$) is shown in the bottom left panel. 
\textit{Right}: \textit{Insight}-HXMT data in 1$-$250 keV and \textit{Fermi}/GBM data in 8$-$200 keV are used in the joint spectral analysis. The CPL model fit ($\alpha$ = $0.73{\pm0.04}$, $E_{\rm cut}$ = $25.14{\pm0.70}$ keV and C-Stat/DOF = $900.34/451$) is shown in the top right panel, while the fit with BB+BB model ($kT_{\rm low}$ = $3.34{\pm0.05}$ keV, $kT_{\rm high}$ = $12.31{\pm0.15}$ keV and C-Stat/DOF = $1032.75/450$) is shown in the bottom right panel.} 
%% 8-200 BIC :  731.65;  CPL  659.66
%% 1-250 BIC : BB+BB 1051.09;  CPL 918.69
\end{figure*}

%\textcolor{orange}{(move next two paragraph to result)}
%The trigger time and spectral characteristics (the fitting model parameters, fit statistics and their flux in 1$-$250 keV) of each saturated and unsaturated burst are listed in Table \ref{bust_table}. For those bursts that can be fit well with complex models, fit results with simple models (i.e., BB, PL) are not listed in Table \ref{bust_table}. We also fit the spectrum of the FRB 200428-Associated Burst, whose result is essentially in agreement with \cite{2021NatAs...5..378L}. In the present paper, we list the parameters of the FRB 200428-Associated Burst from \cite{2021NatAs...5..378L}. 

%As examples of the spectral fitting, the CPL model fit of a saturated burst and an unsaturated burst are shown in Figure \ref{fig:SpecFit}. For the saturated burst as shown in the right panel of Figure \ref{fig:SpecFit}, the parameter $f$ of $const$ of HE is $0.39_{-0.05}^{+0.04}$ \footnote{All errors presented in this paper are for 1 $\sigma$ confidence level.} due to saturation effect, and the parameters $f$ of $const$ of LE and ME are frozen to be 1 since they do not have saturation problem for this burst. 

Following the spectral fitting methods mentioned above, we did the spectral fitting with five spectral models to the \textit{Insight}-HXMT data of 75 bursts from SGR J1935+2154. 
The trigger time, adequately fit spectral model parameters, fit statistics and unabsorbed flux in 1$-$250 keV of each burst as well as their 1 $\sigma$ errors are listed in Table \ref{bust_table}. 
%The errors in 1 $\sigma$ confidence level of model parameters and fluxes are also listed in Table \ref{bust_table}. 
Our fit results of the FRB 200428-Associated Burst agree with those reported in \cite{2021NatAs...5..378L}. We list the published results in Table \ref{bust_table}. Figure \ref{fig:SpecFit} presents the burst spectra of a saturated burst and an unsaturated burst, in the left and right panels, respectively. The relative data loss ratio of HE data is $\sim 39\%$ of the saturated burst %(the right panel of Figure \ref{fig:SpecFit}). Both 
, while LE and ME data of this burst are not affected by the saturation and their constants are fix at one.  
 %$0.39_{-0.05}^{+0.04}$ \footnote{All errors presented in this paper are for 1 $\sigma$ confidence level.}
 
We summarize the results of different models in Table \ref{tab: model_Results}. The number of bursts that can be adequately fit with CPL, BB+BB, BB+PL, PL and BB models are 30, 26, 19, 33 and 19, respectively; while the number of fitting as preferred models of CPL, BB+BB, BB+PL, PL or BB are 11, 3, 1, 18 and 9, respectively. We note that $\sim$ 15 \% (11/75) of the bursts in our sample can be best fit with the CPL model and $\sim$ 24 \% (18/75) can be best fit with the PL model, which means that these bursts do not contain a significant BB component.
Meanwhile, some bursts are equally well described by two or more models. For those bursts, we select the model with minimum BIC value to calculate the burst flux and fluence. The number of bursts with CPL, BB+BB, BB+PL, PL or BB model having minimum BIC values are 24, 13, 3, 22 and 13, respectively.  
%but none model is preferred over other models.

For those 26 bursts, including one saturated, adequately fit by BB+BB model, we study the characteristics of two BB components. The left panel of Figure \ref{fig:kT_index} shows the distributions of temperatures of the low (pink) and high (orange) BB components. The low and high temperature follows a Gaussian trend with the mean value of $\sim2.91$ keV and $\sim12.14$ keV respectively. The best fit parameters of Gaussian fit are listed in Table \ref{parameter_table}. We find no correlation between the temperatures of two BB components (the upper left panel of Figure \ref{fig:par_relation}). The emission area of each BB component can be calculated from the normalization of the spectral fit by assuming that SGR J1935+2154 is 9\,kpc away. 

We study the correlations between parameters of the two BB components. In order to quantify the significance of the correlation, we perform the Spearman's rank order correlation test. We consider a correlation is significant if the chance probability $P<5.7\times10^{-7}$ (about $5\,\sigma$ in a normal distribution). If the chance probability $5.7\times10^{-7}<P<2.7\times10^{-3}$ (about $3\,\sigma$), then the correlation is marginally significant \textcolor{red}{\citep{van_der_Horst_2012}}.
%(van der Horts 2012). 
We further fit a power-law model to the data points of significant and marginally significant correlations \footnote{The PL fit is obtained from linear fitting in logarithmic scale.}, and present results of the Spearman's test as well as the PL fit in Table \ref{correlations_table}. 
%The low and high temperatures are not correlated. 
The emission area of hot and cool BB are marginally correlated. The flux of the cool BB is significantly correlated with that of the hot BB (the lower left panel of Figure \ref{fig:par_relation}). 
As presented in the lower right panel of Figure \ref{fig:par_relation}, for each BB component, the emission area is marginally anti-correlated with the temperature. The power law indices of these two trends are consistent within error.
%the temperatures of both BB components show a clear anti-correlation with the emission area, with the Spearman's rank order correlation coefficient $\rho= -0.90$ and the chance from a random data set $P$=5.10E-20. 
We further fit a power-law model to the data points of two BB components, which yields a power law index of $-3.45{\pm0.23}$. %\textcolor{orange}{(LIN: add kT vs kT in Table 3)}

There are 30 bursts with CPL model as the adequate fitting, seven of which are saturated due to their brightness. The photon index ($\Gamma$) distribution of CPL model is shown in the right panel of Figure \ref{fig:kT_index} (see blue lines), for which the Gaussian mean value is $1.04{\pm0.07}$ (Tabel \ref{parameter_table}). The $E_{\rm peak}$ and $E_{\rm cut}$ distributions are shown in Figure \ref{fig:Epeak} and Figure \ref{fig:Ecut}, respectively. The peak energy ($E_{\rm peak}$) ranges from 12.78 keV to 50.32 keV with an average of 33.56 keV, while the $E_{\rm cut}$ range is from 9.20 keV and 81.80 keV, with the average value of 35.70 keV. We find that there are no significant correlations between photon index ($\alpha$) and $E_{\rm cut}$ with burst flux or fluence (see Figure \ref{fig:index_f_cpl} and Figure \ref{fig:Ecut_f_cpl}).
\textcolor{red}{There are also no clear correlations between $E_{\rm peak}$ and burst flux or fluence, which are shown in Figure \ref{fig:Epeak_f}}.

%which is shown in the left panel of Figure \ref{fig:Epeak_f}. 
%Interestingly, we note that there seem to be two $E_{\rm peak}$-fluence trends in the right panel of Figure \ref{fig:Epeak_f}. For some bursts including those saturated bursts, $E_{\rm peak}$ seems to become harder as the fluence increases. However, many bursts with lower fluence ($<10^{-7}$ erg cm$^{-2}$) seem to be away from this trend.

For 19 bursts with the adequate model of BB+PL, the distribution of the BB+PL model parameters are shown in Figure \ref{fig:kT_index}. The BB temperature and photon index of the BB+PL model follow a Gaussian distribution with the best fit mean values of $7.72{\pm0.39}$ keV and $1.80{\pm0.02}$, respectively (see Table \ref{parameter_table}).

Finally, some bursts in our sample can be adequately fit with simple models (i.e. BB or PL). The mean temperature for 19 bursts with BB model is $11.75{\pm0.71}$ keV, and the mean photon index for 33 PL bursts is $1.76{\pm0.02}$ (see Figure \ref{fig:kT_index}). 
These values with their statistical errors are also shown in Table \ref{parameter_table}. We note that those bursts that can be fit better with simple models are quite dim, with the highest fluence of about $2.5\times10^{-8}$ erg cm$^{-2}$. Interestingly, we find that the photon index of the simple PL model is similar to the PL index of the BB+PL model, and the temperature kT of the simple BB model is similar to the high temperature of the BB+BB model.

\subsection{Joint spectral fit} \label{sec_joint_fit}

In Figure \ref{fig:JointFit}, we present an example of joint spectral analysis of \textit{Fermi}/GBM and \textit{Insight}-HXMT. We first analyze the joint spectrum in 8$-$200\,keV. Our results are consistent with the spectral fit with GBM only data within 1 $\sigma$ confidence level \citep{Lin_2020}. The constants of GBM, HE, ME and LE are 0.82, 0.83, 1 and 1 for the CPL fit in 8$-$200 keV; The constants of GBM, HE, ME and LE are 0.84, 0.94, 1 and 1 for the BB+BB fit in 8$-$200 keV. The difference between HE and LE/ME is due to saturation effect. In the narrow energy range, if only using the GBM data, the BIC value of the CPL model fit is smaller than BIC of BB+BB model. However, the difference is not enough to conclude that CPL is significantly preferred. We also fit the joint spectrum in 1$-$250\,keV. The model parameters agree with the values listed in Table \ref{bust_table}, while quite different from the results from narrow energy band fit. The model parameters (e.g., photon index, $E_{\rm cut}$, low and high temperature) of the broader band are smaller than that of the narrow band (see Figure \ref{fig:JointFit} for more details). Moreover, using data in the broader energy range, CPL model is significantly preferred over BB+BB model with a reduction of 139 in the BIC value. 
%As shown in Figure \ref{fig:JointFit}, we select burst \#74 
%for which the trigger time is 2020-05-20T21:47:07.480 UTC.
%in our catalog as an example for joint spectral analyses,  
%/gecamfs/home/GSDC/source_files/caice/LC
%We take a burst which is detected by both \textit{Insight}-HXMT and \textit{Fermi}/GBM, 
%as an example to show the joint spectra fit results with CPL and BB+BB models. 
%The joint spectral analysis of \textit{Fermi}/GBM (whose energy range used for this SGR spectral analysis is fixed to 10-200 keV ) and 8$-$200 keV data of \textit{Insight}-HXMT is basically consistent with the result of the GBM-only fit \citep{Lin_2020} within a confidence level of 1 $\sigma$. 
%However, joint spectral fit results with data in the broader energy range 1$-$250 keV of \textit{Insight}-HXMT and \textit{Fermi}/GBM are significantly different from the results of GBM-only fitting but consistent with the results of  \textit{Insight}-HXMT only fitting (see Table \ref{bust_table}). 

% in Discussion?
The above facts demonstrate that the cross calibration of energy responses between \textit{Insight}-HXMT and \textit{Fermi}/GBM in the common energy range (i.e. 8-200 keV) are adequately well and the low energy band (i.e. 1-10 keV) of \textit{Insight}-HXMT plays an important role in determining the spectral shape.
%Indeed, we notice that for this burst, it is not easy to be determined whether BB+BB or CPL model is preferred according to the BIC statistics using GBM data alone \citep{Lin_2020}. However, it shows that the CPL model is statistically better than BB+BB with the broad energy band (1-250 keV) \textit{Insight}-HXMT data. 

\section{DISCUSSION} \label{sec:discussion}

\begin{figure}
\centering
\begin{tabular}{c}
\includegraphics[width=0.50\textwidth]{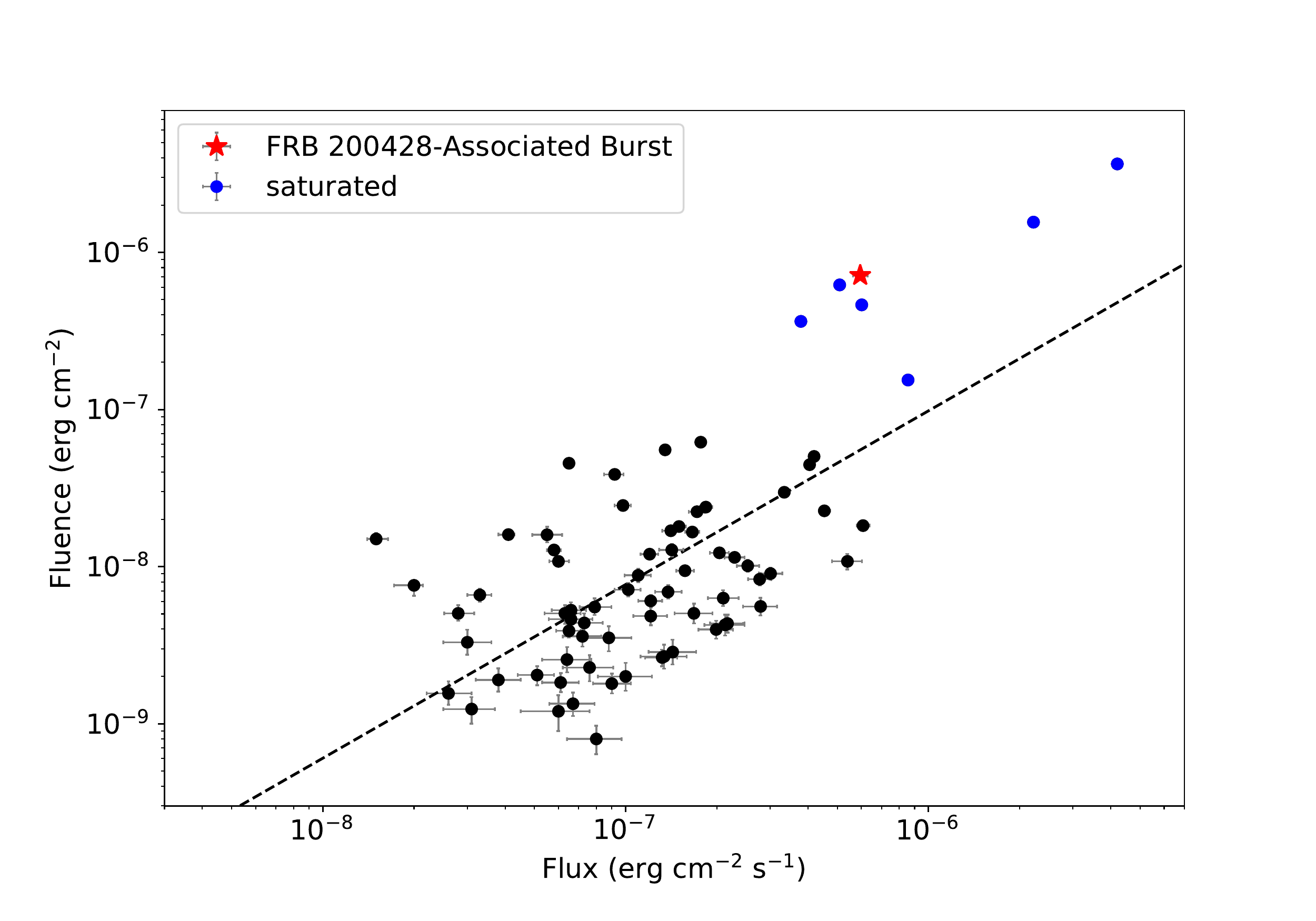} \\
\end{tabular}
\caption{\label{fig:fluence_flux} The scatter plot of the fluence versus flux in 1$-$250 keV of each burst. The red star and blue points represent the FRB 200428-Associated Burst and saturated bursts, respectively. There is a strong correlation between these two parameters. The black dashed line is the PL fit with an index of $1.11{\pm0.13}$. }
\end{figure}

\begin{figure}
\centering
\begin{tabular}{c}
\includegraphics[width=0.5\textwidth]{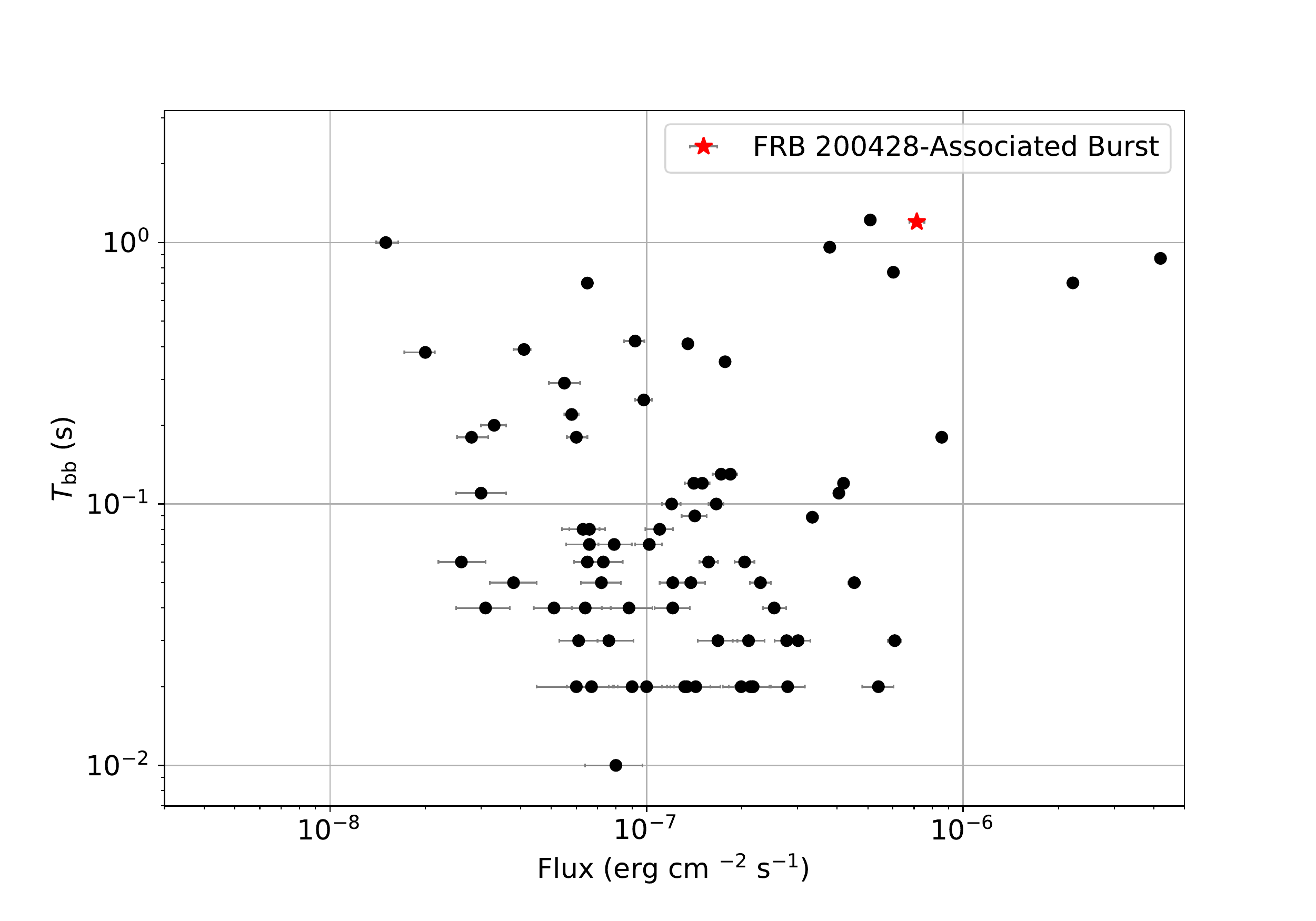}\\
\end{tabular}
\caption{\label{fig:flux_TT} The scatter plot of $T_{\rm bb}$ and energy flux (1-250 keV). The red star is the FRB 200428-Associated Burst.}
\end{figure}

\begin{figure}
\centering
\begin{tabular}{c}
\includegraphics[width=0.5\textwidth]{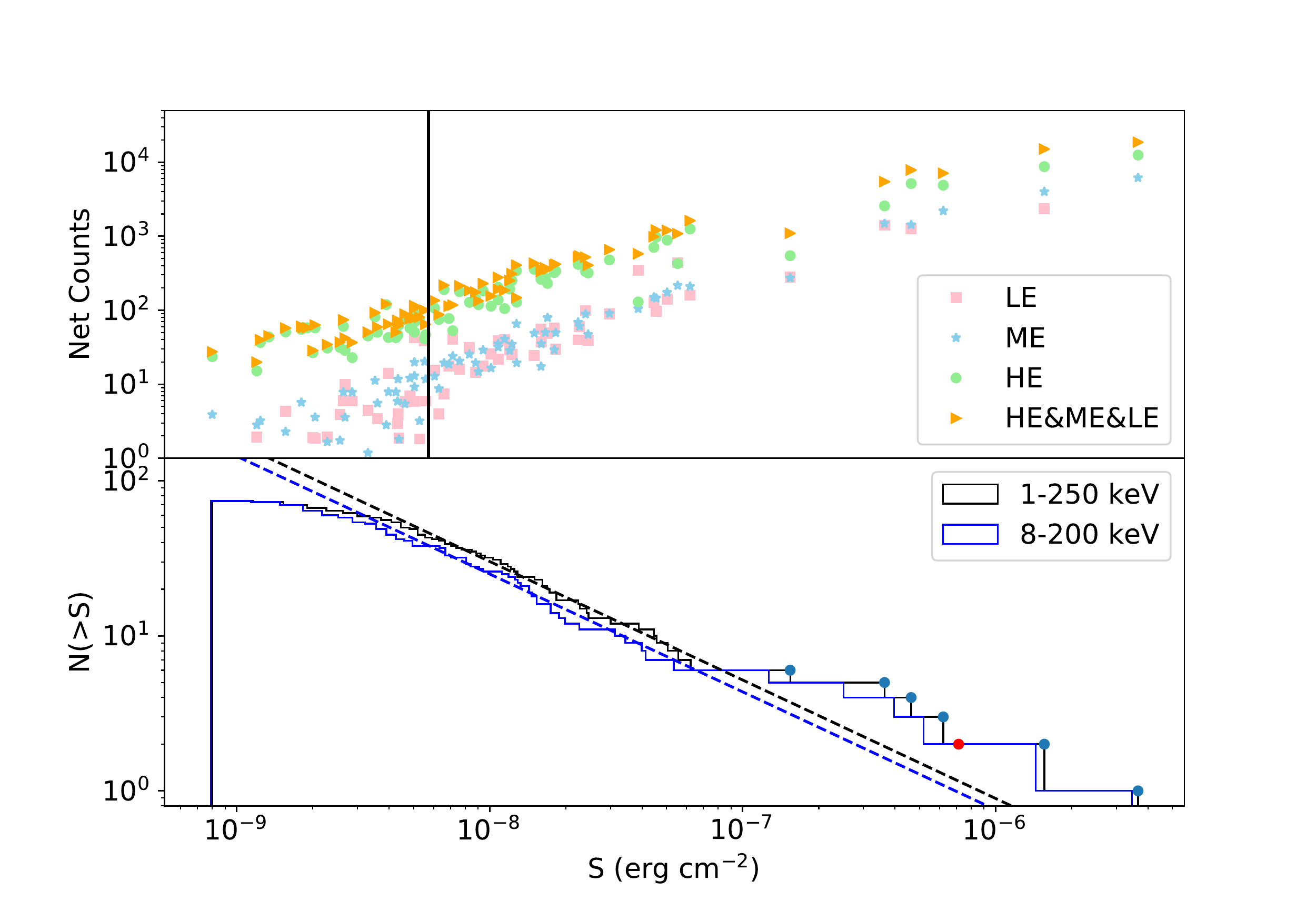} \\
\end{tabular}
\caption{\label{fig:fluenceDist1} \textit{Top}: The relation between the net counts and fluence. The orange triangles are the total net counts of HE, ME and LE. The green points, blue stars and pink squares represent the net counts of HE, ME and LE, respectively. The black solid line is the minimum fluence (1$-$250 keV) used for the PL fits. \textit{Bottom}: The cumulative fluence distributions of SGR J1935+2154 bursts. The black and gray lines represent the different energy ranges of 1$-$250 keV and 8$-$200 keV. The black and gray dashed lines are the best PL fit to the distribution of the range $5.73\times10^{-9}$ to $1.09\times10^{-7}$ erg cm$^{-2}$ for 1$-$250 keV and $5.46\times10^{-9}$ to $1.04\times10^{-7}$ erg cm$^{-2}$ for 8$-$200 keV, respectively. The indices of the PL fit are $0.764{\pm0.004}$ for 1$-$250 keV and $0.760{\pm0.007}$ for 8$-$200 keV. The blue and red points represent the saturated bursts and the FRB 200428-Associated Burst, respectively.
}
\end{figure}

\begin{figure}
\centering
\begin{tabular}{cc}
\includegraphics[width=0.50\textwidth]{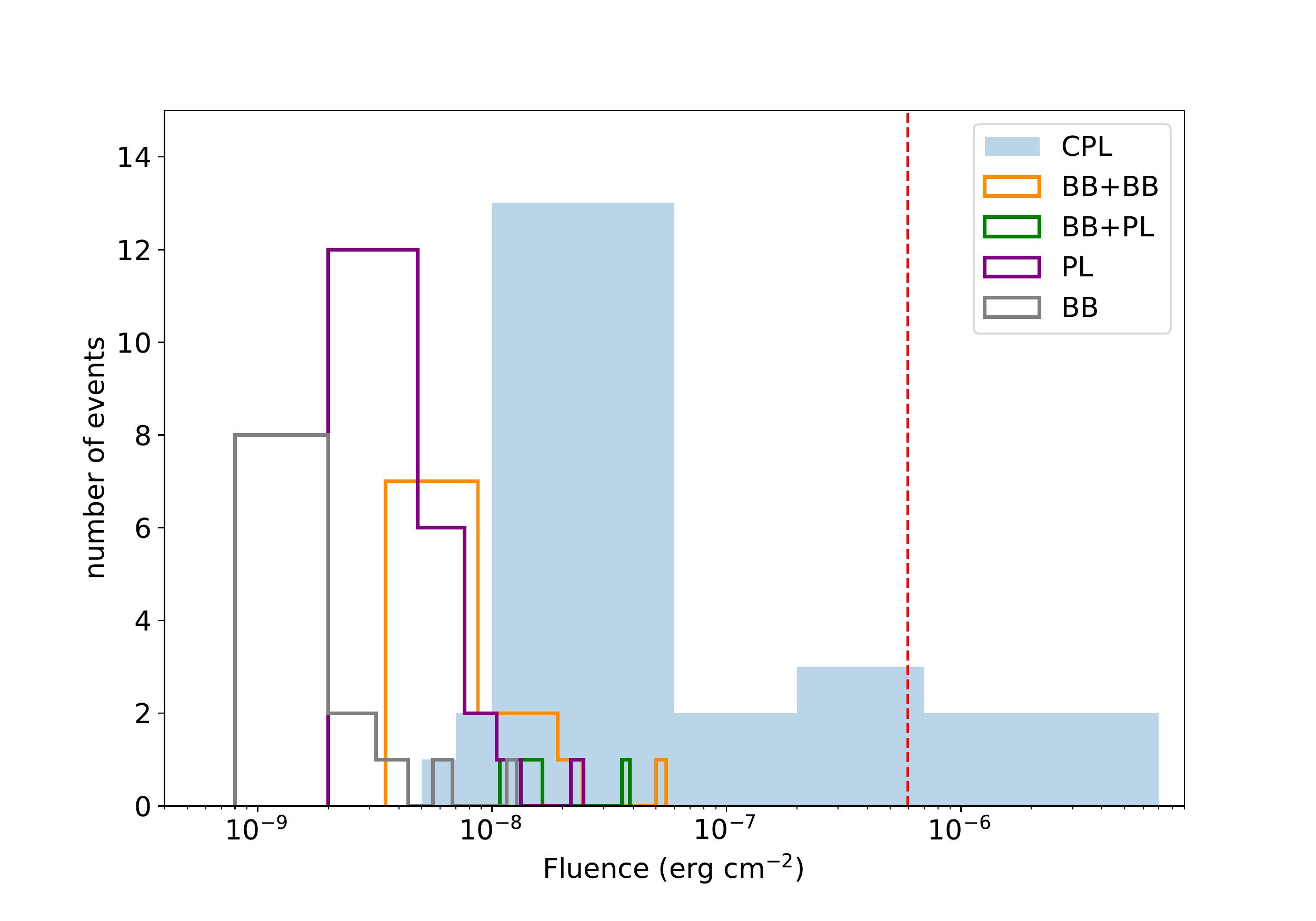} \\
\end{tabular}
\caption{\label{fig:Model_fluence} The distribution of the fluence in 1$-$250 keV for each model. The blue, orange, green, purple and gray lines represent CPL, BB+BB, BB+PL, PL and BB models, of which the bursts are fit with the minimum Bayesian Information Criterion value. The red line is the FRB 200428-Associated Burst.}
\end{figure}

\begin{figure}
\centering
\begin{tabular}{c}
\includegraphics[width=0.5\textwidth]{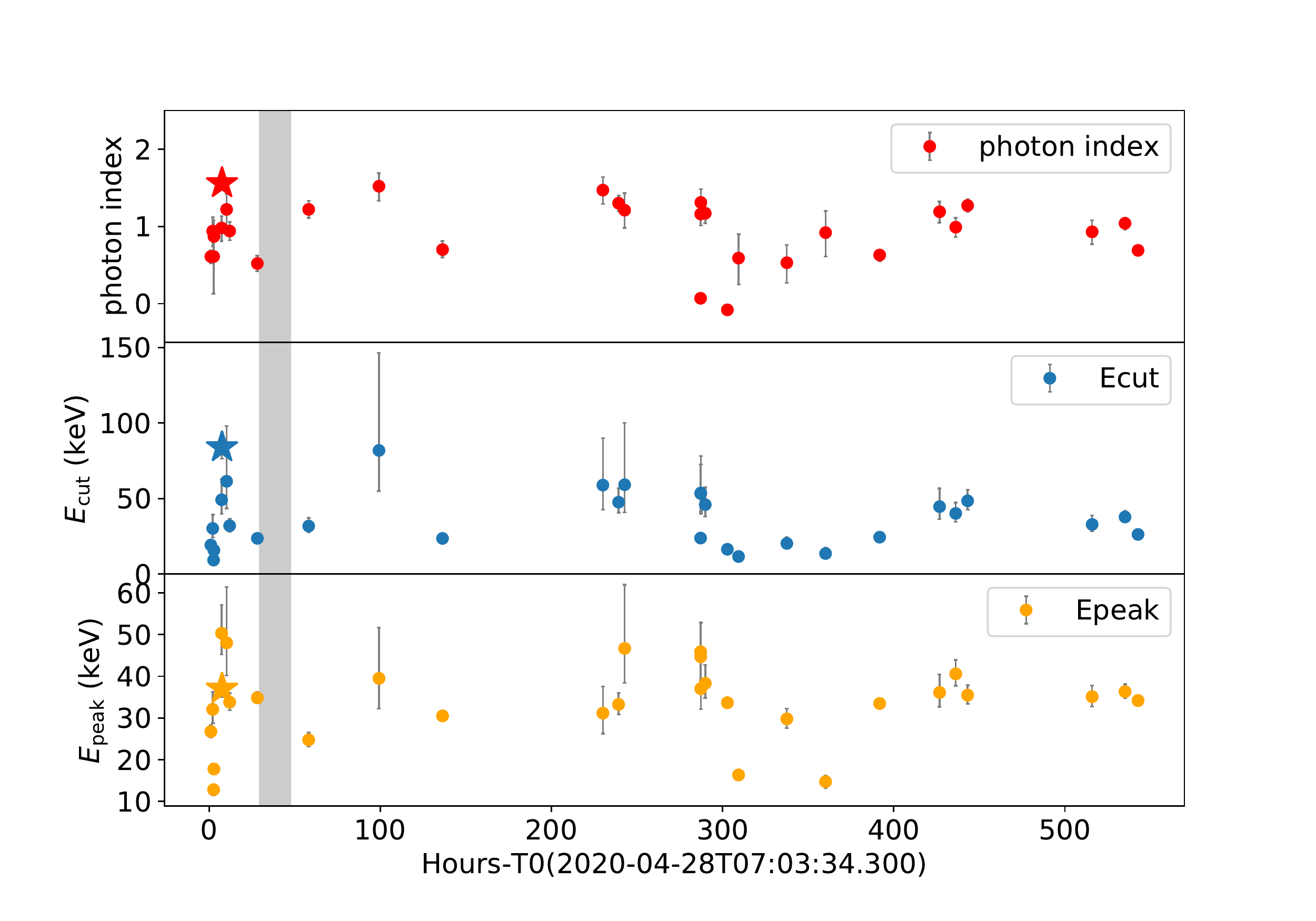} \\
\end{tabular}
\caption{\label{fig:Time_model} The evolution of the parameters of the CPL model. Photon index, $E_{\rm cut}$ and $E_{\rm peak}$ are marked as red, blue and orange points. The stars represent the FRB 200428-Associated Burst. The gray shadow is the time interval of the incomplete monitoring of SGR J1935+2154.}
\end{figure}

\begin{figure}
\centering
\begin{tabular}{c}
\includegraphics[width=0.50\textwidth]{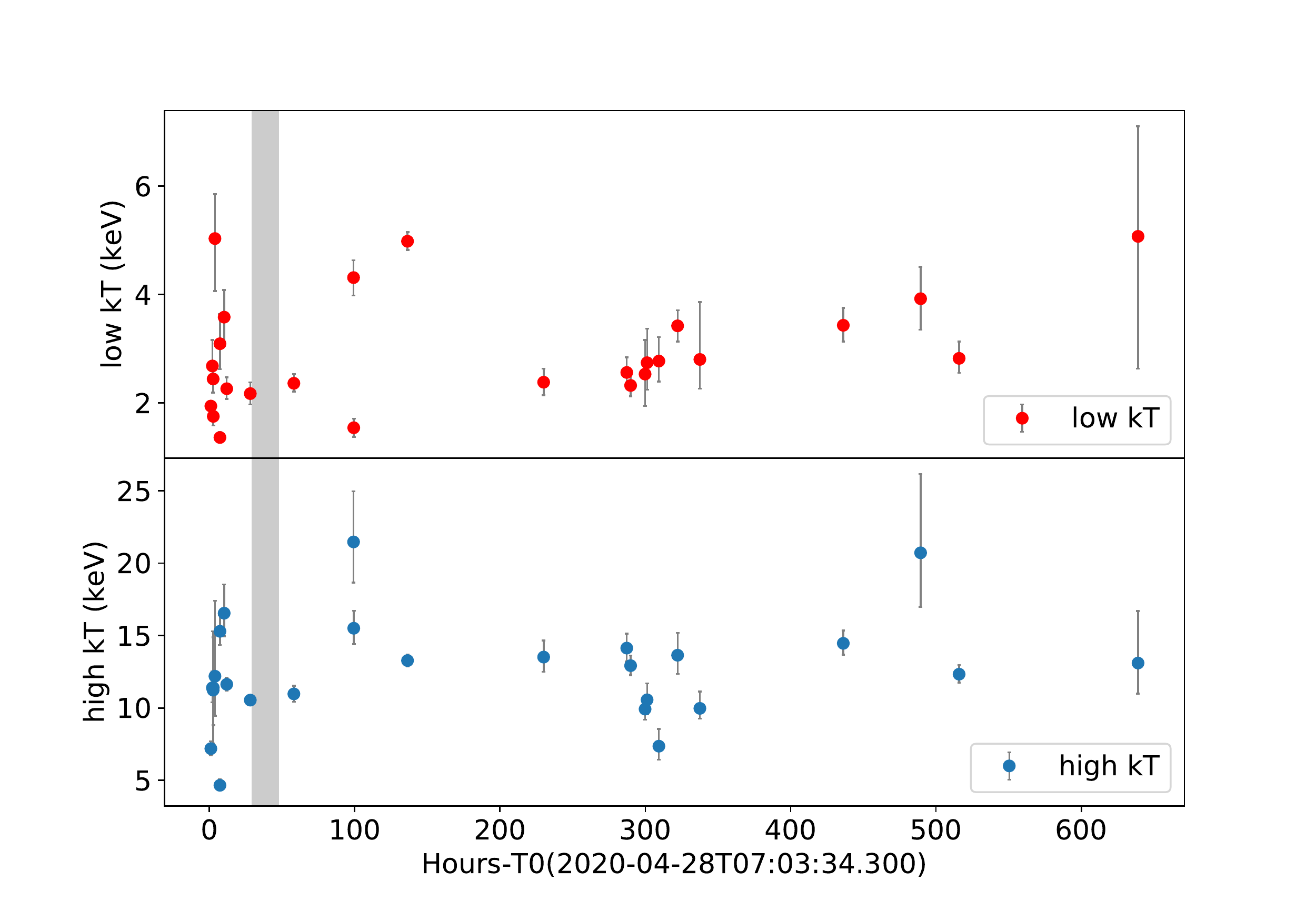}  \\
\end{tabular}
\caption{\label{fig:Time_model1} The evolution of the parameters of the BB+BB model. Low kT and high kT are marked as red and blue points. The gray shadow is the time gap of the monitoring of SGR J1935+2154.}
\end{figure}

\begin{figure}
\centering
\begin{tabular}{cc}
\includegraphics[width=0.50\textwidth]{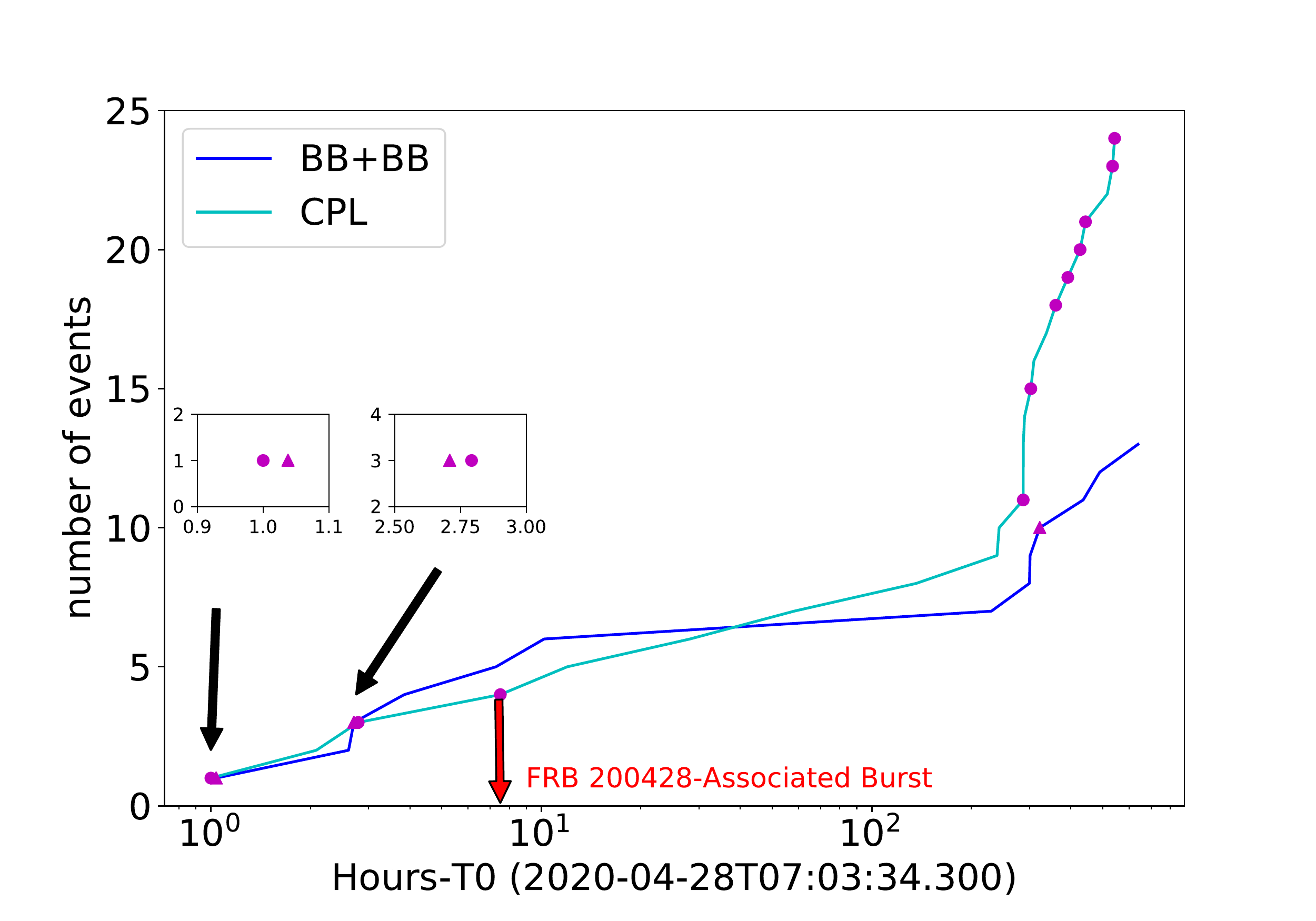} \\
\end{tabular}
\caption{\label{fig:Model_TT} The evolution of the fitting models for CPL and BB+BB. The blue and green lines mark BB+BB models (13) and CPL models (24), for which the bursts are fit with the minimum BIC value. The triangles and points that represent the preferred models are BB+BB models (3) and CPL models (11) using Bayesian Information Criterion of $\Delta>10$ (see Section \ref{Spectral analyses}).}
\end{figure}

\begin{figure}
\centering
\begin{tabular}{c}
\includegraphics[width=0.5\textwidth]{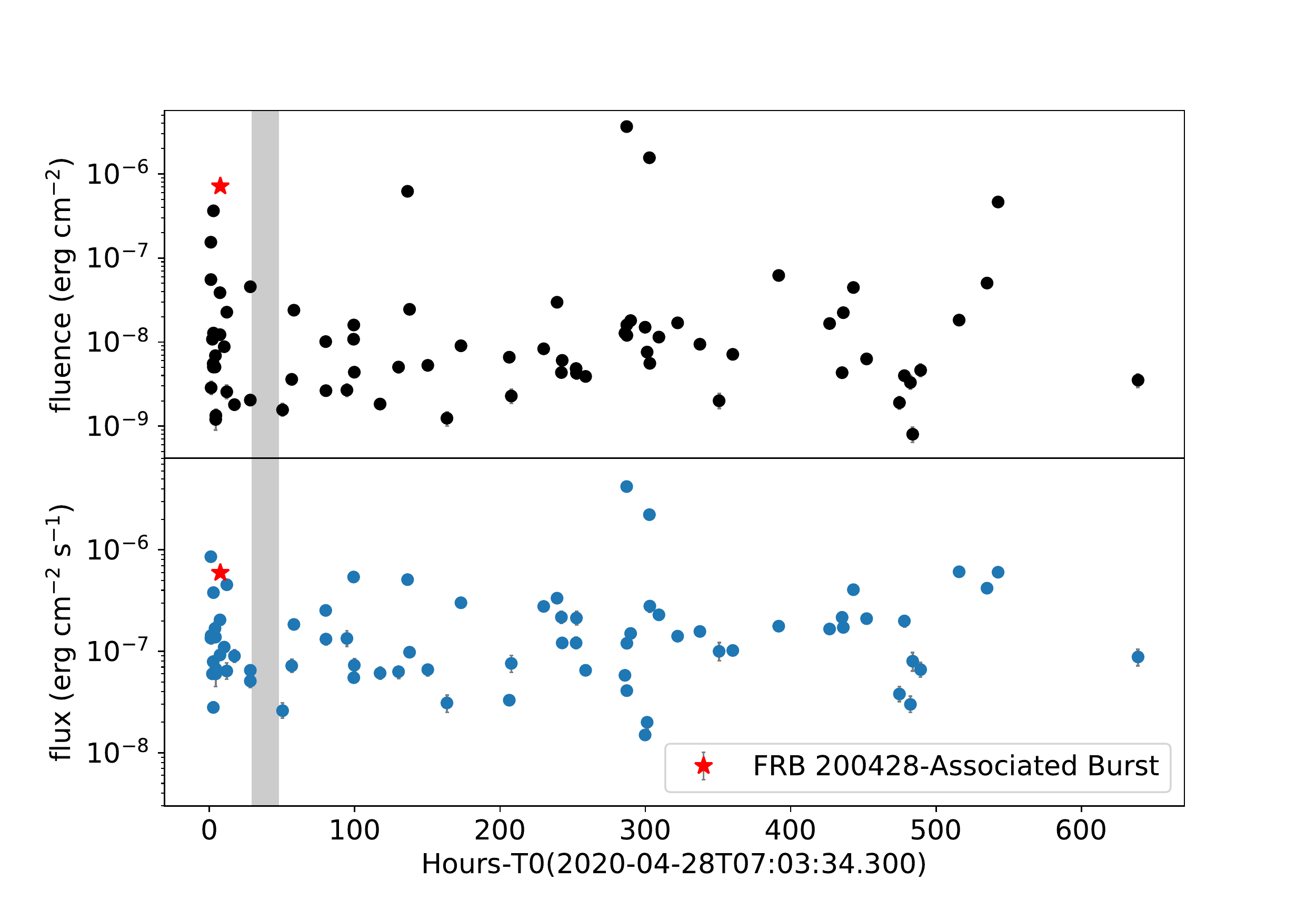} \\
\end{tabular}
\caption{\label{fig:Time_fluence} The evolution of fluence (top panel) and flux (bottom panel) in 1$-$250 keV. The stars represent the FRB 200428-Associated Burst. The gray shadow is the time gap of the monitoring of SGR J1935+2154 of \textit{Insight}-HXMT. }
\end{figure}

\begin{figure}
\centering
\begin{tabular}{c}
\includegraphics[width=0.50\textwidth]{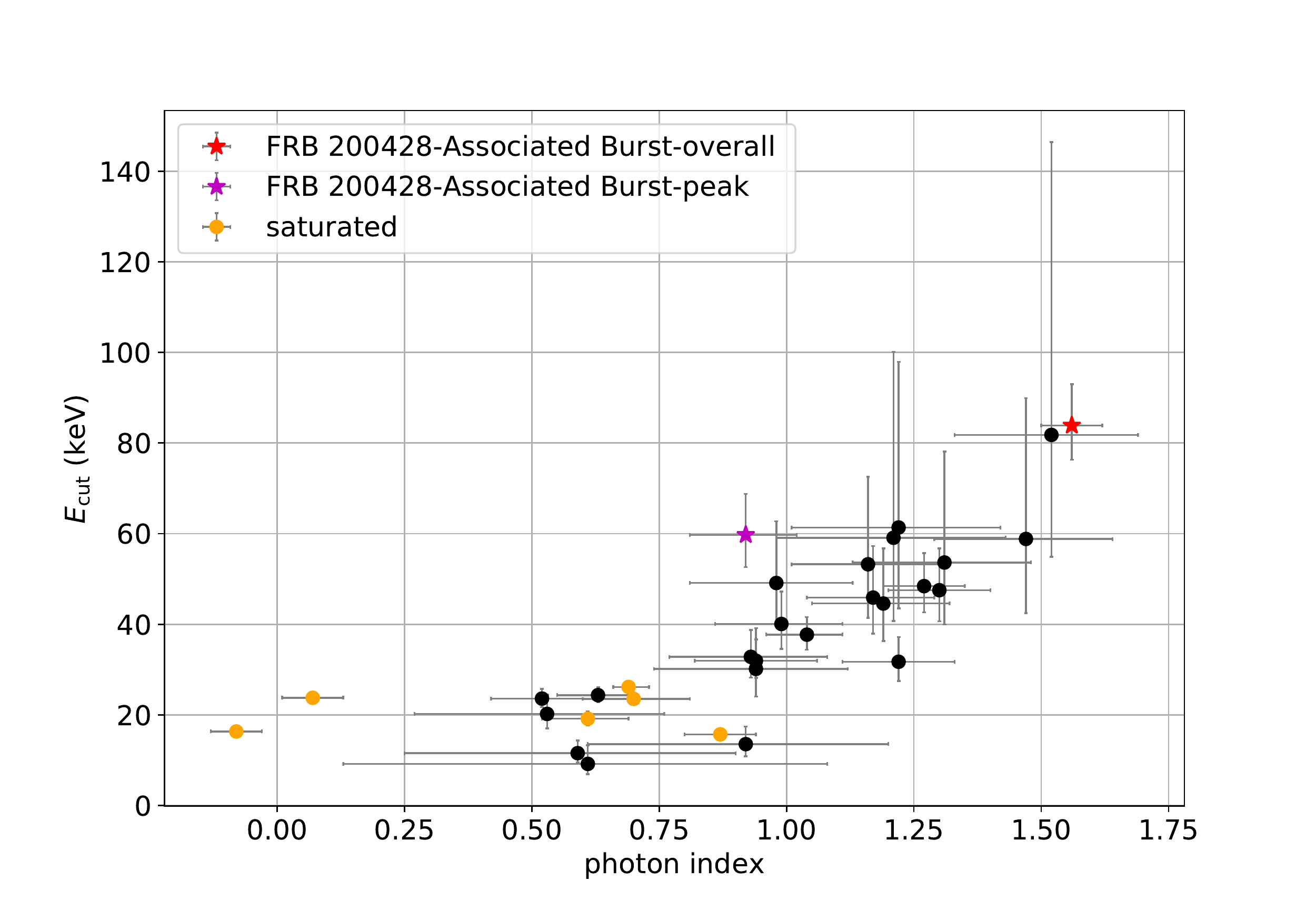} \\
\end{tabular}
\caption{\label{fig:Index_Ecut} The scatter plot of $E_{\rm cut}$ and photon index. Orange points represent the saturated bursts. The red and purple stars are the time-integral and peak spectra of FRB 200428-Associated Burst, respectively.}
\end{figure}

\begin{figure}
\centering
\begin{tabular}{c}
\includegraphics[width=0.50\textwidth]{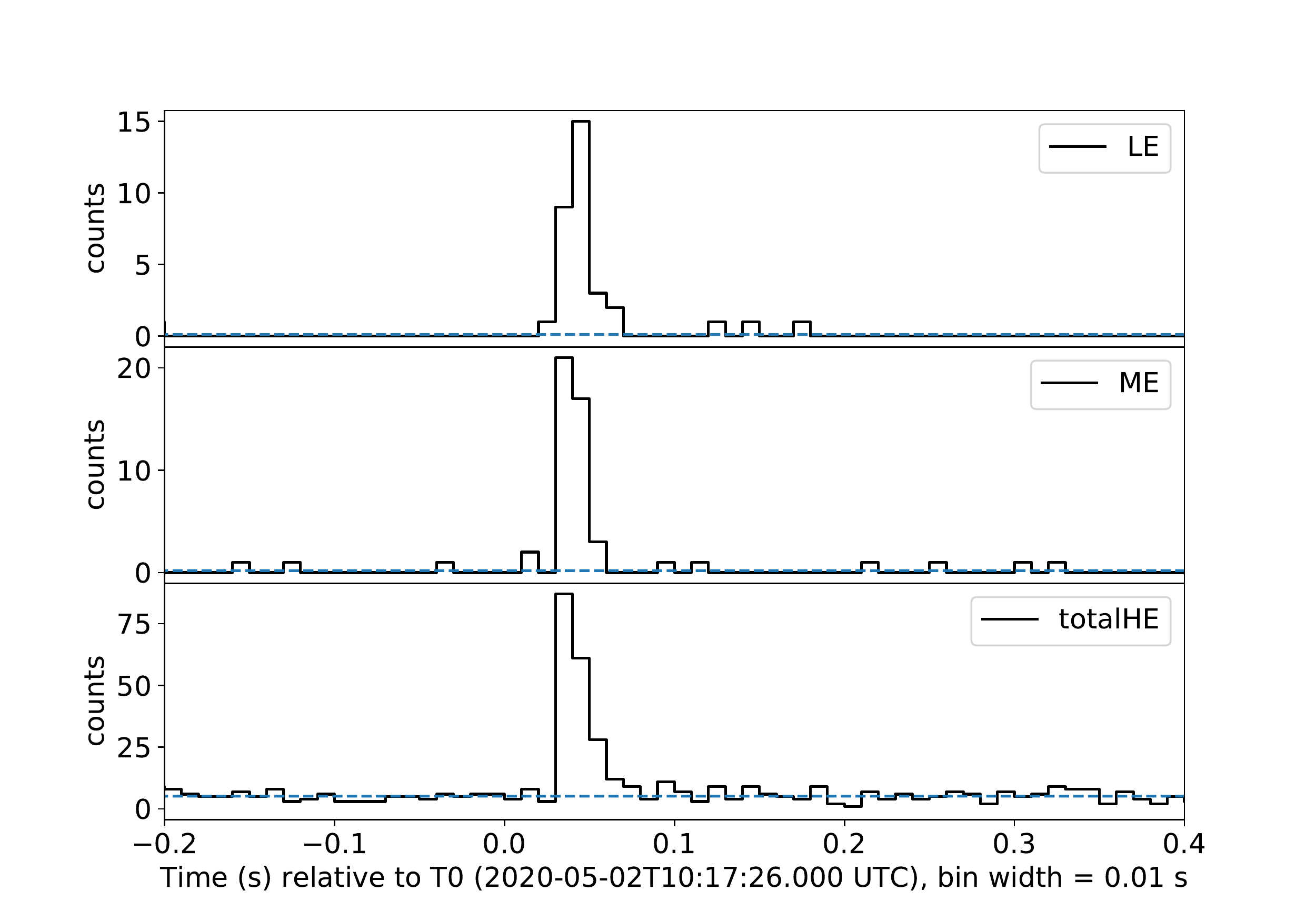} \\
\end{tabular}
\caption{\label{fig:burstLC} The light curves of the burst whose spectrum resembles that of the FRB 200428-Associated Burst. 
The light curves observed with LE (1$-$10 keV), ME (10$-$30 keV) and HE (28$-$250 keV) with a time resolution of 10 ms are shown in the top, middle and bottom panels, respectively.}
\end{figure}

As shown above, we have analyzed the time-integrated spectral properties of
75 bursts from SGR J1935+2154 observed with \textit{Insight}-HXMT. The fluence of this burst sample ranges from $8\times10^{-10}$ erg cm$^{-2}$ to $3.46\times10^{-6}$ erg cm$^{-2}$ in the energy range of 8$-$200 keV, with the minimum fluence about an order of magnitude dimmer than that of observations by \textit{Fermi}/GBM or \textit{Swift}/BAT (e.g., \cite{2020ApJ...893..156L,Lin_2020}. The total fluence emitted in our burst sample is $6.81\times10^{-6}$ erg cm$^{-2}$ (8$-$200 keV), corresponding to $6.57\times10^{40}$ erg (for a distance of 9 kpc). The FRB 200428-Associated Burst is brighter than $\sim$96 \% (72/75) of events in our sample.

\subsection{Comparison of Burst Spectra with that of GBM} 

The parameters of the CPL and BB+BB models derived from \textit{Insight}-HXMT are somewhat different from that of the GBM observations. We find that the mean temperature of the harder BB component ($\sim$ 12.14 keV) is consistent with the results from GBM observations during 2020, while the mean temperature of the softer BB component ($\sim$ 2.91 keV) is lower than that of GBM ($\sim$ 4.50 keV) \citep{Lin_2020}. 
The mean value of spectral photon index is $\sim$ 1.04, softer than the index measured by GBM ($\sim$ 0.10), and the average $E_{\rm peak}$ is $\sim$ 34.14 keV, higher than GBM ($\sim$ 26.30 keV) \citep{Lin_2020}. 
Such differences could be caused by the evolution of the spectral properties of bursts from SGR J1935+2154.
However, as shown in section \ref{sec_joint_fit} of \textit{Insight}-HXMT, the spectral parameters derived from the boarder band (1$-$250 keV) is different from the result of the narrower band (8$-$200 keV). Therefore, the differences between the spectral parameters given by \textit{Insight}-HXMT and \textit{Fermi}/GBM are also likely due to the different energy ranges between these two instruments. Nevertheless, there is no doubt that wide energy bandwidth is very important to accurately measure the whole burst spectrum and the 1-10 keV data of \textit{Insight}-HXMT is necessary to constrain the low energy shape of the burst spectrum of SGR J1935+2154.

\subsection{Burst Energies} 
For \textit{Insight}-HXMT detected 75 bursts from SGR J1935+2154, we find that the fluence is correlated with flux (both in the 1$-$250 keV), as shown in Figure \ref{fig:fluence_flux}. To quantify this correlation, we compute the Spearman's rank order correlation coefficient, $\rho= 0.57$, and the chance from a random data set, $P = 4.86\times10^{-8}$. A power law fitted to the mean values of the data using the least-squares technique yields an index of $1.11{\pm0.13}$. This correlation is not a surprise because the fluence is derived from the (averaged) flux and burst duration, and the distribution of duration is relatively narrow. Thus we check the relation between averaged flux and duration, but do not find a correlation between them (see Figure \ref{fig:flux_TT}).
%There are six saturated bursts with longer duration that are more brighter. 

The cumulative distribution of the fluence ($S$) of all 75 bursts is shown in the bottom panel of Figure \ref{fig:fluenceDist1}. Since there are observational selection effects in the low fluence region (less than about $5\times10^{-9}$ erg cm$^{-2}$) and statistical fluctuation due to small numbers in the high fluence end (greater than about $1\times10^{-7}$ erg cm$^{-2}$), we focus on the distribution of those bursts between $5\times10^{-9}$ erg cm$^{-2}$ to $1\times10^{-7}$ erg cm$^{-2}$ (in 1$-$250 keV), which is well fit by a power law with
%with about fifty percent bursts (exclude the saturated bursts) of which the range is from . 
a best-fit index of $0.764{\pm0.004}$. 
For a direct comparison with \textit{Fermi}/GBM (8-200 keV) results \citep{Lin_2020}, we also fit the fluence in 8-200 keV of \textit{Insight}-HXMT, and get a best-fit PL index of $0.760{\pm0.007}$, which is basically the same as the result derived with fluence in 1-250 keV. This PL index is also comparable to the result reported for the previous active episodes from SGR J1935+2154 by \textit{Fermi}/GBM \citep{2020ApJ...893..156L,Lin_2020} and other magnetars \citep{ 1996Natur.382..518C, 2015ApJS..218...11C}. 
\cite{2020NICER} reported a PL fit with PL index of $\sim$0.5 to the fluence distribution of SGRJ1935+2154 bursts during burst storms, but this measurement is done in a relatively narrower energy band (0.5$-$10 keV).

Thanks to its high sensitivity, \textit{Insight}-HXMT unveiled many bursts with fluence less than $1\times10^{-7}$ erg cm$^{-2}$, below which the measurements of \textit{Fermi}/GBM suffer observational effects and the burst energy distribution deviates from PL \citep{Lin_2020}. Thus, this \textit{Insight}-HXMT measurement indicates that the bursts with lower fluence (less than $1\times10^{-7}$ erg cm$^{-2}$) also follow the same PL distribution as that of higher fluence bursts. 

As mentioned before, there are apparent deviations seen in both the high and low fluence regions (bottom panel of Figure \ref{fig:fluenceDist1}). For the high fluence region (greater than about $1\times10^{-7}$ erg cm$^{-2}$) which includes bright bursts (all are saturated), we calculate the expected cumulative event number by extrapolating the PL fit and find that these numbers in the six bins are 6, 2, 2, 1, 1, 1, 0, well consistent with the number of the detected events. Therefore, the deviation from the PL fit is very probably due to the statistical fluctuations of small number of bursts. 

As for the ramping feature in the low fluence region, we caution that this deviation from PL distribution could be easily misleading. It is the net counts (rather than the fluence) of each telescope that determine whether the burst could be triggered or not (see paper \uppercase\expandafter{\romannumeral1}). We note that, since the fluence depends on the duration and spectral shape of a burst, bursts with basically the same net counts around the trigger threshold show a wide distribution of fluence extending below $\sim$ $5\times10^{-9}$ erg cm$^{-2}$
%and with the same fluences show the different net counts 
(see the upper panel of Figure \ref{fig:fluenceDist1}).
%Therefore, those bursts with similar net counts around the trigger threshold have a wide distribution of fluence are further from the PL fit,
%which mainly reveals that some bursts with the same fluence do not trigger our search due to different spectra resulting in fewer net counts. 
In principle, the detection efficiency of weak bursts is heavily affected by the instrumental sensitivity, search algorithm and the burst properties (e.g., duration, spectra). Therefore the deviation of PL in the low fluence end is caused by those factors.

Bursts could be grouped according to the good fit spectral model (see Table \ref{bust_table}), i.e. which model gives the minimum BIC. We investigate the fluence distribution for bursts in each model group, as shown in Figure \ref{fig:Model_fluence}. We find that the high fluence bursts are almost always fitted with the CPL model and weak bursts are always fitted with simple models (i.e. BB and PL). We note that it is possible that these weak bursts could also have complex spectral components (such as CPL, BB+BB or PL+BB ) but just have insufficient statistics to support the identification of complex models rather than simple ones.

\subsection{Burst Evolution} 
To investigate how the spectral properties of bursts evolve throughout this burst episode of SGR J1935+2154, we plot the burst photon index, peak energies and cutoff energies of bursts derived with the CPL models in Figure \ref{fig:Time_model}, and find that there is no significant evolution of these parameters. The high kT and low kT of the BB+BB model across this episode also have no significant evolution (Figure \ref{fig:Time_model1}).

We present the temporal evolution of the preferred spectrum model (BB+BB or CPL) of the bursts in our sample in Figure \ref{fig:Model_TT}. Interestingly, those bursts with CPL as preferred model mostly occurred in the later part of this burst episode where the burst rate is significantly lower than the active period (see Figure 5 in Paper I).
%the dedicated 33-day observation with \textit{Insight}-HXMT. 
We also check the fluence and flux of bursts (Figure \ref{fig:Time_fluence}) and find there is no apparent evolution in the flux and fluence over time. 
We note that the FRB 200428-Associated Burst has high (but not the highest) flux and fluence with the preferred CPL spectral model and occurred in the early stage where the burst rate is very high.

\begin{figure}
\centering
\begin{tabular}{c}
\includegraphics[width=0.50\textwidth]{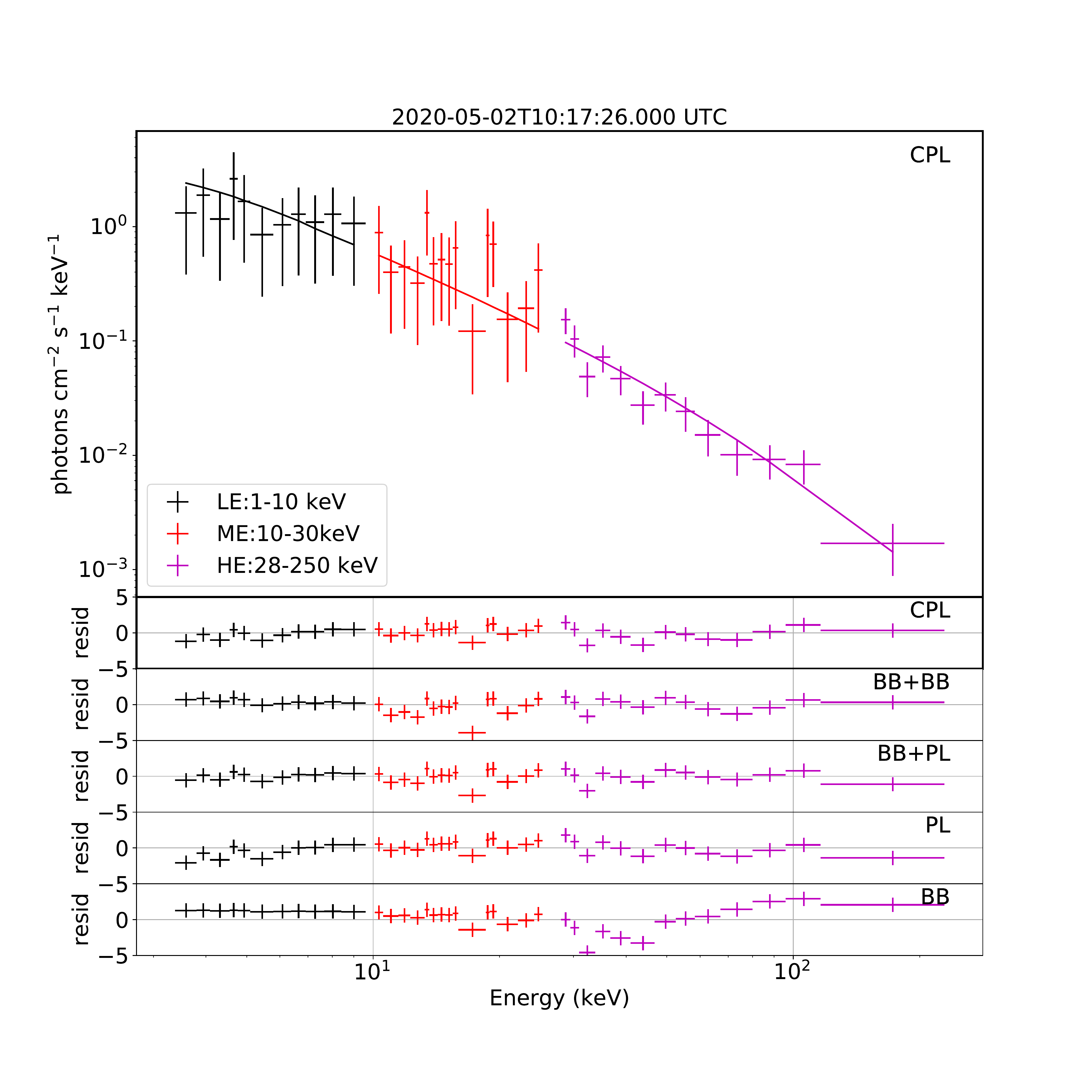} \\
\end{tabular}
\caption{\label{fig:burstSpec} The spectrum of the SGR J1935+2154 burst whose spectrum resembles that of FRB 200428-Associated Burst. Data from LE, ME and HE of \textit{Insight}-HXMT are represented in different colors (LE: black, ME red, HE purple). The X-ray spectrum of the burst described by the CPL model is shown in the top panel. The five bottom panels are the residuals of the data from the individual models of CPL, BB+BB, BB+PL, PL and BB. 
}
\end{figure}

\subsection{Burst Association with Radio Burst} 
In this burst sample of \textit{Insight}-HXMT, there is only one burst associated with a radio burst, i.e. the FRB200428-Associated Burst, which has a longer duration (~1.2 s, \cite{2021NatAs...5..378L}) than other bursts in the sample. 
If consider the full period of the FRB 200428-Associated Burst (~1.2 s), its $E_{\rm peak}$ ($\sim$ 37 keV) is not special (Figure \ref{fig:Epeak}), but the $E_{\rm cut}$ ($\sim$ 84 keV) and photon index ($\sim$ 1.56) are slightly higher than other bursts in our sample (the right panels of Figure \ref{fig:kT_index} and Figure \ref{fig:Ecut}) \citep{2021NatAs...5..378L}. 

Interestingly, there is one burst in our sample (burst \#28) whose photon index ($\sim$ 1.52) and $E_{\rm cut}$ ($\sim$ 82 keV) resemble that of the time-integrated spectrum of the FRB 200428-Associated Burst, as shown in Figure \ref{fig:Index_Ecut}. 
%At the trigger time of this burst (2020-05-02T10:17:26.000 UTC), the \textit{Fermi}/GBM was occulted by the Earth, thus there is no joint analysis between \textit{Insight}-HXMT and \textit{Fermi}/GBM.
The light curve and spectrum of this burst are shown in Figure \ref{fig:burstLC} and Figure \ref{fig:burstSpec}, respectively. 
This burst has a duration of about 0.02 s, which is much less than the full burst but comparable to the peak part of the FRB 200428-Associated Burst.  

\cite{2021NatAs...5..378L} identified that the FRB 200428-Associated Burst has two narrow peaks with a separation of $\sim$ 30 ms, consistent with the separation between the two bursts in FRB 200428. If only consider the bright part of FRB200428-Associated Burst, which lasts for about 0.06 s, although both the photon index ($\sim$ 0.9) and $E_{\rm cut}$ ($\sim$ 60 keV) are not atypical (Figure \ref{fig:Index_Ecut}), its $E_{\rm peak}$ ($\sim$ 65 keV) is somewhat high (which is also reported in \cite{2021Ridnaia}). No other burst in this \textit{Insight}-HXMT sample has such a high $E_{\rm peak}$ (Figure \ref{fig:Epeak}).

%Yet, the photon index ($\sim$ 0.92), $E_{\rm cut}$ ($\sim$ 59.74 keV) and $E_{\rm peak}$ ($\sim$ 64.64 keV) of the peak spectrum of FRB 200428-Associated Burst are shown in Figure 2 in \cite{2021NatAs...5..378L}. 
%The $E_{\rm peak}$ is rare in our burst samples (see Figure \ref{fig:Epeak}). 
%There is no burst with the same spectrum as the peak of the FRB 200428-Associated Burst.

We check the radio observational campaign of SGR J1935+2154 and found that there were radio observations (e.g., CHIME/Pulsar, Northern Cross radio telescope) of this source during this burst but without any detection of radio bursts from SGR1935+2154 \citep{2020ATel13838....1T,2020ATel13739....1N}.

\section{Summary} \label{sec:Summary}
In this paper, we report the detailed results of time-integrated spectral analyses of the 75 bursts from SGR J1935+2154 detected by \textit{Insight}-HXMT during the dedicated 33-day ToO observation. 
The wide energy range (1$-$250 keV) and high sensitivity of \textit{Insight}-HXMT allow for accurate spectral characterization of magnetar bursts. 

Five spectral models are adopted to fit the spectra: the sum of two blackbody functions (BB+BB), cutoff power-law (CPL), blackbody and power law functions (BB+PL), power law (PL), and single blackbody (BB). The \textit{Insight}-HXMT and \textit{Fermi}/GBM joint fits show that the preferred model can be more easily identified when \textit{Insight}-HXMT data in broader energy band are used in the joint fitting. For about $\sim15\%$ of the 75 bursts, the CPL model is preferred. The photon index and $E_{\rm peak}$ of CPL model for 30 bursts center at the mean value of 1.04 and 34.14 keV, respectively. The temperatures of low and high BB components of the BB+BB model center around $\sim2.9$ keV and $\sim12.1$ keV, respectively. 

The burst fluences range from $\sim 10^{-9}$ to $\sim 10^{-6}$ erg cm$^{-2}$ in the energy range of 8$-$200 keV. Although these bursts detected by \textit{Insight}-HXMT in this observation campaign is less energetic, the cumulative distribution of fluence follows the same power law trend with that of brighter bursts reported by \textit{Fermi}/GBM. We find that the deviation of the weaker bursts from the power law of the cumulative fluence distribution is not only due to the instrument sensitivity and burst search algorithm, but also to the difference of the duration and spectral shape of bursts. We also find that the bursts with preferred CPL model mostly occurred in the later epoch of this activity period.

An interesting single-pulse burst is found to be similar with the time-integrated spectrum of FRB 200428-Associated Burst. However, it is different from the spectrum of the peak of the FRB 200428-Associated Burst which is directly related to the FRB. The time-resolved analysis of bursts is also crucial to study the detailed physics. We leave it for the future work.

\clearpage

\begin{longrotatetable}
\centering
\begin{deluxetable*}{lcccccccccccccc}
\tablecaption{SGR J1935+2154 burst list detected by \textit{Insight}-HXMT. \label{bust_table}}
\tablewidth{2pt}
\tabletypesize{\scriptsize}
\tablehead{
\multicolumn{2}{c}{Burst Information} & \multicolumn{3}{c}{CPL or PL} & 
\multicolumn{3}{c}{BB+BB or BB} & \multicolumn{3}{c}{BB+PL} & \multicolumn{2}{c}{Two Components} & \multicolumn{1}{c}{Selected}\\
\hline
\colhead{ID} & 
\colhead{Trigger Time} & \colhead{$\Gamma$} & 
\colhead{$E_{\rm peak}$} &  \colhead{C-Stat/DoF$^1$} & \colhead{$kT_{\rm low}$} & 
\colhead{$kT_{\rm high}$} &  \colhead{C-Stat/DoF$^2$} & \colhead{$\Gamma$} & \colhead{$kT$} & \colhead{C-Stat/DoF$^3$} &\colhead{$Flux^4_{1}$} &\colhead{$Flux^4_{2}$}  &\colhead{Model$^5$}\\ 
\colhead{} & \colhead{(UTC)} &  \colhead{} & 
\colhead{(keV)} & \colhead{} & \colhead{(keV)} &
\colhead{(keV)} & \colhead{} & \colhead{} & \colhead{(keV)} & \colhead{}  &\colhead{} &\colhead{} &\colhead{}}
%($10^{-7}$ erg $cm^{-2}$ $s^{-1}$)
\startdata
1$^a$	&	2020-04-28T08:03:34.300	&	$0.61_{-0.09}^{+0.08}$	&	$26.73_{-1.43}^{+1.53}$	&	$91.1/79$	&	\nodata	&	\nodata	&	\nodata	&	\nodata	&	\nodata	&	\nodata	&	$8.56_{-0.26}^{+0.26}$	&	\nodata	&	CPL	\\
2	&	2020-04-28T08:05:50.080	&	\nodata	&	\nodata	&	\nodata	&	$1.94_{-0.08}^{+0.08}$	&	$7.19_{-0.47}^{+0.5}$	&	$96.59/107$	&	\nodata	&	\nodata	&	\nodata	&	$0.92_{-0.04}^{+0.04}$	&	$0.43_{-0.02}^{+0.02}$	&	BB+BB	\\
3	&	2020-04-28T08:14:45.985	&	$2.1_{-0.17}^{+0.16}$	&	\nodata	&	$22.99/21$	&	$4.95_{-0.49}^{+0.56}$	&	\nodata	&	$28.93/21$	&	\nodata	&	\nodata	&	\nodata	&	$1.43_{-0.24}^{+0.28}$	&	\nodata	&	PL	\\
4	&	2020-04-28T09:08:44.280	&	$0.94_{-0.2}^{+0.18}$	&	$32.05_{-3.28}^{+4.2}$	&	$85.97/87$	&	$2.68_{-0.34}^{+0.48}$	&	$11.39_{-1.0}^{+1.13}$	&	$89.5/86$	&	$1.65_{-0.12}^{+0.11}$	&	$6.03_{-0.67}^{+0.85}$	&	$86/86$	&	$0.6_{-0.04}^{+0.05}$	&	\nodata	&	CPL	\\
5	&	2020-04-28T09:40:10.980	&	$0.61_{-0.48}^{+0.47}$	&	$12.78_{-1.36}^{+1.28}$	&	$49.45/29$	&	$2.44_{-0.25}^{+0.26}$	&	$11.23_{-3.7}^{+4.06}$	&	$41.41/28$	&	$1.72_{-0.32}^{+0.27}$	&	$2.57_{-0.28}^{+0.34}$	&	$41.44/28$	&	$0.63_{-0.08}^{+0.1}$	&	$0.16_{-0.03}^{+0.04}$	&	BB+BB	\\
6	&	2020-04-28T09:46:05.300	&	\nodata	&	\nodata	&	\nodata	&	$1.75_{-0.17}^{+0.18}$	&	$11.4_{-2.59}^{+3.47}$	&	$29.19/30$	&	\nodata	&	\nodata	&	\nodata	&	$0.2_{-0.02}^{+0.03}$	&	$0.08_{-0.02}^{+0.02}$	&	BB+BB	\\
7$^b$	&	2020-04-28T09:51:04.634	&	$0.87_{-0.07}^{+0.07}$	&	$17.74_{-0.75}^{+0.76}$	&	$715.90/595$ 	&	\nodata	&	\nodata	&	\nodata	&	\nodata	&	\nodata	&	\nodata	&	$3.79_{-0.05}^{+0.06}$	&	\nodata	&	CPL	\\
8	&	2020-04-28T09:51:39.394	&	$2.06_{-0.16}^{+0.15}$	&	\nodata	&	$10.08/13$	&	\nodata	&	\nodata	&	\nodata	&	\nodata	&	\nodata	&	\nodata	&	$1.42_{-0.13}^{+0.13}$	&	\nodata	&	PL	\\
9	&	2020-04-28T10:54:23.850	&	\nodata	&	\nodata	&	\nodata	&	$5.03_{-0.97}^{+0.82}$	&	$12.19_{-2.73}^{+5.2}$	&	$16.1/18$	&	$1.77_{-0.41}^{+0.31}$	&	$6.44_{-0.63}^{+0.76}$	&	$16.91/18$	&	$1.15_{-0.2}^{+0.22}$	&	$0.53_{-0.11}^{+0.13}$	&	BB+BB	\\
10	&	2020-04-28T11:12:58.520	&	$1.82_{-0.09}^{+0.09}$	&	\nodata	&	$12.86/19$	&	\nodata	&	\nodata	&	\nodata	&	\nodata	&	\nodata	&	\nodata	&	$1.38_{-0.13}^{+0.15}$	&	\nodata	&	PL	\\
11	&	2020-04-28T11:24:28.120	&	$1.72_{-0.24}^{+0.21}$	&	\nodata	&	$6.5/5$	&	$6.97_{-1.02}^{+1.39}$	&	\nodata	&	$6.47/5$	&	\nodata	&	\nodata	&	\nodata	&	$0.6_{-0.15}^{+0.16}$	&	\nodata	&	BB	\\
12	&	2020-04-28T11:30:36.180	&	\nodata	&	\nodata	&	\nodata	&	$11.21_{-1.46}^{+1.7}$	&	\nodata	&	$14.86/21$	&	\nodata	&	\nodata	&	\nodata	&	$0.67_{-0.11}^{+0.12}$	&	\nodata	&	BB	\\
13	&	2020-04-28T14:20:52.519	&	\nodata	&	\nodata	&	\nodata	&	$1.36_{-0.09}^{+0.1}$	&	$4.66_{-0.33}^{+0.4}$	&	$90.28/68$	&	$2.29_{-0.08}^{+0.08}$	&	$1.89_{-0.11}^{+0.12}$	&	$83.59/68$	&	$0.44_{-0.05}^{+0.04}$	&	$0.48_{-0.05}^{+0.05}$	&	BB+PL	\\
14	&	2020-04-28T14:20:57.900	&	$0.98_{-0.17}^{+0.15}$	&	$50.32_{-5.11}^{+6.82}$	&	$54.98/50$	&	$3.09_{-0.47}^{+0.55}$	&	$15.29_{-0.94}^{+1.12}$	&	$50.83/49$	&	\nodata	&	\nodata	&	\nodata	&	$0.71_{-0.11}^{+0.12}$	&	$1.33_{-0.09}^{+0.09}$	&	BB+BB	\\
15$^*$	&	2020-04-28T14:34:24.000	&	$1.56_{-0.06}^{+0.06}$	&	$37.00_{-3.32}^{+3.99}$	&	-	&	$1.63_{-0.04}^{+0.04}$	&	$14.46_{-0.24}^{+0.25}$	&	-	&	$1.93_{-0.04}^{+0.04}$	& $11.32_{-0.56}^{+0.55}$	&	\nodata	&	$5.95_{-0.32}^{+0.34}$	&	\nodata	&	CPL	\\
16	&	2020-04-28T17:15:26.237	&	$1.22_{-0.21}^{+0.2}$	&	$48.01_{-7.8}^{+13.39}$	&	$32.54/19$	&	$3.58_{-0.53}^{+0.5}$	&	$16.54_{-1.6}^{+1.98}$	&	$29.43/18$	&	$1.59_{-0.13}^{+0.12}$	&	$4.97_{-0.99}^{+2.12}$	&	$35.58/18$	&	$0.52_{-0.09}^{+0.1}$	&	$0.58_{-0.06}^{+0.05}$	&	BB+BB	\\
17	&	2020-04-28T19:00:29.948	&	$1.65_{-0.22}^{+0.19}$	&	\nodata	&	$4.57/5$	&	\nodata	&	\nodata	&	\nodata	&	\nodata	&	\nodata	&	\nodata	&	$0.64_{-0.11}^{+0.13}$	&	\nodata	&	PL	\\
18	&	2020-04-28T19:01:59.850	&	$0.94_{-0.12}^{+0.12}$	&	$33.78_{-1.95}^{+2.17}$	&	$34.65/35$	&	$2.26_{-0.19}^{+0.21}$	&	$11.63_{-0.45}^{+0.47}$	&	$31.18/34$	&	$2.16_{-0.1}^{+0.13}$	&	$10.7_{-0.61}^{+0.61}$	&	$36.74/34$	&	$4.53_{-0.19}^{+0.2}$	&	\nodata	&	CPL	\\
19	&	2020-04-29T00:17:40.942	&	\nodata	&	\nodata	&	\nodata	&	$12.85_{-1.46}^{+1.74}$	&	\nodata	&	$14.58/15$	&	\nodata	&	\nodata	&	\nodata	&	$0.9_{-0.12}^{+0.14}$	&	\nodata	&	BB	\\
20	&	2020-04-29T11:12:39.397	&	$1.96_{-0.2}^{+0.2}$	&	\nodata	&	$28.11/13$	&	$10.59_{-1.2}^{+1.37}$	&	\nodata	&	$26.77/13$	&	\nodata	&	\nodata	&	\nodata	&	$0.51_{-0.07}^{+0.07}$	&	\nodata	&	BB	\\
21	&	2020-04-29T11:13:57.650	&	$0.52_{-0.1}^{+0.1}$	&	$34.83_{-1.29}^{+1.4}$	&	$73.7/76$	&	$2.17_{-0.2}^{+0.21}$	&	$10.54_{-0.32}^{+0.33}$	&	$69.52/75$	&	$1.96_{-0.09}^{+0.1}$	&	$9.63_{-0.39}^{+0.39}$	&	$73.25/75$	&	$0.65_{-0.02}^{+0.02}$	&	\nodata	&	CPL	\\
22	&	2020-04-30T09:25:22.750	&	\nodata	&	\nodata	&	\nodata	&	$13.01_{-1.78}^{+2.12}$	&	\nodata	&	$23.38/19$	&	\nodata	&	\nodata	&	\nodata	&	$0.26_{-0.04}^{+0.05}$	&	\nodata	&	BB	\\
23	&	2020-04-30T15:41:53.947	&	$1.53_{-0.17}^{+0.17}$	&	\nodata	&	$15.08/16$	&	$15.66_{-1.92}^{+2.29}$	&	\nodata	&	$21.3/16$	&	\nodata	&	\nodata	&	\nodata	&	$0.72_{-0.1}^{+0.11}$	&	\nodata	&	PL	\\
24	&	2020-04-30T17:12:52.837	&	$1.22_{-0.11}^{+0.11}$	&	$24.74_{-1.57}^{+1.72}$	&	$38.68/42$	&	$2.36_{-0.15}^{+0.17}$	&	$10.97_{-0.54}^{+0.57}$	&	$41.85/41$	&	\nodata	&	\nodata	&	\nodata	&	$1.84_{-0.08}^{+0.09}$	&	\nodata	&	CPL	\\
25	&	2020-05-01T15:05:56.635	&	$1.86_{-0.08}^{+0.09}$	&	\nodata	&	$29.07/31$	&	\nodata	&	\nodata	&	\nodata	&	\nodata	&	\nodata	&	\nodata	&	$2.53_{-0.2}^{+0.23}$	&	\nodata	&	PL	\\
26	&	2020-05-01T15:15:20.876	&	$1.9_{-0.13}^{+0.13}$	&	\nodata	&	$28.66/16$	&	$8.62_{-0.74}^{+0.82}$	&	\nodata	&	$23.34/16$	&	\nodata	&	\nodata	&	\nodata	&	$1.32_{-0.16}^{+0.16}$	&	\nodata	&	BB	\\
27	&	2020-05-02T05:40:53.151	&	$1.85_{-0.15}^{+0.14}$	&	\nodata	&	$6.79/9$	&	\nodata	&	\nodata	&	\nodata	&	\nodata	&	\nodata	&	\nodata	&	$1.34_{-0.22}^{+0.25}$	&	\nodata	&	PL	\\
28	&	2020-05-02T10:17:26.000	&	$1.52_{-0.19}^{+0.17}$	&	$40.07_{-7.4}^{+12.65}$	&	$28.68/36$	&	$4.31_{-0.33}^{+0.33}$	&	$21.39_{-2.82}^{+3.49}$	&	$29.2/35$	&	$1.72_{-0.15}^{+0.12}$	&	$4.88_{-0.58}^{+0.76}$	&	$22.31/35$	&	$1.4_{-0.37}^{+0.37}$	&	$3.94_{-0.47}^{+0.51}$	&	BB+PL	\\
29	&	2020-05-02T10:25:25.777	&	\nodata	&	\nodata	&	\nodata	&	$1.54_{-0.17}^{+0.17}$	&	$15.5_{-1.09}^{+1.2}$	&	$70.25/66$	&	$1.82_{-0.12}^{+0.16}$	&	$15.74_{-4.46}^{+4.08}$	&	$68.41/66$	&	$0.09_{-0.03}^{+0.03}$	&	$0.46_{-0.05}^{+0.06}$	&	BB+PL	\\
30	&	2020-05-02T10:46:20.850	&	$1.35_{-0.19}^{+0.18}$	&	\nodata	&	$8.01/9$	&	\nodata	&	\nodata	&	\nodata	&	\nodata	&	\nodata	&	\nodata	&	$0.73_{-0.1}^{+0.11}$	&	\nodata	&	PL	\\
31	&	2020-05-03T04:30:59.050	&	\nodata	&	\nodata	&	\nodata	&	$12.53_{-1.51}^{+1.75}$	&	\nodata	&	$29.81/16$	&	\nodata	&	\nodata	&	\nodata	&	$0.61_{-0.08}^{+0.09}$	&	\nodata	&	BB	\\
32	&	2020-05-03T17:12:55.600	&	$1.65_{-0.23}^{+0.22}$	&	\nodata	&	$21.57/27$	&	\nodata	&	\nodata	&	\nodata	&	\nodata	&	\nodata	&	\nodata	&	$0.63_{-0.09}^{+0.08}$	&	\nodata	&	PL	\\
33$^c$	&	2020-05-03T23:25:13.250	&	$0.70_{-0.10}^{+0.11}$	&	$30.49_{-1.14}^{+1.08}$	&	$306.84/281$	&	$4.98_{-0.16}^{+0.17}$	&	$13.27_{-0.39}^{+0.42}$	&	$308.77/280$	&	\nodata	&	\nodata	&	\nodata	&	$5.09_{-0.06}^{+0.07}$	&	\nodata	&	CPL	\\
34	&	2020-05-04T00:48:07.343	&	$1.73_{-0.06}^{+0.06}$	&	\nodata	&	$49.25/42$	&	\nodata	&	\nodata	&	\nodata	&	\nodata	&	\nodata	&	\nodata	&	$0.98_{-0.06}^{+0.06}$	&	\nodata	&	PL	\\
35	&	2020-05-04T13:20:00.700	&	$1.37_{-0.18}^{+0.16}$	&	\nodata	&	$13.69/11$	&	$17.93_{-2.18}^{+2.66}$	&	\nodata	&	$18.29/11$	&	\nodata	&	\nodata	&	\nodata	&	$0.66_{-0.09}^{+0.08}$	&	\nodata	&	PL	\\
36	&	2020-05-05T02:30:28.450	&	\nodata	&	\nodata	&	\nodata	&	$15.23_{-2.45}^{+3.11}$	&	\nodata	&	$16.35/24$	&	\nodata	&	\nodata	&	\nodata	&	$0.31_{-0.06}^{+0.06}$	&	\nodata	&	BB	\\
37	&	2020-05-05T12:09:29.750	&	$1.82_{-0.15}^{+0.14}$	&	\nodata	&	$22.72/37$	&	\nodata	&	\nodata	&	\nodata	&	\nodata	&	\nodata	&	\nodata	&	$3.01_{-0.26}^{+0.28}$	&	\nodata	&	PL	\\
38	&	2020-05-06T21:25:16.350	&	\nodata	&	\nodata	&	\nodata	&	$10.17_{-0.66}^{+0.7}$	&	\nodata	&	$84.25/68$	&	\nodata	&	\nodata	&	\nodata	&	$0.33_{-0.03}^{+0.03}$	&	\nodata	&	BB	\\
39	&	2020-05-06T22:48:21.550	&	$1.63_{-0.26}^{+0.24}$	&	\nodata	&	$4.5/4$	&	\nodata	&	\nodata	&	\nodata	&	\nodata	&	\nodata	&	\nodata	&	$0.76_{-0.14}^{+0.15}$	&	\nodata	&	PL	\\
40	&	2020-05-07T21:05:41.345	&	$1.47_{-0.18}^{+0.17}$	&	$31.15_{-4.94}^{+6.38}$	&	$37.78/31$	&	$2.38_{-0.24}^{+0.25}$	&	$13.51_{-1.01}^{+1.14}$	&	$26.69/30$	&	\nodata	&	\nodata	&	\nodata	&	$1.38_{-0.2}^{+0.22}$	&	$1.39_{-0.12}^{+0.13}$	&	BB+BB	\\
41	&	2020-05-08T06:17:16.589	&	$1.3_{-0.1}^{+0.1}$	&	$33.24_{-2.38}^{+2.75}$	&	$58.77/46$	&	\nodata	&	\nodata	&	\nodata	&	$1.83_{-0.06}^{+0.06}$	&	$6.02_{-0.58}^{+0.7}$	&	$64.73/45$	&	$3.34_{-0.13}^{+0.14}$	&	\nodata	&	CPL	\\
42	&	2020-05-08T09:17:05.185	&	$1.82_{-0.16}^{+0.15}$	&	\nodata	&	$20.41/15$	&	\nodata	&	\nodata	&	\nodata	&	\nodata	&	\nodata	&	\nodata	&	$2.17_{-0.27}^{+0.3}$	&	\nodata	&	PL	\\
43	&	2020-05-08T09:49:21.134	&	$1.21_{-0.23}^{+0.22}$	&	$46.68_{-8.29}^{+15.31}$	&	$25.29/23$	&	\nodata	&	\nodata	&	\nodata	&	$1.82_{-0.18}^{+0.25}$	&	$11.22_{-3.8}^{+3.27}$	&	$26.22/22$	&	$1.21_{-0.11}^{+0.11}$	&	\nodata	&	CPL	\\
44	&	2020-05-08T19:23:36.028	&	$1.81_{-0.13}^{+0.12}$	&	\nodata	&	$18.34/13$	&	\nodata	&	\nodata	&	\nodata	&	\nodata	&	\nodata	&	\nodata	&	$1.21_{-0.15}^{+0.16}$	&	\nodata	&	PL	\\
45	&	2020-05-08T19:37:25.270	&	$2.04_{-0.26}^{+0.26}$	&	\nodata	&	$12.69/16$	&	\nodata	&	\nodata	&	\nodata	&	\nodata	&	\nodata	&	\nodata	&	$2.13_{-0.31}^{+0.34}$	&	\nodata	&	PL	\\
46	&	2020-05-09T01:56:38.750	&	\nodata	&	\nodata	&	\nodata	&	$14.78_{-1.16}^{+1.3}$	&	\nodata	&	$28.52/32$	&	\nodata	&	\nodata	&	\nodata	&	$0.65_{-0.06}^{+0.06}$	&	\nodata	&	BB	\\
47	&	2020-05-10T05:00:28.195	&	$2.27_{-0.08}^{+0.09}$	&	\nodata	&	$100.96/80$	&	$8.39_{-0.4}^{+0.42}$	&	\nodata	&	$100.41/80$	&	\nodata	&	\nodata	&	\nodata	&	$0.58_{-0.03}^{+0.03}$	&	\nodata	&	BB	\\
48$^d$	&	2020-05-10T06:12:01.622	&	$0.07_{-0.06}^{+0.06}$	&	$45.88_{-0.53}^{+0.53}$	&	$410.74/378$	&	\nodata	&	\nodata	&	\nodata	&	\nodata	&	\nodata	&	\nodata	&	$42.03_{-0.31}^{+0.31}$	&	\nodata	&	CPL	\\
49	&	2020-05-10T06:16:41.100	&	$1.16_{-0.15}^{+0.15}$	&	$44.68_{-5.6}^{+8.17}$	&	$32.35/42$	&	$2.56_{-0.24}^{+0.28}$	&	$14.13_{-0.92}^{+1.0}$	&	$30.1/41$	&	$1.82_{-0.11}^{+0.13}$	&	$11.93_{-2.19}^{+1.92}$	&	$37.98/41$	&	$0.41_{-0.03}^{+0.02}$	&	\nodata	&	CPL	\\
50	&	2020-05-10T06:20:09.400	&	$1.31_{-0.18}^{+0.17}$	&	$37.01_{-4.89}^{+7.06}$	&	$27.9/23$	&	\nodata	&	\nodata	&	\nodata	&	$1.89_{-0.11}^{+0.13}$	&	$8.76_{-1.75}^{+2.03}$	&	$28.23/22$	&	$1.2_{-0.08}^{+0.08}$	&	\nodata	&	CPL	\\
51	&	2020-05-10T08:55:46.300	&	$1.17_{-0.13}^{+0.12}$	&	$38.29_{-3.47}^{+4.39}$	&	$27.44/30$	&	$2.32_{-0.2}^{+0.23}$	&	$12.92_{-0.66}^{+0.7}$	&	$31.97/29$	&	\nodata	&	\nodata	&	\nodata	&	$1.5_{-0.08}^{+0.08}$	&	\nodata	&	CPL	\\
52	&	2020-05-10T18:53:01.040	&	\nodata	&	\nodata	&	\nodata	&	$2.53_{-0.59}^{+0.63}$	&	$9.92_{-0.74}^{+0.82}$	&	$79.92/65$	&	$1.82_{-0.19}^{+0.29}$	&	$8.87_{-0.71}^{+0.73}$	&	$84.35/65$	&	$0.03_{-0.0}^{+0.01}$	&	$0.12_{-0.01}^{+0.01}$	&	BB+BB	\\
53	&	2020-05-10T20:16:22.000	&	\nodata	&	\nodata	&	\nodata	&	$2.74_{-0.5}^{+0.63}$	&	$10.56_{-1.01}^{+1.13}$	&	$72.28/50$	&	$1.72_{-0.21}^{+0.29}$	&	$8.84_{-0.99}^{+1.11}$	&	$74.59/50$	&	$0.05_{-0.02}^{+0.01}$	&	$0.15_{-0.02}^{+0.01}$	&	BB+BB	\\
54$^e$	&	2020-05-10T21:51:16.221	&	$-0.08_{-0.05}^{+0.05}$	&	$33.64_{-0.39}^{+0.40}$	&	$532.96/529$	&	\nodata	&	\nodata	&	\nodata	&	\nodata	&	\nodata	&	\nodata	&	$22.22_{-0.18}^{+0.18}$	&	\nodata	&	CPL	\\
55	&	2020-05-10T22:08:09.000	&	$1.7_{-0.17}^{+0.17}$	&	\nodata	&	$8.84/12$	&	\nodata	&	\nodata	&	\nodata	&	\nodata	&	\nodata	&	\nodata	&	$2.79_{-0.35}^{+0.37}$	&	\nodata	&	PL	\\
56	&	2020-05-11T04:22:52.560	&	$0.59_{-0.34}^{+0.31}$	&	$16.32_{-1.2}^{+1.15}$	&	$16.08/16$	&	$2.77_{-0.38}^{+0.44}$	&	$7.36_{-0.94}^{+1.19}$	&	$16.59/15$	&	$2.35_{-0.18}^{+0.18}$	&	$4.65_{-0.33}^{+0.38}$	&	$16.7/15$	&	$2.29_{-0.17}^{+0.18}$	&	\nodata	&	CPL	\\
57	&	2020-05-11T17:15:43.320	&	\nodata	&	\nodata	&	\nodata	&	$3.42_{-0.29}^{+0.29}$	&	$13.64_{-1.3}^{+1.54}$	&	$27.12/31$	&	\nodata	&	\nodata	&	\nodata	&	$0.89_{-0.08}^{+0.09}$	&	$0.52_{-0.04}^{+0.05}$	&	BB+BB	\\
58	&	2020-05-12T08:35:19.700	&	$0.53_{-0.26}^{+0.23}$	&	$29.77_{-2.19}^{+2.46}$	&	$17.72/17$	&	$2.8_{-0.54}^{+1.06}$	&	$9.97_{-0.71}^{+1.16}$	&	$20.27/16$	&	$1.98_{-0.17}^{+0.19}$	&	$7.96_{-0.81}^{+0.87}$	&	$22.45/16$	&	$1.57_{-0.1}^{+0.11}$	&	\nodata	&	CPL	\\
59	&	2020-05-12T21:47:43.340	&	$1.29_{-0.25}^{+0.25}$	&	\nodata	&	$18.97/15$	&	$20.46_{-3.57}^{+4.88}$	&	\nodata	&	$23.84/15$	&	\nodata	&	\nodata	&	\nodata	&	$1.00_{-0.19}^{+0.22}$	&	\nodata	&	PL	\\
60	&	2020-05-13T07:12:57.543	&	$0.92_{-0.31}^{+0.28}$	&	$14.7_{-1.53}^{+1.43}$	&	$32.84/36$	&	\nodata	&	\nodata	&	\nodata	&	\nodata	&	\nodata	&	\nodata	&	$1.02_{-0.1}^{+0.1}$	&	\nodata	&	CPL	\\
61	&	2020-05-14T14:49:22.000	&	$0.63_{-0.08}^{+0.07}$	&	$33.46_{-1.0}^{+1.05}$	&	$88.06/80$	&	\nodata	&	\nodata	&	\nodata	&	\nodata	&	\nodata	&	\nodata	&	$1.77_{-0.04}^{+0.05}$	&	\nodata	&	CPL	\\
62	&	2020-05-16T01:50:23.542	&	$1.19_{-0.14}^{+0.13}$	&	$36.1_{-3.42}^{+4.36}$	&	$37.45/34$	&	\nodata	&	\nodata	&	\nodata	&	\nodata	&	\nodata	&	\nodata	&	$1.66_{-0.09}^{+0.09}$	&	\nodata	&	CPL	\\
63	&	2020-05-16T10:26:32.309	&	$1.61_{-0.16}^{+0.14}$	&	\nodata	&	$12.81/19$	&	\nodata	&	\nodata	&	\nodata	&	\nodata	&	\nodata	&	\nodata	&	$2.16_{-0.26}^{+0.28}$	&	\nodata	&	PL	\\
64	&	2020-05-16T11:16:17.000	&	$0.99_{-0.13}^{+0.12}$	&	$40.57_{-2.85}^{+3.33}$	&	$37.06/39$	&	$3.43_{-0.3}^{+0.32}$	&	$14.46_{-0.78}^{+0.89}$	&	$32.72/38$	&	\nodata	&	\nodata	&	\nodata	&	$0.78_{-0.09}^{+0.08}$	&	$0.94_{-0.05}^{+0.06}$	&	BB+BB	\\
65	&	2020-05-16T18:12:52.080	&	$1.27_{-0.08}^{+0.08}$	&	$35.47_{-2.1}^{+2.39}$	&	$63.17/64$	&	\nodata	&	\nodata	&	\nodata	&	\nodata	&	\nodata	&	\nodata	&	$4.05_{-0.13}^{+0.14}$	&	\nodata	&	CPL	\\
66	&	2020-05-17T03:18:10.320	&	$1.95_{-0.17}^{+0.16}$	&	\nodata	&	$22.03/19$	&	\nodata	&	\nodata	&	\nodata	&	\nodata	&	\nodata	&	\nodata	&	$2.1_{-0.23}^{+0.26}$	&	\nodata	&	PL	\\
67	&	2020-05-18T01:54:21.550	&	\nodata	&	\nodata	&	\nodata	&	$17.8_{-2.51}^{+3.12}$	&	\nodata	&	$188/23$	&	\nodata	&	\nodata	&	\nodata	&	$0.38_{-0.06}^{+0.07}$	&	\nodata	&	BB	\\
68	&	2020-05-18T05:17:57.715	&	$1.86_{-0.11}^{+0.11}$	&	\nodata	&	$17.62/14$	&	\nodata	&	\nodata	&	\nodata	&	\nodata	&	\nodata	&	\nodata	&	$1.99_{-0.25}^{+0.27}$	&	\nodata	&	PL	\\
69	&	2020-05-18T09:27:59.151	&	$1.22_{-0.31}^{+0.26}$	&	\nodata	&	$4.88/11$	&	\nodata	&	\nodata	&	\nodata	&	\nodata	&	\nodata	&	\nodata	&	$0.3_{-0.05}^{+0.06}$	&	\nodata	&	PL	\\
70	&	2020-05-18T11:00:41.150	&	\nodata	&	\nodata	&	\nodata	&	$12.96_{-2.24}^{+2.95}$	&	\nodata	&	$14.56/12$	&	\nodata	&	\nodata	&	\nodata	&	$0.8_{-0.16}^{+0.17}$	&	\nodata	&	BB	\\
71	&	2020-05-18T16:28:18.300	&	\nodata	&	\nodata	&	\nodata	&	$3.92_{-0.57}^{+0.59}$	&	$20.71_{-3.73}^{+5.44}$	&	$47.85/28$	&	$1.39_{-0.36}^{+0.26}$	&	$4.08_{-0.89}^{+1.67}$	&	$485/28$	&	$0.35_{-0.09}^{+0.1}$	&	$0.31_{-0.05}^{+0.06}$	&	BB+BB	\\
72	&	2020-05-19T18:57:36.300	&	$0.93_{-0.16}^{+0.15}$	&	$35.1_{-2.38}^{+2.68}$	&	$31.64/29$	&	$2.82_{-0.27}^{+0.31}$	&	$12.33_{-0.58}^{+0.63}$	&	$28.9/28$	&	\nodata	&	\nodata	&	\nodata	&	$6.08_{-0.29}^{+0.31}$	&	\nodata	&	CPL	\\
73	&	2020-05-20T14:10:49.780	&	$1.04_{-0.08}^{+0.07}$	&	$36.32_{-1.59}^{+1.76}$	&	$76.14/71$	&	\nodata	&	\nodata	&	\nodata	&	\nodata	&	\nodata	&	\nodata	&	$4.19_{-0.12}^{+0.13}$	&	\nodata	&	CPL	\\
74$^f$	&	2020-05-20T21:47:07.480	&	$0.69_{-0.03}^{+0.04}$	&	$34.17_{-0.73}^{+0.74}$	&	$466.60.1/314$	&	\nodata	&	\nodata	&	\nodata	&	\nodata	&	\nodata	&	\nodata	&	$6.02_{-0.07}^{+0.07}$	&	\nodata	&	CPL	\\
75	&	2020-05-24T22:05:03.480	&	\nodata	&	\nodata	&	\nodata	&	$5.07_{-2.44}^{+2.03}$	&	$13.1_{-2.12}^{+3.59}$	&	$18/24$	&	\nodata	&	\nodata	&	\nodata	&	$0.34_{-0.13}^{+0.13}$	&	$0.54_{-0.09}^{+0.1}$	&	BB+BB	\\
% can't leave the space
\enddata
\tablecomments{$^1$ C-Stat for the CPL model fit or PL fit. \\
$^2$ C-Stat for the BB+BB model fit or BB fit. \\
$^3$ C-Stat for the BB+PL model fit. \\
$^4$ flux in 1$-$250 keV of different component of complex models ($10^{-7}$ erg $cm^{-2}$ $s^{-1}$ ). \\
$^5$ The model we selected was used to calculate flux. A, B, C, D, E represent CPL, BB+BB, BB+PL, PL, BB models. \\
$^*$ The parameters of the FRB 200428-Associated Burst are from \cite{2021NatAs...5..378L}. (CPL: $factor_{\rm ME} = 0.98_{-0.06}^{+0.07}$, $factor_{\rm HE} = 0.68_{-0.07}^{+0.07}$; BB+PL: $factor_{\rm ME} = 1.05_{-0.07}^{+0.08}$ , $factor_{\rm HE} = 0.54_{-0.06}^{+0.07}$ )\\
$^a$ Burst of which the HE data suffered from saturation with constant of $factor_{\rm HE} = 0.39_{-0.05}^{+0.04}$. \\
$^b$ Burst of which the LE and HE data suffered from saturation with constant of $factor_{\rm LE} = 0.72_{-0.05}^{+0.05}$ and $factor_{\rm HE} = 0.64_{-0.04}^{+0.04}$. \\
$^c$ Burst of which the HE data suffered from saturation with constant of $factor_{\rm HE} = 0.44_{-0.03}^{+0.02}$ for the CPL model (BB+BB: $factor_{\rm HE} = 0.51_{-0.04}^{+0.03}$). \\
$^d$ Burst of which the HE data suffered from saturation with constant of  $factor_{\rm HE} = 0.15_{-0.01}^{+0.01}$. \\
$^e$ Burst of which the LE and HE data suffered from saturation with constant of  $factor_{\rm LE} = 1.05_{-0.05}^{+0.06}$ and $factor_{\rm HE} = 0.26_{-0.01}^{+0.01}$. \\
$^f$ Burst of which the HE data suffered from saturation with constant of $factor_{\rm HE} = 0.57_{-0.02}^{+0.03}$. \\}
\end{deluxetable*}
\end{longrotatetable}

%\appendix 
%\label{Appendix}

\clearpage
\section*{Acknowledgements}
\textcolor{red}{We thank the anonymous reviewer for very helpful comments and suggestions.}
We are grateful to Prof. Shri Kulkarni for suggesting the one-month observation after the initial ToO of SGR J1935+2154.
This work is supported by the National Key R\&D Program of China (2021YFA0718500).
The authors thank supports from the Strategic Priority Research Program on Space Science, the Chinese Academy of Sciences (Grant No. XDA15360300, XDA15052700, XDB23040400),  %% Xiongshaolin
and the National Natural Science Foundation of China under Grants (No. U1838201, U1838202, U1838113, U2038106, U1938201, 12133007, 11961141013).

\bibliography{ref}
\bibliographystyle{aasjournal}

\end{document}